\documentclass[12pt]{article}

\usepackage{scicite}
\usepackage{graphicx}
\usepackage{etoolbox}
\usepackage{hyperref}
\usepackage{times}
\usepackage{bm}
\usepackage{makecell}
\usepackage{color}
\usepackage{amssymb}
\usepackage{amsmath}
\usepackage{braket}
\usepackage[table]{xcolor}

\usepackage{soul}

\usepackage{titling}

\newenvironment{sciabstract}{%
\begin{quote} \bf}
{\end{quote}}

\usepackage{graphicx}

\graphicspath{ {figures/} }

\makeatletter
\let\saved@includegraphics\includegraphics
\AtBeginDocument{\let\includegraphics\saved@includegraphics}
\makeatother

\begin{document}

\title{Emergent hydrodynamics in a strongly interacting dipolar spin ensemble}

\author{
C.~Zu,$^{1,2,*}$
F.~Machado,$^{1,*}$ 
B.~Ye,$^{1,*}$ 
S.~Choi,$^1$
B.~Kobrin,$^{1,2}$
T.~Mittiga,$^{1,2}$\\
S.~Hsieh,$^{1,2}$
P.~Bhattacharyya,$^{1,2}$ 
M.~Markham, $^{3}$
D.~Twitchen, $^{3}$
A.~Jarmola, $^{1,4}$\\
D.~Budker, $^{1,5}$
C.~R.~Laumann, $^{6}$
J. E. Moore,$^{1,2}$
N. Y. Yao$^{1,2,\dag}$\\
\\
\normalsize{$^{1}$Department of Physics, University of California, Berkeley, CA 94720, USA}\\
\normalsize{$^{2}$Materials Science Division, Lawrence Berkeley National Laboratory, Berkeley, CA 94720, USA}\\
\normalsize{$^{3}$Element Six, Harwell, OX11 0QR, United Kingdom}\\
\normalsize{$^{4}$U.S. Army Research Laboratory, Adelphi, Maryland 20783, USA}\\
\normalsize{$^{5}$Helmholtz Institut Mainz, Johannes Gutenberg Universitat Mainz, 55128 Mainz, Germany}\\
\normalsize{$^{6}$Department of Physics, Boston University, Boston, MA 02215, USA}\\
\normalsize{$^*$These authors contributed equally to this work.}\\
\normalsize{$^\dag$To whom correspondence should be addressed; E-mail:  norman.yao@berkeley.edu}
}

\date{}

\baselineskip24pt
\maketitle

\vspace{5mm}

\begin{sciabstract}
Conventional wisdom holds that macroscopic classical phenomena naturally emerge from microscopic quantum laws \cite{ljubotina2017spin, bulchandani2018bethe, sommer2011universal, moll2016evidence, bandurin2016negative,crossno2016observation,cepellotti2015phonon,castro2016emergent}. However, despite this mantra, building direct connections between these two descriptions has remained an enduring scientific challenge. In particular, it is difficult to quantitatively predict the emergent ``classical'' properties of a system (e.g.~diffusivity, viscosity, compressibility) from a generic microscopic quantum Hamiltonian~ \cite{andreev2011hydrodynamic, castro2016emergent, bertini2016transport, bulchandani2017solvable,leviatan2017quantum,wurtz2018quantum,ye2019emergent,de2018hydrodynamic,bertini2020finite, jenkins2020imaging, ku2020imaging, kumar2017superballistic}. Here, we introduce a hybrid solid-state spin platform, where the underlying disordered, dipolar quantum Hamiltonian gives rise to the emergence of unconventional  spin diffusion at nanometer length scales. In particular, the combination of positional disorder and on-site random fields leads to diffusive dynamics that are Fickian yet  non-Gaussian  ~\cite{chechkin2017brownian,stylianidou2014cytoplasmic,kim2013simulation,chakraborty2020disorder,chubynsky2014diffusing,postnikov2020brownian,barkai2020packets}. Finally, by tuning the underlying parameters within the spin Hamiltonian via a combination of static and driven fields, we demonstrate direct control over the emergent spin diffusion coefficient. Our work opens the door to investigating hydrodynamics in many-body quantum spin systems.

\end{sciabstract}

Even in the absence of a precise microscopic description, \emph{classical} hydrodynamics provides a powerful framework for characterizing the macroscopic behavior of local, conserved quantities, such as energy. 
Understanding whether and how it emerges in the late-time dynamics of strongly-interacting \emph{quantum} systems remains an essential open question \cite{hartle2011quasiclassical,halliwell1999decoherent,greenbaum2005hydrodynamic}. 
For a quantum system coupled to an environment, it is  unsurprising that the late-time behavior becomes classical; indeed, dephasing from the bath destroys the intrinsic quantum coherences of the system. 
However, even for an isolated, many-particle quantum system, conventional wisdom holds that late-time dynamics usually exhibit an \emph{emergent} classical description; understanding how to prove this fact and the required conditions has remained an enduringly hard question~\cite{andreev2011hydrodynamic, castro2016emergent, bertini2016transport, bulchandani2017solvable,leviatan2017quantum,wurtz2018quantum,ye2019emergent,de2018hydrodynamic, bertini2020finite}.
At the same time, it has also motivated seminal advances: On the theoretical front, precise analytic insights have been obtained in the context of integrable systems using generalized hydrodynamics and non-integrable systems using perturbative approaches~\cite{de2018hydrodynamic,bertini2020finite,narozhny2015hydrodynamics,moore2011kubo,vznidarivc2019nonequilibrium,friedman2020diffusive}. 
On the experimental front, tremendous progress in time-resolved measurement techniques has enabled the direct observation of emergent classical diffusion in several classes of quantum systems~\cite{sommer2011universal,moll2016evidence,schemmer2019generalized,zhang1998first, pagliero2020optically, boutis2004spin, leppelmeier1968measurement, jv1977measurement,eberhardt2007direct,dikarov2016direct}. 

There are, however, a wide variety of classical dynamical ``universality classes'' other than diffusion: aside from the simple case of free (ballistic) behavior, two well-known classes are Kardar-Parisi-Zhang dynamics and Sinai diffusion \cite{agarwal2015anomalous,vznidarivc2016diffusive,kardar1986dynamic,kim1995bethe,gopalakrishnan2019kinetic,sinai1983limiting}.
%
Here, we report time-resolved experiments on a closed quantum system, which exhibits an unconventional approach to late-time diffusion characterized by a long-lived, non-Gaussian polarization profile.

\begin{figure*}
         \centering
         \includegraphics[width = 4.5in]{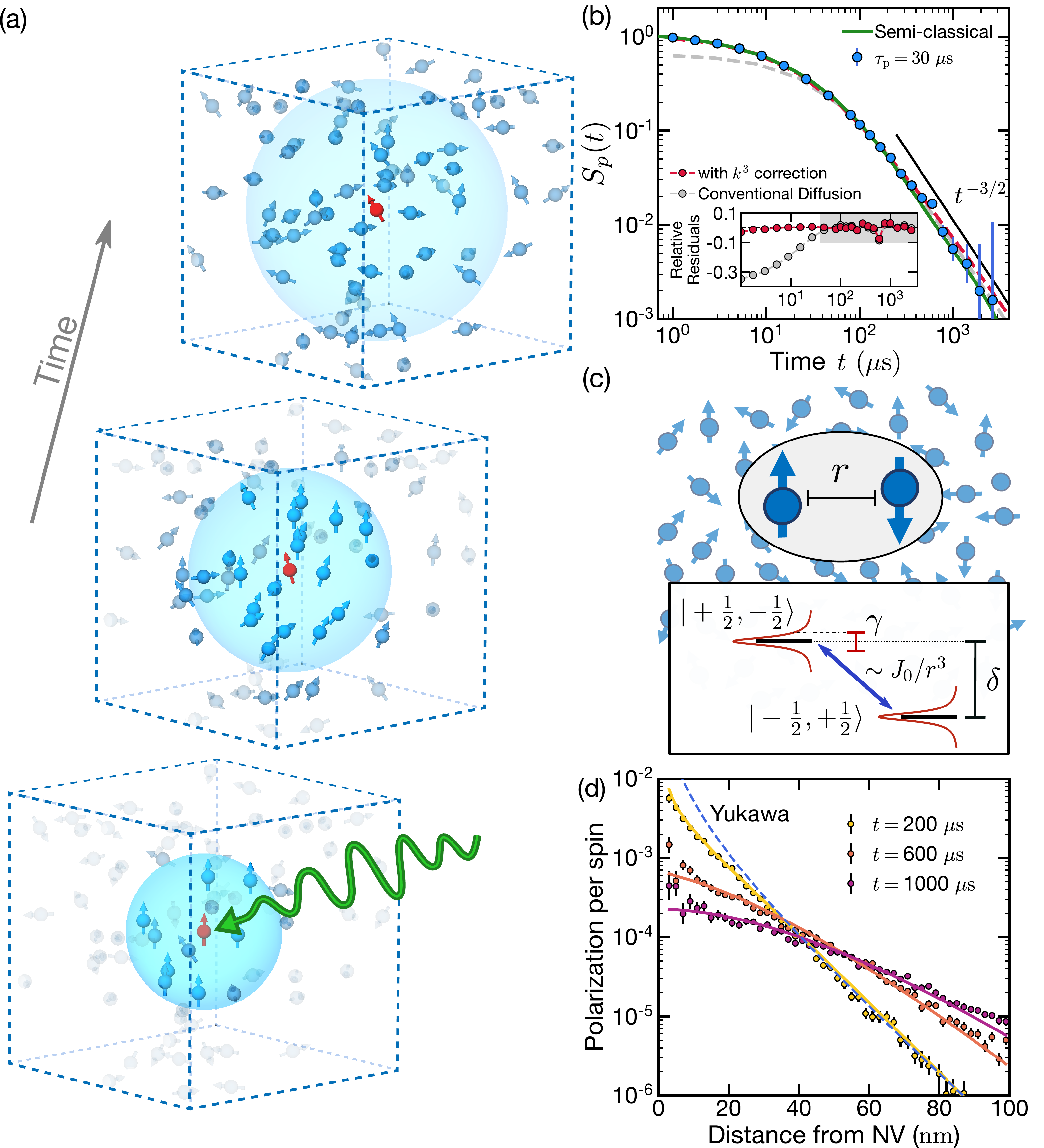}         \caption{\footnotesize
         {\bf Nanoscale spin diffusion in a long-range interacting quantum system.}      
           {\bf a,} Schematic depicting the emergence of hydrodynamics in a strongly interacting dipolar spin ensemble. 
           Optical pumping (green arrow) of the NV center (red) enables it to behave as a spin sink for nearby P1 centers (blue), resulting in the preparation of a local, inhomogenous spin-polarization profile. 
          Dynamics then lead to the spreading of this profile as a function of time. 
      {\bf b,} Dynamics of the survival probability $S_p(t)$  of the $\nu=1/3$ P1 subgroup in sample S2 at $T = 25$~K following a polarization period of $\tau_{\textrm{p}} = 30~\mu$s. 
      After an initial transient, $S_p(t)$ approaches a robust power-law decay $\sim t^{-3/2}$, indicating diffusion. 
      The late-time dynamics are accurately described by the conventional diffusion equation (gray dashed line).  
      (inset) Relative residuals when fitting with (red) or without (grey) an additional long-range correction $C_{\mathrm{lr}} k^3$.
      In the hydrodynamical regime (grey shaded region) both models capture the data equally well.
      Corresponding fits appear in full panel.
        {\bf c,} Illustration of our semi-classical description for the spin-polarization dynamics.
      Each pair of spins exchanges polarization via the dipolar interaction.
      The presence of other nearby P1 spins leads to an energy mismatch $\delta$ and a homogeneous broadening $\gamma$. 
       {\bf d,} 
       Initializing with unit polarization, a robust non-Gaussian polarization profile emerges from the semi-classical model for all experimentally accessible time-scales. 
       The  crossover from a Yukawa to Gaussian polarization profile is accurately captured by including the disorder-induced dynamical modification (see Methods), $C_{\mathrm{dyn}} k^2\partial_t P_k$, in the diffusion equation with $C_{\mathrm{dyn}} =  204 \pm 45~\mathrm{nm^3}$ (see Methods). 
      }
         \label{fig:fig1}
     \end{figure*}

%
%
%
Our experimental platform consists of two strongly-interacting species of dipolar electronic spins in diamond: substitutional nitrogen defects (P1 centers) and nitrogen-vacancy (NV) color centers \cite{doherty2013nitrogen,hall2016detection}.
%
By controlling the relative density of these two species, we demonstrate the ability to prepare inhomogeneous spatial profiles of a conserved spin density, as well as to locally probe the resulting nanoscale spin dynamics (Fig.~\ref{fig:fig1}). 
%
These dynamics can be tuned via three independent controls: 1) the initial spin polarization, 2) the average spacing between spins, and 3) the magnitude of the on-site random fields. 

Exploring this phase space leads us to an understanding of how the details of the microscopic spin Hamiltonian modify conventional diffusion.  By tracking the local autocorrelation function of the spin polarization, $S_p(t)$, we observe the emergence of a long-time, diffusive power-law, $S_p (t) \sim t^{-3/2}$,  for over an order of magnitude in time (Fig.~\ref{fig:fig1}b).  
However, the details of this autocorrelation function over a broad range of timescales indicate that, following local initialization, the spin polarization distribution remains non-Gaussian throughout the time-scales accessible in the experiment; this originates from the presence of strong disorder in our system, which leads to a \emph{distribution} of local diffusion coefficients and a Yukawa-like spin polarization profile (Fig.~\ref{fig:fig1}d). 
Indeed, we find that the disorder-averaged spin polarization is well described by the crossover from a Yukawa to Gaussian form,  which is naturally obtained by incorporating disorder-induced dynamical effects into the standard diffusion equation.

\begin{figure*}
         \centering
         \includegraphics[width = 6in]{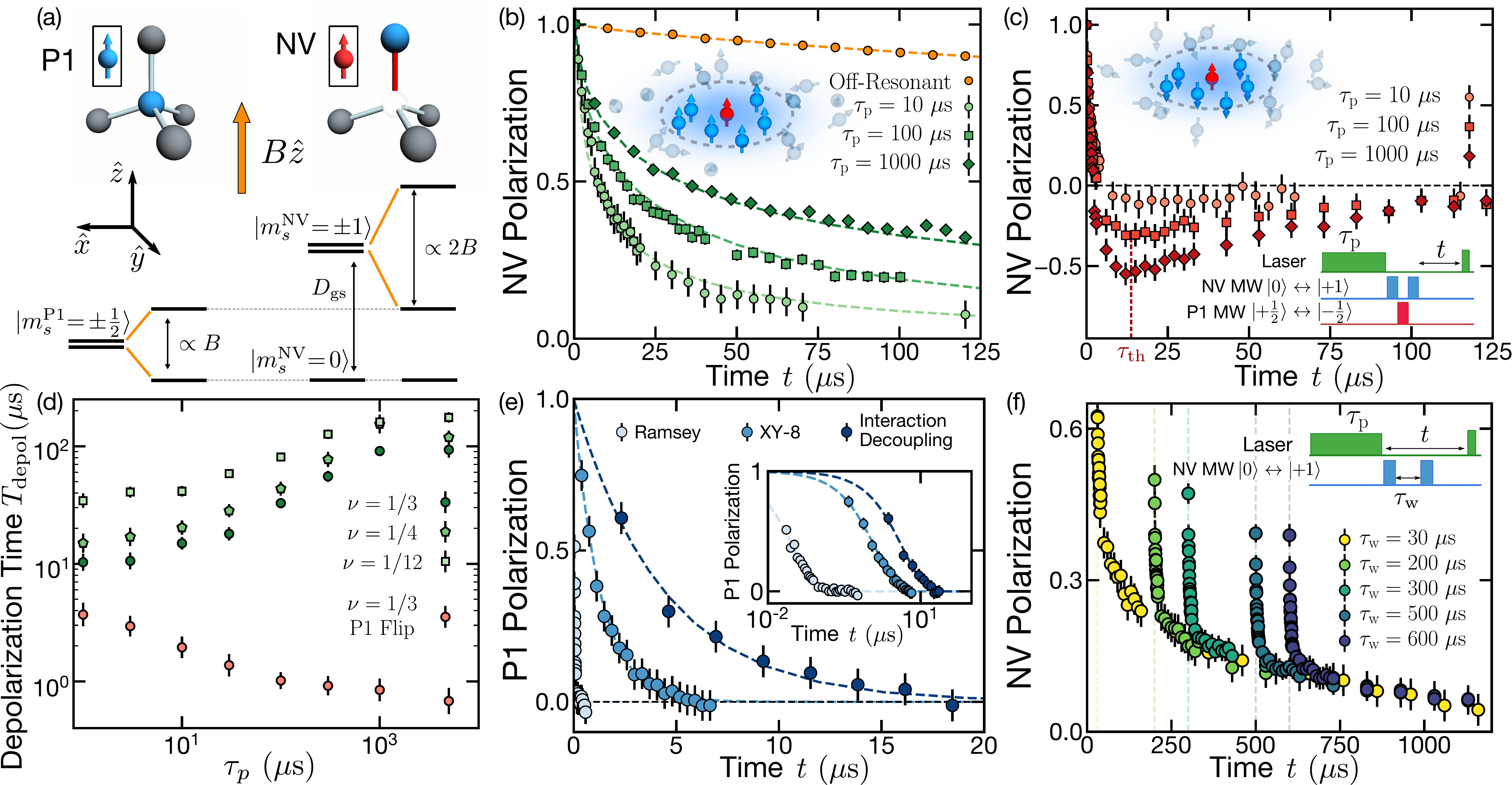}
         \caption{\footnotesize
           {\bf Probing local spin-polarization dynamics using the NV center.}
           {\bf a,} In the absence of a magnetic field, the P1's spin-1/2 sub-levels are degenerate, while the NV's spin-1 sublevels exhibit a zero field splitting, $D_{\mathrm{gs}} = (2\pi)\times 2.87~\mathrm{GHz}$.
           By applying an external magnetic field, the P1 and NV center can be brought into resonance. 
           {\bf b,} 
           When the NV and P1 are off-resonant (orange), $B = 360$~G, the NV exhibits a stretched exponential decay $\sim e^{-(t/T_1^{\mathrm{NV}})^{0.8}}$ (dashed line) with  $T_1^{\mathrm{NV}}=2.3\pm0.1$~ms, consistent with spin-phonon relaxation.
           When the NV is resonant with the $\nu=1/3$ group of P1s (green), $B = 511$~G,  depolarization occurs significantly more rapidly and is strongly dependent upon the polarization time $\tau_{\mathrm{p}}$;
           a longer $\tau_{\mathrm{p}}$ leads to a larger local polarization of P1 centers (inset) and a correspondingly longer NV relaxation time. 
           Dashed green lines correspond to the NV dynamics as captured by our semi-classical model [Eqn.~(\ref{eq:Semi}), see Methods]. 
           {\bf c,}
           NV depolarization dynamics with an anti-polarized $\nu=1/3$ P1 ensemble (top inset).
           Depolarization occurs in two distinct steps:
           an initial decay, $t \lesssim \tau_{\mathrm{th}} \sim 12~\mathrm{\mu s}$, corresponding to local equilibration with the P1 ensemble, followed by late-time diffusion.
           (bottom inset) Pulse sequence describing the preparation of the anti-polarized P1 ensemble. 
           {\bf d,}
           Depolarization time $T_{\mathrm{depol}}$ (extracted as the $1/e$ decay time of the initial polarization) as a function of laser polarization time $\tau_{\mathrm{p}}$ for different effective P1 densities $\nu$. 
           The anti-polarized case for $\nu=1/3$ is denoted as P1 Flip [panel (c) above].
           {\bf e,}
           P1 spin coherence  time, $T_2$, for different dynamical decoupling sequences, Ramsey [$0.032 \pm 0.005~\mathrm{\mu s}$], XY-8 [$1.27 \pm 0.02~\mathrm{\mu s}$] and an interaction decoupling sequence [$4.4 \pm 0.1~\mathrm{\mu s}$ using DROID \cite{choi2020robust,zhou2020quantum}] (see Methods); coherence times are extracted from single exponential decays  (dashed blue lines). 
           Crucially, when the dipolar interactions are canceled, we observe an enhancement of $T_2$, demonstrating that the dynamics are generated by coherent interactions. 
           (inset) Data plotted in semi-log.
           {\bf f,}
           Depolarization dynamics for $\tau_{\mathrm{p}} = 1000~\mathrm{\mu s}$ with variable NV-shelving time, $\tau_{\mathrm{w}}$ (inset). 
           After unshelving, the NV exhibits a fast decay (corresponding to local thermalization $t \lesssim \tau_{\mathrm{th}}$), followed by slow dynamics that capture the ensemble's emergent spin diffusion.
           The  $\tau_{\mathrm{w}}$-independent collapse of the late-time  data confirms the NV's role as a local probe of the P1's polarization dynamics.
           All  data are taken using sample S1 at room temperature $T \sim 300~\mathrm{K}$.
                      }
         \label{fig:fig2}
     \end{figure*}

%
\emph{Hybrid spin platform}---We choose to work with samples containing a P1 density $\sim100$~ppm and an NV density $\sim0.5$~ppm, leading to a geometry where each spin-1 NV center is surrounded by a strongly-interacting ensemble of spin-1/2 P1 centers (Fig.~\ref{fig:fig1}a). 
In this geometry, the NV center naturally plays the role of both a polarization source and a local probe for nearby P1 centers.
These roles rely upon two ingredients. First, the NV center can be optically initialized (to $|m_s^{\textrm{NV}}=0\rangle$) and read out using green laser illumination, which does not affect the P1 center. Second, the NV and P1 centers can coherently exchange spin polarization when brought into resonance via an external magnetic field (Fig.~\ref{fig:fig2}a)  \cite{hall2016detection}; this polarization exchange is driven by the  $\Delta m_s = \pm 2$ components of the dipolar interaction: 
\begin{align}
    H_{\textrm{NV-P1}}&= - \sum_{i} \frac{J_0}{r_{\textrm{NV},i}^3}\left( A_{i} \left[S^+ P^+_i + S^- P^-_i \right] + B_{i} S^zP^z_i \right)~,   \label{eq:NVP1}
\end{align}
where $J_0 = (2\pi)\times52~\mathrm{MHz}\cdot\mathrm{nm^3}$ characterizes the strength of the dipolar interaction, $r_{\textrm{NV},i}$ is the distance between the NV center and the $i^{\textrm{th}}$ P1 center,  $A_{i}$ and $B_{i}$ are $\mathcal{O}(1)$ coefficients that capture the angular dependence of the dipolar interaction \cite{SI}, while $S^{\pm}$ and $P^{\pm}$ are raising and lowering operators for the NV and P1, respectively.
We note that $H_{\textrm{NV-P1}}$ corresponds to the energy-conserving terms of the dipolar interaction, upon restricting our attention to the NV spin subspace $\{ |0 \rangle , |-1\rangle  \}$  (Fig.~\ref{fig:fig2}a).

In addition, the P1 centers also exhibit dipolar interactions among themselves driven by the $\Delta m_s =  0$ component of the dipolar interaction: 
\begin{align}
    H_{\textrm{P1-P1}}&=  -\sum_{i<j} \frac{J_0}{r_{i,j}^3}\left( \tilde{A}_{i,j} \left[P^+_i P^-_j + P^-_i P^+_j\right] + \tilde{B}_{i,j} P^z_iP^z_j \right)   \label{eq:P1P1}
\end{align}
where  $\tilde{A}_{i,j}$, $\tilde{B}_{i,j}$ are the analogous angular coefficients \cite{SI}. 
%


When the NV and P1 are off-resonant, we observe an NV depolarization timescale, $T_{\mathrm{depol}}=2.3\pm 0.1~\mathrm{ms}$, consistent with room-temperature, spin-phonon relaxation   (Fig.~\ref{fig:fig2}b)~ \cite{doherty2013nitrogen}.
%
By applying a magnetic field, $B = 511$~G, along the NV axis, the NV's $\ket{0} \leftrightarrow \ket{-1}$ transition becomes resonant with the P1's $\ket{-\frac{1}{2}}\leftrightarrow \ket{+\frac{1}{2}}$ transition (Fig.~\ref{fig:fig2}a), and we find that $T_{\mathrm{depol}}$ decreases by over two orders of magnitude to $8.9 \pm 0.6~\mathrm{\mu s}$ (Fig.~\ref{fig:fig2}b) \cite{hall2016detection}.
We emphasize that the reduced $T_\mathrm{depol}$ should not be thought of as extrinsic decoherence, but rather as a consequence of coherent interactions between the two spin ensembles (Fig.~\ref{fig:fig2}e); as such, it naturally sets the time-scale on which the NV transfers polarization to the P1 ensemble or measures it.

\emph{Local spin polarization---}%
By continuously repolarizing the NV center via green laser excitation, one can use $H_{\textrm{NV-P1}}$ to transfer spin polarization to nearby P1 centers; this polarization is further spread-out among the P1s by $H_{\textrm{P1-P1}}$.
The duration of the laser excitation, $\tau_{\mathrm{p}}$, then controls the amplitude, shape and width of the local spin polarization.
A longer $\tau_{\mathrm{p}}$ leads to a larger local P1 polarization, which acts as a ``frozen core'' around the NV center (inset, Fig.~\ref{fig:fig2}b), suppressing  dipolar spin exchange from $H_{\textrm{NV-P1}}$ \cite{khutsishvili1962spin}.
This suppression suggests that $T_\mathrm{depol}$, measured \emph{after} P1 polarization, should be significantly enhanced.
This is indeed borne out by the data. 
As shown in Fig.~\ref{fig:fig2}b,d, $T_\mathrm{depol}$ is extended by an order of magnitude as a function of increasing $\tau_{\mathrm{p}}$.
The increase saturates as $\tau_{\mathrm{p}}$ approaches the spin-phonon relaxation time  and the polarization process reaches a steady state (Fig.~\ref{fig:fig2}d)~\cite{barklie198113c}.

\emph{Probing nanoscale spin dynamics---}In order to study the long-time dynamics associated with the dipolar-induced spreading of our initial polarization profiles, it is essential  to distinguish between early-time local equilibration and late-time emergent dynamics. 
This is especially poignant in disordered  three-dimensional dipolar ensembles, where  relaxation can occur anomalously slowly \cite{rovny2018observation, kucsko2018critical}.
To this end, we introduce an experimental technique which allows us to explicitly observe local thermalization.
In particular, after polarizing for $\tau_{\mathrm{p}}$, we  utilize a microwave $\pi$-pulse to shelve the NV population from $\ket{0}$ into the highly off-resonant $\ket{+1}$ state (bottom inset, Fig.~\ref{fig:fig2}c).
Next, we perform a global microwave $\pi$-pulse on the $\ket{-\frac{1}{2}}\leftrightarrow \ket{+\frac{1}{2}}$ P1-transition, flipping the ensemble's spin polarization.
Finally, we unshelve the NV population, effectively preparing an initial condition where the NV is antipolarized relative to the P1 ensemble (top inset, Fig.~\ref{fig:fig2}c).

The dynamics starting from this antipolarized configuration are markedly distinct.
First, the NV polarization quickly changes sign and reaches a negative value, indicating local thermalization with the oppositely oriented P1 ensemble.
Second, the larger the antipolarization (controlled by $\tau_{\mathrm{p}}$), the faster the NV initially decays (Fig.~\ref{fig:fig2}c,d). 
Crucially, this allows us to extract a characteristic time-scale for local thermalization, $\tau_\textrm{th} \sim 12~\mu$s.

Returning to the polarized case (i.e.~without the additional P1 $\pi$-pulse, Fig.~\ref{fig:fig2}f), we can now leverage the  shelving technique to experimentally isolate the emergent late-time dynamics.
In particular, we polarize for time $\tau_\mathrm{p}$, shelve the NV and then wait for a variable time $\tau_\textrm{w}$ to allow the P1 polarization to spread independent of the NV. 
Upon unshelving the NV, we observe a two-step relaxation process, as shown in Fig.~\ref{fig:fig2}f. 
After an initial step of rapid local equilibration, the late-time dynamics exhibit a $\tau_\textrm{w}$-independent collapse. 
Crucially, this demonstrates that for $t > \tau_\textrm{th}$, the NV polarization functions as a \emph{local} probe of the amplitude of the P1 polarization profile, $P(t, \bm{r})$; alternatively, one can also think of the NV's polarization as an autocorrelation function that captures the survival probability  of the P1's polarization dynamics~\cite{Hunt_1956}. 



\emph{Observation of emergent diffusion}---At late times, the conservation of total polarization and the dynamical exponent $z=2$ determine the characteristic  behavior of the survival probability in $d$ dimensions, $S_p(t)~\sim~t^{-d/2}$; the simplest hydrodynamic model capturing this corresponds to Gaussian diffusion:
\begin{equation}\label{eq:diff}
  \partial_t P(t, \bm{r}) = D \nabla^2 P(t, \bm{r}) - \frac{P(t, \bm{r})}{T_1} + Q(t,\bm{r}),
\end{equation}
where $D$ is the diffusion coefficient. The latter two terms in Eqn.~(\ref{eq:diff}) are motivated by our experiment: $Q(t,\bm{r})$ is a source term that characterizes the P1 polarization process, while $T_1$ is an extrinsic relaxation time, after which the experimental signal becomes suppressed.
%
In order to maximize the experimental window for observing emergent hydrodynamics, we work at low temperatures $T = 25$~K, where the NV's $T_1^{\mathrm{NV}}$ time extends by an order of magnitude, and the P1's $T_1$ time extends by a factor of three (see Methods and Extended Data) \cite{jarmola2012temperature}.
The source $Q(t,\bm{r})$ contains contributions from each of the randomly distributed NVs, whose finite density produces an overall  uniform  background polarization that decays exponentially in time.
Isolating the nanoscale polarization dynamics from this background (see Methods), we observe a robust power-law decay of the survival probability, $S_p(t)~\sim t^{-3/2}$, for over a decade in time  following local equilibration, demonstrating the emergence of spin diffusion  [Fig.~\ref{fig:fig1}b]~\cite{Hunt_1956}.
Extracting the corresponding diffusion coefficient from $S_p(t) = P_{\textrm{total}}/(4 \pi D t)^{3/2}$ requires one additional piece of information, namely, the total amount of spin polarization transferred to the P1 ensemble.
Fortunately, this is naturally determined by  combining the height of the measured polarization background with the density of NVs, which we calibrate independently  using a spin-locking experiment \cite{belthangady2013dressed, SI}.
This enables us to experimentally extract the spin-diffusion  coefficient: $D = 0.35 \pm 0.05~\mathrm{nm^2}/\mathrm{\mu s}$ [Table~\ref{tab:DiffCoeff}].

\emph{An unconventional approach to diffusion}---While the hydrodynamic model in Eqn.~\eqref{eq:diff} captures the correct dynamical scaling and thus one key aspect of our observations, it assumes that the dynamics follow  Gaussian diffusion at all times.
However, disorder induces important modifications to this picture and leads to a novel dynamical correction. 
In a homogeneous system, the long-wavelength description of conventional diffusion follows from the derivative  expansion, $\partial_t P_{\bm{k}}(t) = -(Dk^2 + C k^4 + \cdots)P_{\bm{k}}(t)$, where $P_{\bm{k}}(t)$ is the Fourier component of the polarization with wavevector ${\bm{k}}$.
%
%
%
%
However, our system is far from homogeneous. 
Around each P1 center there is a  distinct local environment, arising from both  positional disorder and the presence of on-site random fields (generated from the Ising portion of the dipolar interaction with other  nearby P1s).  This leads to a spatially-dependent local diffusion coefficient.
As an initial polarization profile spreads, its dynamics naturally average over an increasing number of local P1 environments. 
This generates a \emph{dynamical} modification to the diffusion equation, whose leading contribution  is $C_{\mathrm{dyn}} k^2 \partial_t$ (see Methods):
\begin{align}
    \label{eq:LongR_diff}
    \partial_t P_{\bm{k}}(t) = - \left[D k^2 +  C_{\mathrm{dyn}} k^2 \partial_t + C k^4  +  \cdots \right] P_{\bm{k}} (t).
\end{align}
%
%
%
Although this term has the same scaling dimension as the $Ck^4$ term, it induces two
striking modifications to the diffusive dynamics. 
First, the early time polarization profile follows a Yukawa-like form  $\sim \frac{1}{r} e^{-r/\ell}$, and only crosses over to a Gaussian at late times \cite{yukawa1935interaction}. 
Second, the relationship between the height of the polarization profile, as captured by $S_p(t)$, and the width, as captured by $\sim \sqrt{Dt}$, is fundamentally altered; more precisely, in order to faithfully extract $D$ from $S_p(t)$, one must account for the non-Gaussianity of the polarization profile.

In order to connect our nanoscale spin dynamics to these disorder-induced hydrodynamical features, we consider a semi-classical description of the polarization evolution. 
%
%
Starting from the microscopic Hamiltonian, we estimate the rate of polarization transfer, $\Gamma_{ij}$, between any pair of P1 spins via Fermi's golden rule (Fig.~\ref{fig:fig1}c)  \cite{SI, kreuzer1981nonequilibrium, fischetti1998theory, fischetti1999master}:
\begin{align} \label{eq:Semi}
\Gamma_{ij} = \left ( \frac{ J_0 \tilde{A}_{i,j}  }{r_{ij}^3} \right )^2 \frac{2\gamma}{\gamma^2 + (\delta_i - \delta_j)^2}~.
\end{align}
%
Each of the relevant parameters is independently measured: $\gamma~\sim 0.5~\mathrm{\mu s^{-1}}$ represents the interaction-induced linewidth and is characterized by the spin-echo decoherence time of the NV center (Extended Data, Figure E6);  $\delta_i$ represents the strength of the on-site random fields and is drawn from a distribution with width $W\sim (2\pi) \times 4.5~\mathrm{MHz}$, characterized by the NV linewidth (Extended Data, Figure E7).
The analogous polarization transfer rate between NV and P1 spins is obtained by replacing $\tilde{A}_{i,j}$ with $A_{i}$. 
Using our semi-classical model, we perform extensive  numerical simulations accounting for both the P1 polarization process and the subsequent dynamics \cite{SI}. 
Averaging over both positional disorder and on-site random fields, we find excellent agreement with the experimentally measured $S_p(t)$ for over three orders of magnitude in time (Fig.~\ref{fig:fig1}b).

Crucially, our model also provides direct access to the spatial polarization profile, which remains robustly non-Gaussian throughout the time-scale of the experiment, indicative of unconventional diffusion.
Remarkably, the polarization profile precisely exhibits the predicted Yukawa to Gaussian crossover (Fig.~\ref{fig:fig1}d) and enables us to extract the coefficient of the dynamical modification [Eqn.~(\ref{eq:LongR_diff})] as $C_{\mathrm{dyn}} = 204 \pm 45~\mathrm{nm^2}$. 
A few remarks are in order.
First, this coefficient defines a physical length scale, $\ell = \sqrt{C_{\mathrm{dyn}} } = 14.3 \pm 1.6~\mathrm{nm}$, which sets the decay of the Yukawa form $\sim \frac{1}{r} e^{-r/\ell}$  of the polarization profile (Table~E1 in the Extended Data). 
%
More intuitively, $\ell$ can be thought of as the length-scale over which the disorder-induced variations of the  local P1 environments start to become averaged out.
%
%
Thus, only when the polarization expands to a characteristic size much larger than $\ell$, will the dynamics approach Gaussian diffusion.

Second, as evinced in Fig.~\ref{fig:fig1}d, for a wide range of intermediate time-scales, the polarization profile is well-described by a simple exponential. 
As aforementioned, this long-lived non-Gaussianity has an important effect: It  modifies the relationship between the survival probability and the diffusion coefficient.
Somewhat remarkably, this modification can be computed analytically and takes the form of a  geometric factor $g = 2\pi^{1/3}$, which then corrects the  diffusion coefficient from $D \rightarrow gD$ (Table~\ref{tab:DiffCoeff}).
Crucially, the mean square displacement of the polarization profile, $\langle r^2\rangle(t) = 6D_{\langle r^2\rangle} t$ ,  provides an independent measure of the diffusion coefficient (see Extended Data)~\cite{spitzer2013principles,einstein1956investigations}.
As highlighted in Table~\ref{tab:DiffCoeff}, only by accounting for the disorder-induced geometric factor do we observe agreement between the diffusion coefficient extracted from $S_p(t)$ and $\langle r^2\rangle(t)$; this agreement directly demonstrates the non-Gaussian nature of the observed dynamics.



\begin{figure*}
         \centering
         \includegraphics[width = 5.0in]{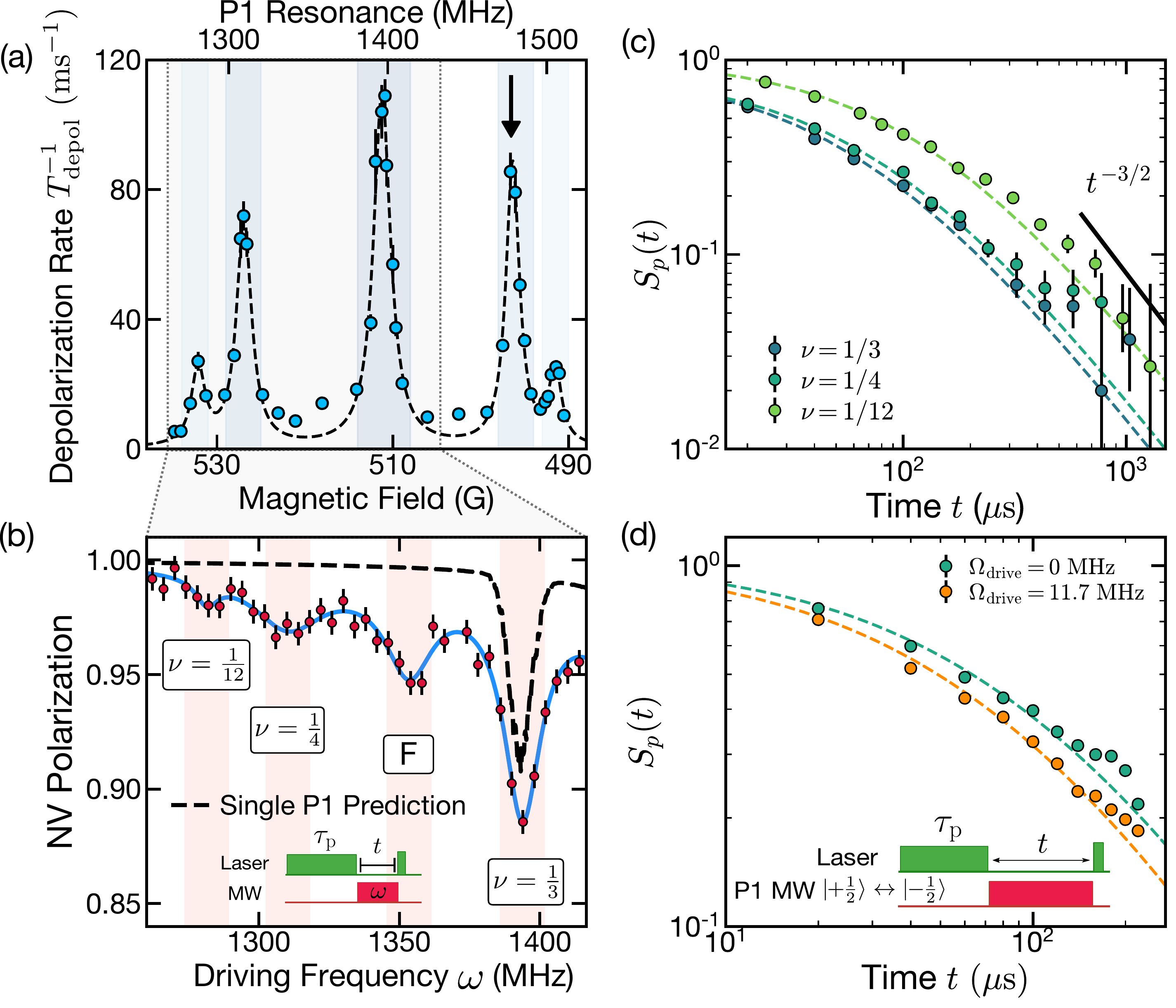}
         \caption{\footnotesize
           {\bf Controlling emergent hydrodynamics by engineering the microscopic Hamiltonian.}
           {\bf a,}
            Depolarization rate, $T^{-1}_{\mathrm{depol}}$, of the NV center as a function of magnetic field after $\tau_\mathrm{p} = 1~\mathrm{\mu s}$. 
           The NV exhibits five distinct resonances corresponding to five different subgroups of P1s with density ratios  $\nu \in \{\frac{1}{12}, \frac{1}{4}, \frac{1}{3}, \frac{1}{4}, \frac{1}{12}\}$.
           For panels (b,d) below, we fix the magnetic field strength, $B = 496.5$~G, wherein the NV is resonant with a $\nu=1/4$ P1 subgroup (indicated by the arrow); the top axis shows the  frequency of the P1 subgroups at this field strength. 
           {\bf b,}
           %
           Fixing a polarization time, $\tau_\mathrm{p}=300~\mathrm{\mu s}$, and an interaction time $t = 3~\mathrm{\mu s}$ (inset), we probe the polarization transfer between the NV and the resonant $\nu=\frac{1}{4}$ P1 subgroup.
           Crucially, polarization exchange depends on the strength of the on-site random field disorder. 
           By driving the \emph{other} P1 subgroups, one can effectively reduce the magnitude of this disorder by ``echoing'' out a portion of the Ising piece of the dipolar interactions. 
            Sweeping the microwave driving frequency, $\omega$, we observe an enhanced NV decay when  it is resonant with the $\nu=\frac{1}{12},\frac{1}{4},\frac{1}{3}$ subgroups as well as an additional ``forbidden'' transition, $F$ (see Methods Section).
            By comparing against numerics of a single P1 spin (dashed black line), we conclude that---aside from the $\nu=\frac{1}{3}$ resonance where an additional hyperfine depolarization channel plays a crucial role---echoing out disorder enhances the coherent many-body interactions and leads to faster dynamics.
            %
       %
        {\bf c,}
           Dynamics of $S_p(t)$ for different effective P1 densities with $\tau_{\mathrm{p}}=100~\mathrm{\mu s}$; control over the P1 density is achieved by tuning the external magnetic field to bring the NV into resonance with the $\nu=\frac{1}{3},\frac{1}{4}$ and $\frac{1}{12}$ P1 subgroups. 
           A smaller P1 density leads to correspondingly slower spin diffusion [Table~\ref{tab:DiffCoeff}].
           {\bf d,}
           Dynamics of $S_p(t)$ for different on-site disorder strengths with $\tau_{\mathrm{p}}=300~\mathrm{\mu s}$.
          %
           %
           Under continuous microwave driving [$\Omega_{\mathrm{drive}}=(2\pi)\times 11.7~\mathrm{MHz}$] of the \emph{other} $\nu=\frac{1}{4}$ P1 subgroup (inset), the effective disorder is suppressed and spin diffusion is enhanced [Table~\ref{tab:DiffCoeff}].
           Dashed lines in (c) and (d) correspond to $S_p(t)$ obtained via Eqn.~(\ref{eq:diff}).
        All experimental data are taken using sample S1 at room temperature $T \sim 300~\mathrm{K}$.
           }
         \label{fig:fig3}
     \end{figure*}

\emph{Microscopic control of emergent spin diffusion}---We now demonstrate the ability to directly translate changes in the underlying microscopic Hamiltonian to changes in the emergent macroscopic behavior.
In order to engineer the Hamiltonian, we  exploit the  hyperfine structure of the P1 defect, enabling control over the effective density and the on-site random field disorder. 
This additional P1 structure is revealed by sweeping the strength of the external magnetic field from $490~\mathrm{G}$ to $540~\mathrm{G}$, where one finds that $T_{\mathrm{depol}}$ exhibits not one, but five distinct resonances (Fig.~\ref{fig:fig3}a) \cite{doherty2013nitrogen,hall2016detection}.
These resonances arise from an interplay between the P1's hyperfine interaction (nuclear spin $I=1$ for $^{14}$N) and its Jahn-Teller orientation, which leads to five spectroscopically distinct subgroups of the P1 ensemble \cite{doherty2013nitrogen,hall2016detection}; each contains a different fraction of the total P1 spins,  with density ratios $\nu = \{\frac{1}{12}, \frac{1}{4}, \frac{1}{3}, \frac{1}{4}, \frac{1}{12}\}$  (Fig.~\ref{fig:fig3}a).
Thus, tuning the external magnetic field provides discrete control over the average spacing between resonant P1 spins, which in turn modifies the average interaction strength~(Fig.~\ref{fig:fig2}d).
As shown in Fig.~\ref{fig:fig3}c, the survival probability for both the $\nu = 1/4$ and $\nu = 1/12$ P1 subgroups exhibits  significantly slower spin diffusion than the $\nu = 1/3$ subgroup.
This is consistent with the presence of weaker interactions arising from the larger spin spacing, and leads to smaller values for the measured diffusion coefficient [Table~\ref{tab:DiffCoeff}].

Finally, one can also experimentally control the strength of the effective on-site random field disorder  via continuous  driving. 
Since these fields are generated by the Ising portion of the interactions between the various P1 subgroups, rapid microwave driving of a single subgroup causes its contributions  to the on-site disorder to become averaged out (Fig.~\ref{fig:fig3}b).
In particular, by bringing the NV into resonance with one of the $\nu = 1/4$ subgroups (black arrow, Fig.~\ref{fig:fig3}a), while driving the other $\nu = 1/4$ subgroup with Rabi frequency $\Omega_{\mathrm{drive}} = (2\pi) \times11.7$~MHz, one  effectively reduces the width of the random field disorder distribution from its undriven value, $W \sim (2\pi) \times 4.5$~MHz, to $W \sim (2\pi) \times  3.4$~MHz. 
The reduced disorder increases the effective rate of long-range hopping and leads to the observation of faster emergent spin diffusion [Table~\ref{tab:DiffCoeff}], as depicted in Fig.~\ref{fig:fig3}d.
In the spirit of translating from the microscopic Hamiltonian to the macroscopic spin dynamics, we can utilize our  semi-classical model to directly account for changes in both the P1 density and the on-site disorder strength. 
Doing so leads to predicted values of the spin diffusion coefficient, which are in excellent agreement with the experimental measurements (see Methods).


\emph{Discussion and outlook}---Our observations demonstrate that hybrid platforms based upon multiple, strongly-interacting species of solid-state spins represent a promising way to study non-equilibrium quantum dynamics in large-scale systems.
By coherently manipulating and measuring the quantum state of local probe spins, one can experimentally distinguish between various dynamical regimes ranging from early-time local thermalization to late-time emergent hydrodynamics. 
Looking forward, our work opens the door to a number of intriguing future directions. 
First, the presence of long-range, power-law interactions can lead to different dynamical  universality classes~\cite{levy1954theorie, shlesinger1995levy}. 
In our system, the polarization dynamics are governed by an effective $\sim 1/r^6$ power-law [Eqn.~(5)].
Interestingly, much like disorder, this particular power-law also leads to an unconventional approach to diffusion, albeit governed by a distinct non-analytic correction $\sim C_{\mathrm{lr}} k^{3}$  \cite{schuckert2020nonlocal} (see Methods); our data (inset, Fig.~\ref{fig:fig1}) do not exhibit clear signatures of this power-law correction and we leave its observation to future work~\cite{zhang1998first,eberhardt2007direct,pagliero2020optically}.
%
%
%
Second, by modifying the polarization process, one can use the NV center as an energy sink without directly removing spin excitations from the P1 ensemble. 
This would correspond to locally cooling the P1 system toward its many-body ground state; in both two and three dimensions, the ground state of disordered, dipolar quantum spins remains an open question and the subject of intense study \cite{pollet2010supersolid,yao2018quantum,zou2017frustrated}. 
%
Third, the ability to experimentally isolate local equilibration dynamics naturally points to the study of many-body localization and Floquet thermalization \cite{nandkishore2015many,yao2015many,yao2016quasi,ponte2015many,serbyn2014interferometric,schreiber2015observation,mori2016rigorous,abanin2017effective,else2017prethermal,else2019long,machado2019exponentially,peng2019observation}.
In long-range interacting systems, the precise criteria for delocalization remain unknown \cite{yao2014many,nandkishore2017many}, while in Floquet systems, the late-time dynamics involve a complex interplay between heating and hydrodynamic behaviour \cite{ye2019emergent,machado2020long,peng2019observation}.
%
%
%
Finally, the presence of a Yukawa-like polarization profile in our system is reminiscent of an open question in the biochemical sciences, namely, what is the underlying mechanism behind the wide-spread emergence of Fickian yet non-Gaussian diffusion in complex fluids \cite{chechkin2017brownian,stylianidou2014cytoplasmic,kim2013simulation,chakraborty2020disorder,chubynsky2014diffusing,postnikov2020brownian,barkai2020packets}; in such systems, it is notoriously difficult to change the microscopic equations of motion, suggesting the possibility for our platform to be utilized as a controllable ``simulator'' of soft, heterogeneous materials. 
A direct route for exploring this question is to leverage sub-diffraction imaging techniques in order to measure correlation functions between spatially separated NVs \cite{rittweger2009sted, maurer2010far,chen2013wide,pfender2014single,arai2015fourier}.
%

\vspace{3mm}

\emph{Acknowledgments}---We gratefully acknowledge the
insights of and discussions with P. Stamp, A. Jayich, M. Dupont.  This
work was supported as part of the Center for Novel Pathways to Quantum Coherence in Materials, an Energy
Frontier Research Center funded by the U.S. Department
of Energy, Office of Science, Basic Energy Sciences under Award No. DE-AC02-05CH11231. A.J. acknowledges
support from the Army Research Laboratory under Cooperative Agreement no. W911NF-16-2-0008. S.H. acknowledges support from the National Science Foundation Graduate Research Fellowship under grant no. DGE1752814.  N.Y.Y. acknowledges support from the David and Lucile Packard foundation and the W. M. Keck foundation. 
This work of D.B. was supported in part by the EU FET
OPEN Flagship Project ASTERIQS.

\bibliography{ref_v14.bib}

\newpage 
 
 \makeatletter
\newcommand*{\centerfloat}{%
  \parindent \z@
  \leftskip \z@ \@plus 1fil \@minus \textwidth
  \rightskip\leftskip
  \parfillskip \z@skip}
\makeatother

\renewcommand{\arraystretch}{2}

\begin{table*}[h!]
  \centerfloat
 \begin{tabular}{c||c|c|c|c|c} 
    \hline \hline
    \makecell{Effective density} &  \multicolumn{2}{c|}{$\nu = 1/3$} & \multicolumn{2}{c|}{$\nu=1/4$} & $\nu = 1/12$ \\ \hline 
    \makecell{Sample}  & \makecell{S1} & \makecell{S2} & \makecell{S1 \\  ~$\Omega_{\mathrm{drive}} = 0~\mathrm{MHz}$}~& ~\makecell{S1 \\  $\Omega_{\mathrm{drive}} = 11.7~\mathrm{MHz}$}~ & S1 \\ \hline \hline
    \makecell{$D ~ [\mathrm{nm^2/\mu s}]$} &  $0.28\pm 0.06$ & $0.35\pm 0.05$ & $0.25\pm 0.06$ & $0.33\pm0.09$ &$0.11\pm 0.03$ \\ \hline
    \makecell{$gD ~ [\mathrm{nm^2/\mu s}]$} &  $0.82\pm 0.17$ &  $1.03\pm 0.13$ & $0.74 \pm 0.18~$ & $0.95 \pm 0.26$ &  $0.33 \pm 0.08$  \\ \hline
    \makecell{$D_{\langle r^2\rangle} ~ [\mathrm{nm^2/\mu s}]$}  & $0.98\pm0.03$  & $1.09\pm 0.02$ & $0.66\pm 0.04$ &  $0.95\pm 0.02$ &  $0.21\pm 0.03$ \\
    \hline \hline 
  \end{tabular}
  \caption{
  To demonstrate the generality of our observations, we investigate the emergence of spin diffusion for a variety of samples and conditions.
  Across all samples, temperatures, P1 densities and disorder strengths, we  observe that the late-time spin dynamics exhibit excellent agreement with emergent diffusion. 
  Accounting for the appropriate non-Gaussian geometric factor, $g = 2\pi^{1/3}$,  yields  agreement between the diffusion coefficient extracted from the survival probability and that extracted from the growth of $\langle r^2\rangle$.
     Samples S1 and S2 both contain a P1 density of $\sim 110$~ppm, while their NV densities are $\sim 0.7$~ppm and $\sim 0.3$~ppm, respectively \cite{SI}. 
     Measurements on S1 are performed at room temperature, while measurements on S2 are taken at $T=25~\mathrm{K}$. 
    For sample S1, we also consider two additional tuning parameters: (i) different effective P1 densities, $\nu \in \{\frac{1}{3},\frac{1}{4},\frac{1}{12}\}$, tuned via the hyperfine structure (Fig.~\ref{fig:fig3}c and Methods), and (ii) different disorder strengths, $W$, tuned via continuous microwave driving (Fig.~\ref{fig:fig3}d).
    We emphasize that the reported uncertainties include propagated uncertainties from other experimentally extracted parameters (e.g. $T_1$ and $\rho_\mathrm{NV}$). 
    Despite overlapping error bars, a detailed analysis (see Methods and Extended data Fig.~E4) confirms that the driven diffusion coefficient is statistically larger than the undriven case.
    }
     \label{tab:DiffCoeff}
\end{table*}

    \newpage
     
\newpage 

\newpage

\end{document}


\title{Supplementary Information for:\\
Emergent hydrodynamics in a strongly interacting dipolar spin ensemble}

\author{Chong Zu}
\thanks{These authors contributed equally to this work.}
\affiliation{Department of Physics, University of California, Berkeley, CA 94720, USA}

\author{Francisco Machado}
\thanks{These authors contributed equally to this work.}
\affiliation{Department of Physics, University of California, Berkeley, CA 94720, USA}

\author{Bingtian Ye}
\thanks{These authors contributed equally to this work.}
\affiliation{Department of Physics, University of California, Berkeley, CA 94720, USA}

\author{Soonwon Choi}
\affiliation{Department of Physics, University of California, Berkeley, CA 94720, USA}

\author{Bryce Kobrin}
\affiliation{Department of Physics, University of California, Berkeley, CA 94720, USA}
\affiliation{Materials Science Division, Lawrence Berkeley National Laboratory, Berkeley, CA 94720, USA}

\author{Thomas Mittiga}
\affiliation{Department of Physics, University of California, Berkeley, CA 94720, USA}
\affiliation{Materials Science Division, Lawrence Berkeley National Laboratory, Berkeley, CA 94720, USA}

\author{Satcher Hsieh}
\affiliation{Department of Physics, University of California, Berkeley, CA 94720, USA}
\affiliation{Materials Science Division, Lawrence Berkeley National Laboratory, Berkeley, CA 94720, USA}

\author{Prabudhya Bhattacharyya}
\affiliation{Department of Physics, University of California, Berkeley, CA 94720, USA}
\affiliation{Materials Science Division, Lawrence Berkeley National Laboratory, Berkeley, CA 94720, USA}

\author{Matthew Markham}
\affiliation{Element Six, Harwell, OX11 0QR, United Kingdom}

\author{Dan Twitchen}
\affiliation{Element Six, Harwell, OX11 0QR, United Kingdom}

\author{Andrey Jarmola}
\affiliation{Department of Physics, University of California, Berkeley, CA 94720, USA}
\affiliation{U.S. Army Research Laboratory, Adelphi, Maryland 20783, USA}

\author{Dmitry Budker}
\affiliation{Department of Physics, University of California, Berkeley, CA 94720, USA}
\affiliation{Helmholtz Institut Mainz, Johannes Gutenberg Universitat Mainz, 55128 Mainz, Germany}

\author{Chris R. Laumann}
\affiliation{Department of Physics, Boston University, Boston, MA 02215, USA}

\author{Joel E. Moore}
\affiliation{Department of Physics, University of California, Berkeley, CA 94720, USA}
\affiliation{Materials Science Division, Lawrence Berkeley National Laboratory, Berkeley, CA 94720, USA}

\author{Norman Y. Yao}
\affiliation{Department of Physics, University of California, Berkeley, CA 94720, USA}
\affiliation{Materials Science Division, Lawrence Berkeley National Laboratory, Berkeley, CA 94720, USA}
\date{\today}

\maketitle
\tableofcontents

\section{Experimental Setup}

We address NV centers using a home-built confocal microscope. 
A 532~nm diode-pumped solid-state laser (Coherent Compass), controlled by an acousto-optic modulator (AOM, Isomet 1250C-848) in a double-pass configuration, is used for both NV spin initialization and detection.
The laser beam is focused onto the sample using an objective lens (Olympus LUCPLFLN, NA 0.6), with diffraction limited spot size $\sim 600$~nm.
The NV fluorescence is collected using the same objective lens, spectrally separated from the laser using a dichroic mirror, further filtered using a 633 nm long-pass filter, and then detected by a avalanche photodetectors (APD, Thorlabs).
A data acquisition card (DAQ, National Instruments USB-6343) is used for fluorescence counting and subsequent data processing. 
The lateral scanning of the laser beam is performed using a two-dimensional galvanometer (Thorlabs GVS212), while the vertical focal spot position is controlled by a piezo-driven positioner (Edmund Optics at room temperature; attocube at cryogenic temperature).
The external magnetic field is applied via a combination of permanent magnet and 3D electromagnetic coils.
Measurements on sample S2 are performed at cryogenenic temperature $T=25$~K using a closed-cycle cryostat (AttoDry 800) to achieve longer NV and P1 intrinsic lifetime.

Three microwave sources (Stanford Research SG384 and SG386) are used to address the NV $\ket{0} \leftrightarrow \ket{-1}$ and resonant P1 $\ket{-\frac{1}{2}} \leftrightarrow \ket{+\frac{1}{2}}$ transitions, NV $\ket{0} \leftrightarrow \ket{+1}$ transition, and other off-resonant P1 subgroups $\ket{-\frac{1}{2}} \leftrightarrow \ket{+\frac{1}{2}}$ transitions respectively.
Microwave signals are combined together together using power combiners (Mini-Circuits), amplified by a broad-band amplifier (Mini-Circuits ZHL-50W-63+), and then delivered to the diamond sample using a coplaner waveguide deposited on a coverslip (sample S1 at room temperature) or a $40$~$\mathrm{\mu}$m diameter copper wire (sample S2 at low temperature).
An arbitrary-wave generator (AWG, Chase Scientific) with $2$~GHz sampling rate is used to control the timing and phase of microwave pulses (I/Q modulation) in the experiment.
The AOM, DAQ, microwave sources and AWG are gated by a programmable multi-channel pulse generator (SpinCore PulseBlasterESR-PRO 500) with $2$~ns temporal resolution.

\section{Hamiltonian in interaction frame}

The Hamiltonian of the entire system is separated into two parts, the on-site energy of each defect (NV and P1) and the interaction between the different defects, given by the dipole-dipole interaction. Let's first consider the NV-P1 interaction in the laboratory frame,
\begin{equation}
  H_{dip} = -\frac{J_0}{r^3} ( 3 (\hat{S}\cdot \hat{n})(\hat{P}\cdot \hat{n}) - \hat{S} \cdot \hat{P})
\end{equation}
where $J_0 = (2\pi)~52~$MHz$\cdot$nm$^3$, $\hat{S}$ and $\hat{P}$ are the spin operators of the two defects. In our case we will label the NV center by spin 1 operators $\hat{S}$ and the P1 centers by spin 1/2 operators $\hat{P}$. Moreover our dynamics only focuses on $m_s=0,-1$ of the NV, so we restrict the Hilbert space to those two levels. In extent:
\begin{align*}
  S_z = \begin{bmatrix}
    0 & 0 \\
    0 & -1
  \end{bmatrix}\quad , \quad
  S_x = \frac{1}{\sqrt{2}}\begin{bmatrix}
    0 & 1 \\
    1 & 0
  \end{bmatrix}\quad , \quad
  S_y = \frac{1}{\sqrt{2}}\begin{bmatrix}
    0 & -i \\
    i & 0
  \end{bmatrix}\\
  P_z = \frac{1}{2}\begin{bmatrix}
    1 & 0 \\
    0 & -1
  \end{bmatrix}\quad, \quad
  P_x = \frac{1}{2}\begin{bmatrix}
    0 & 1 \\
    1 & 0
  \end{bmatrix}\quad, \quad
  P_y = \frac{1}{2}\begin{bmatrix}
    0 & -i \\
    i & 0
  \end{bmatrix}
\end{align*}

We can also define the raising and lowering operators for both spin systems:
\begin{align*}
  P_+ = \begin{bmatrix}
    0 & 1\\
    0 & 0
  \end{bmatrix} = P_x + iP_y\quad, \quad
  P_- = \begin{bmatrix}
    0 & 0\\
    1 & 0
  \end{bmatrix} = P_x - iP_y\\
  S_+ = \begin{bmatrix}
    0 & 1\\
    0 & 0
  \end{bmatrix} = \frac{1}{\sqrt{2}} (S_x + iS_y)\quad, \quad
  S_- = \begin{bmatrix}
    0 & 0\\
    1 & 0
  \end{bmatrix} = \frac{1}{\sqrt{2}}(S_x - iS_y)\\
\end{align*}
Now we write the interaction in terms of the raising and lowering operators as:
\begin{align*}
  S_x = \frac{1}{\sqrt{2}}(S_+ + S_-) \quad,\quad S_y= \frac{1}{i\sqrt{2}}(S_+ - S_-)\\
  P_x = \frac{1}{2}(P_+ + P_-) \quad,\quad P_y =\frac{1}{i2}(P_+ - P_-)
\end{align*}
and expand the dipole interaction as:
\begin{align*}
  H_{dip} &= -\frac{J_0}{r^3} \times \Big\{3\ \left[S_zn_z + n_x\frac{(S_+ + S_-)}{\sqrt{2}} + n_y \frac{(S_+ - S_-)}{i\sqrt{2}}\right]\left[P_zn_z + n_x\frac{(P_++P_-)}{2} + n_y\frac{(P_+-P_-)}{2i}\right] \\
  -& S_zP_z -\frac{(S_+ + S_-)}{\sqrt{2}}\frac{(P_++P_-)}{2} - \frac{(S_+ - S_-)}{i\sqrt{2}}\frac{(P_+-P_-)}{2i}\Big \}
\end{align*}

Having written down the Hamiltonian in the laboratory frame we now go into the rotating frame of the NV and resonant P1 centers, and only keep the energy conserving terms of the dipolar interaction (rotating-wave approximation).

In the laboratory frame, each defect has a splitting $\Delta$ separating the two levels of interest. Since the splitting is in the $z$ direction (NV-axis), we are interested in the evolution of $|\psi\rangle = e^{-i\delta s_zt} |\phi\rangle$, where $s_z$ can either represent the NV or the P1 spin operator:
\begin{align*}
  i\partial_t e^{-i\delta s_zt}|\phi\rangle= (\delta s_z + H_{dip})  e^{-i\delta s_zt}|\phi\rangle\\
  \delta s_z e^{-i\delta s_zt}|\phi\rangle + ie^{-i\delta s_zt}\partial_t|\phi\rangle= (\delta s_z + H_{dip})  e^{-i\delta s_zt}|\phi\rangle\\
  i\partial_t|\phi\rangle=e^{i\delta s_zt} H_{dip}  e^{-i\delta s_zt}|\phi\rangle = \tilde{H}_{dip} |\phi\rangle
\end{align*}

Now we need to write the spin operators in the rotation frame:
\begin{align*}
   \tilde{s}_z = e^{i\Delta s_z t}s_ze^{-i\Delta s_z t} = \hat{s}_z\\
  \tilde{s}_+ = e^{i\Delta s_z t}s_+e^{-i\Delta s_z t}|0\rangle = e^{-i(\Delta)t} \hat{s}_+\\
  \tilde{s}_- = e^{i\Delta s_z t}s_-e^{-i\Delta s_z t}|1\rangle = e^{+i(\Delta)t} \hat{s}_-
\end{align*}

In the case of the NV and P1 center we have:
\begin{align*}
  \tilde{S}_z = S_z\quad \tilde{S}_+ = e^{-i\Delta_{NV}t}S_+\quad \tilde{S}_- = e^{i\Delta_{NV}t}S_-\\
  \tilde{P}_z = P_z\quad \tilde{P}_+ = e^{i\Delta_{P1}t}P_+\quad \tilde{P}_- = e^{-i\Delta_{P1}t}P_-
\end{align*}

Now we can compute the dipolar interaction in the rotational frame:
\begin{align*}
  \tilde{H}_{dip} &= -\frac{J_0}{r^3} \times \Big\{3\ \left[S_z n_z + n_x\frac{(S_+e^{-i\Delta_{NV}t} + S_-e^{i\Delta_{NV}t})}{\sqrt{2}}
  + n_y \frac{(S_+e^{-i\Delta_{NV}t} - S_-e^{i\Delta_{NV}t})}{i\sqrt{2}} \right] \\
  &\times\left[P_zn_z + n_x\frac{(P_+e^{i\Delta_{P1}t}+P_-e^{-i\Delta_{P1}t})}{2} + n_y\frac{(P_+e^{i\Delta_{P1}t}-P_-e^{-i\Delta_{P1}t})}{2i}\right]\\
  -& S_zP_z -\frac{(S_+e^{-i\Delta_{NV}t} + S_-e^{i\Delta_{NV}t})}{\sqrt{2}}\frac{(P_+e^{i\Delta_{P1}t}+P_-e^{-i\Delta_{P1}t})}{2} \\
  &+ \frac{(S_+e^{-i\Delta_{NV}t} - S_-e^{i\Delta_{NV}t})}{\sqrt{2}}\frac{(P_+e^{i\Delta_{P1}t}-P_-e^{i\Delta_{P1}t})}{2}\Big \}
\end{align*}
which simplifies to:
\begin{align*}
  \tilde{H}_{dip} &= -\frac{J_0}{r^3} \times \Big\{ \\
  &(3n_z^2-1) S_zP_z + \frac{3S_+P_+}{2\sqrt{2}}e^{-i(\Delta_{NV}-\Delta_{P1})t}\left[n_x^2-n_y^2 - 2in_xn_y\right] +\\
  &+ \frac{3S_-P_-}{2\sqrt{2}}e^{i(\Delta_{NV}-\Delta_{P1})}\left[n_x^2-n_y^2 + 2in_xn_y\right]\\
  +& \left(e^{-i(\Delta_{NV}+\Delta_{P1})}\frac{P_-S_+}{2\sqrt{2}}+ e^{i(\Delta_{NV}+\Delta_{P1})}\frac{P_+S_-}{2\sqrt{2}}\right)(3n_x^2+3n_y^2-2)\\
  +& 3n_zS_z\left[n_x\frac{(P_+e^{i\Delta_{P1}t}+P_-e^{-i\Delta_{P1}t})}{2} + n_y\frac{(P_+e^{i\Delta_{P1}t}-P_-e^{-i\Delta_{P1}t})}{2i}\right] +\\
  +& 3n_zP_z\left[n_x\frac{(S_+e^{-i\Delta_{NV}t} + S_-e^{i\Delta_{NV}t})}{\sqrt{2}} + n_y \frac{(S_+e^{-i\Delta_{NV}t} - S_-e^{i\Delta_{NV}t})}{i\sqrt{2}}\right]\Big \}
\end{align*}

In the rotation frame we can drop the last three lines because they always have a time dependence that is much faster than the average interaction strength between spins (rotating-wave approximation).
There are two other possibilities, either for resonant P1 group $\Delta_{NV} = \Delta_{P1}$ on which flip-flip interactions are meaningful because they conserve energy so all of the first line matters, or $\Delta_{NV} \neq \Delta_{P1}$, on which only the Ising term matters.

Summarizing for the interaction between NV and resonant P1 groups:
\begin{equation}\label{eq:NVP1}
  \tilde{H}_{NV-P1} = -\frac{J_0}{r^3}  \left\{(3n_z^2-1) S_zP_z + \frac{3S_+P_+}{2\sqrt{2}}(n_x-in_y)^2+ \frac{3S_-P_-}{2\sqrt{2}}(n_x+in_y)^2\right\}
\end{equation}
For off-resonant P1 groups:
\begin{equation} \label{eq:Ising}
 \tilde{H}_{NV-P1} = -\frac{J_0}{r^3} \times (3n_z^2-1) S_z P_z
\end{equation}

The previous result corresponds to the interaction between NV and P1. In the case of the P1-P1 interaction, the interaction should be spin conserving. This corresponds to taking $\Delta NV = -\Delta P1$ in the previous computation, which leaves us with the Ising term as well as the spin conserving terms. Moreover we need to exchange the $\sqrt{2} \to 2$ from the different definitions of the raising and lowering operator between the spin 1/2 and spin 1 systems. The result is then:
\begin{equation}\label{eq:P1P1}
  \tilde{H}_{P1-P1} = -\frac{J_0}{r^3} \times \left\{ (3n_z^2-1) P^{(1)}_z P^{(2)}_z  + \left(\frac{P^{(1)}_- P^{(2)}_+}{4}+ \frac{P^{(1)}_+ P^{(2)}_-}{4}\right)(1-3n_z^2)\right\}
\end{equation}
while the off-resonant interaction is the same as NV-P1 interaction.

\section{Extraction of defect densities}

In this section we describe the calibration procedures to measure the P1 and NV density in our samples.

\subsection{Extraction of P1 density}
We use $3$ independent methods to extract P1 concentration in our samples.

\subsubsection{NV linewidth}

The magnetic dipolar coupling between NV and its local P1 ensemble leads to a broadening of the NV spin transition. 
When NV is off-resonant with nearby P1 centers, the NV-P1 coupling is governed by Ising-type interactions (Eqn.~\ref{eq:Ising}), which leads to a frequency shift of NV transition that depends on the configuration of local P1 spins.
When NV is resonant with a subgroup of P1 centers, the effective interaction Hamiltonian (Eqn.~\ref{eq:NVP1}) includes a polarization exchange term which leads to an additional broadening of the NV transition.
By characterizing the linewidth of the NV as a function of the resonant group density $\nu$, we can direct extract the P1 density in our sample.

\begin{figure}
    \centering
    \includegraphics[width = 2.8in]{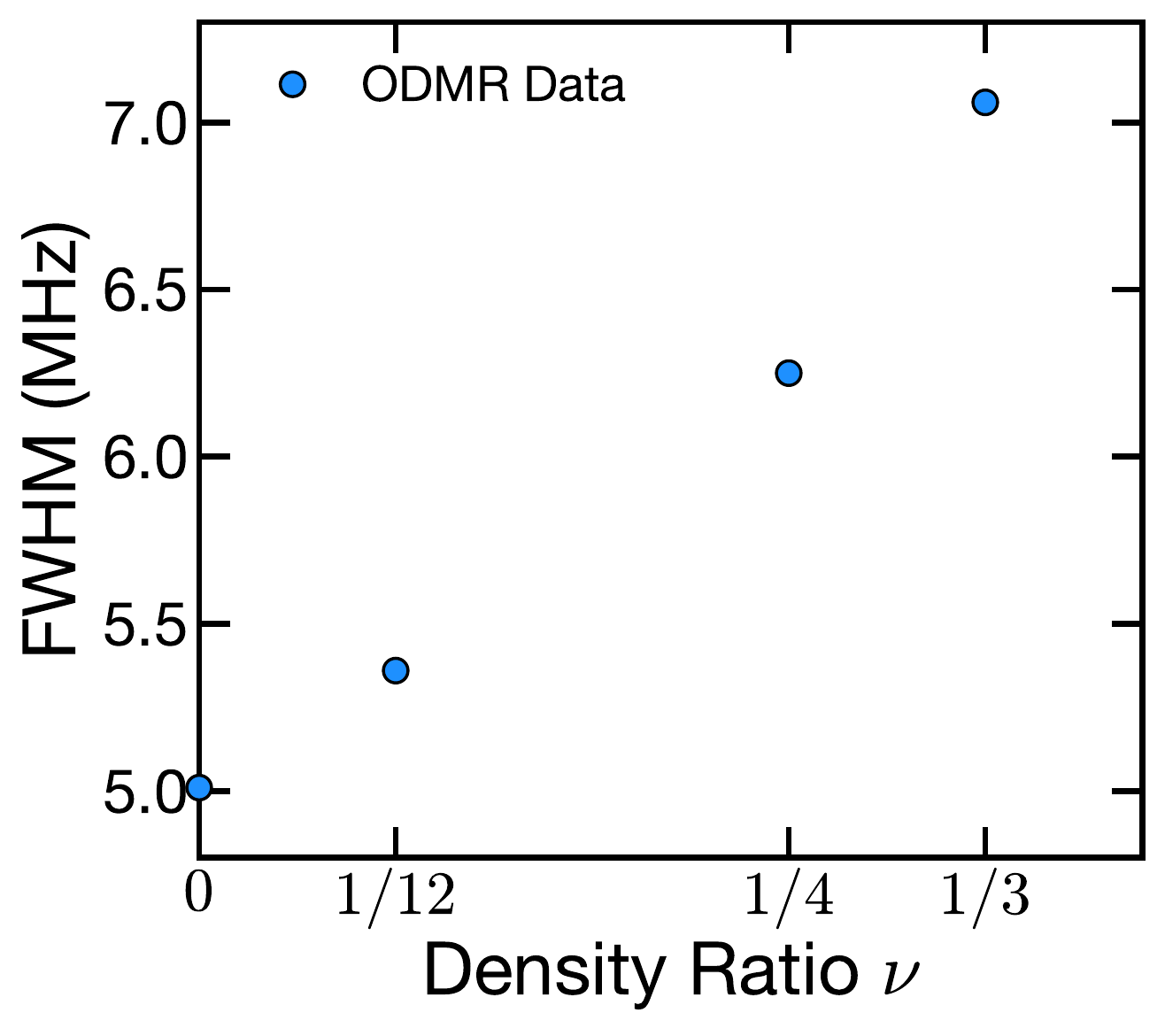}
    \caption{Full-width-half-max (FWHM) of the NV linewidth as a function of the resonant P1 subgroup with density ratio $\nu$. 
    When off-resonant, the Ising interaction between the NV and the P1 centers leads to a finite linewidth.
    Bringing the NV into resonance with a larger number of P1s leads to an increased linewidth due to polarization exchange interactions.}
    \label{fig:FWHM_rhoNV}
\end{figure}

\begin{figure}
    \centering
    \includegraphics[width = 7in]{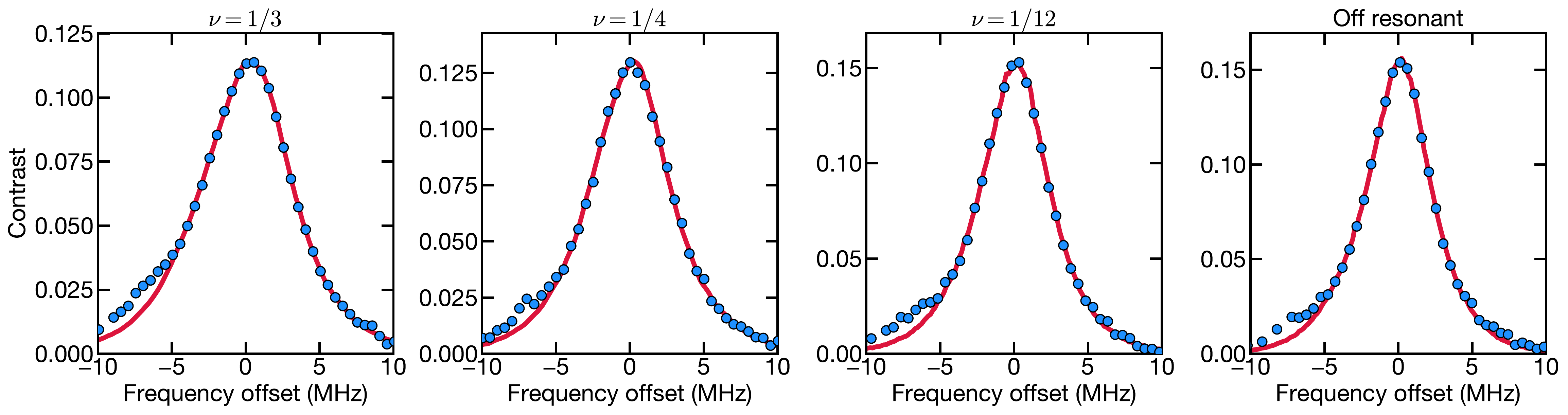}
    \caption{Numerically computed spectra for $(\rho_{\mathrm{P1}}, r_{\mathrm{cut}}) = (110~\mathrm{ppm}, 1.75~\mathrm{nm})$ (red) are in great agreement with the measured ODMR signal for sample S1.}
    \label{fig:rhoP1_Best}
\end{figure}

To this end, we perform pulsed optical-detected magnetic resonance (ODMR) measurement to experimentally measure the NV linewidth at different magnetic fields. 
We choose a slow microwave $\pi$ pulse with duration $2$~$\mu$s to avoid power broadening from microwave drive. 
By controlling the strength of magnetic field, we allow the NV to fully off-resonant with P1 ensemble (at  $B=470.1~\mathrm{G}$) as well as resonant to different P1 subgroups: $\nu=\{\frac{1}{12}, \frac{1}{4}, \frac{1}{3}\}$ at $B=491.6~\mathrm{G}$, $B=496.5~\mathrm{G}$ and $B=511.1~\mathrm{G}$,  respectively.
As a greater ratio of the P1s is resonant with the NV, we observe a remarkable increase in the NV's linewidth (Fig.~\ref{fig:FWHM_rhoNV})
Curiously, across all ODMR spectra we observe an additional small peak which is not captured by the inclusion of P1 spins and $^{14}$N hyperfine, Fig.~\ref{fig:rhoP1_Best}.
This is consistent with the presence of nearby strongly coupled $^{13}$C nuclear spins \cite{smeltzer201113c}.

In parallel, we compute the linewidth of the NV as follows.
In the off-resonant case, we randomly place $N \gg 100$ P1 spins at a density $\rho_{\mathrm{P1}}$ around an NV center at $\bm{r}_{NV}=(0,0,0)$.
Motivated by the spatial extent of the NV wavefunction, we impose a short-range cut-off $r_{\mathrm{cut}}$ between the NV and the P1s. This serves as an effective short-range regulator of the power-law interaction.
Each P1 is simulated classicaly: it is taken to randomly be in either the $|-\frac{1}{2}\rangle$ or $|\frac{1}{2}\rangle$ state.
Owing to their position and state, each P1 induces a small magnetic field at the position of the NV that shifts its energy level.
For each positional realization and spin configuration of P1 centers, we obtain a single shift of the energy levels.
Averaging over both ensembles, we obtain an histogram of the energy shifts which characterizes the linewidth in \emph{a single hyperfine state of the NV}.
To obtain the final observed NV spectrum, we take the aforementioned histogram shift it by the different possible hyperfine interactions $-2.162~\mathrm{MHz}$, $0~\mathrm{MHz}$, and $2.162~\mathrm{MHz}$.
We then scale each hyperfine contribution to capture the nuclear polarization that may occur during optical polarization; these scaling parameters are chosen by a least square best fit to the experimental data (excluding the region where the $^{13}C$ peak lies).

In the on-resonant case, we again randomly place $N \gg 100$ P1 spins at a density $\rho_{\mathrm{P1}}$ around an NV center at $\bm{r}_{NV}=(0,0,0)$.
We then choose a ratio $\nu$ of these spins and mark them as being part of the resonant group.
We would like to solve for the strongly interacting Hamiltonian of these spins under a classical field distribution of the other ones. Unfortunately, owing to the exponentially growing size of the associated Hilbert space we can only diagonalize a system of 1 NV and $N_{\mathrm{P1}} \sim 10$ P1s.
To this end, we pick the $N_{\mathrm{P1}}$ closest P1s to $\bm{r}_{NV}=(0,0,0)$ to simulate quantum mechanically (by computing their interactions via Eqs.~\ref{eq:NVP1} and \ref{eq:P1P1} and diagonalizing the resulting Hamiltonian), while the remaining ones are included as classical spins that generate onsite fields on the quantum spins via Eqn.~\ref{eq:Ising} (as prescribed before).
The linewidth is then characterized by the spectral function of the spin flip operator in the NV:
\begin{align*}
  S(\omega) = \sum_{i,j} |\langle i | \sigma^x | j\rangle|^2 \delta(\omega - (\omega_i-\omega_j))
\end{align*}
where $|i\rangle$ is the eigenstate with energy $\omega_i$ and $\sigma^x = S_+ + S_-$ is the NV flip operator that connects the $\ket{0}$ and $\ket{-1}$ states.

For a single realization, $S(\omega)$ will correspond to a discrete set of delta functions, however, by averaging over positional disorder and spin configurations of the classical P1 spins, we again obtain a histogram that captures the linewidth of a single hyperfine state of the NV (one could include the hyperfine interaction directly, but owing to the very large mismatch between the hyperfine interaction and the NV splitting, its effect is accurately captured by a simple shifting of the energy levels).
To obtain the final linewidth spectrum, a similar analysis to the off-resonant case is performed; after shifting the by the appropriate hyperfine energies, each contribution is scaled according to a least-square fit against the experimental ODMR data.

\begin{figure}
    \centering
    \includegraphics[width = 7in]{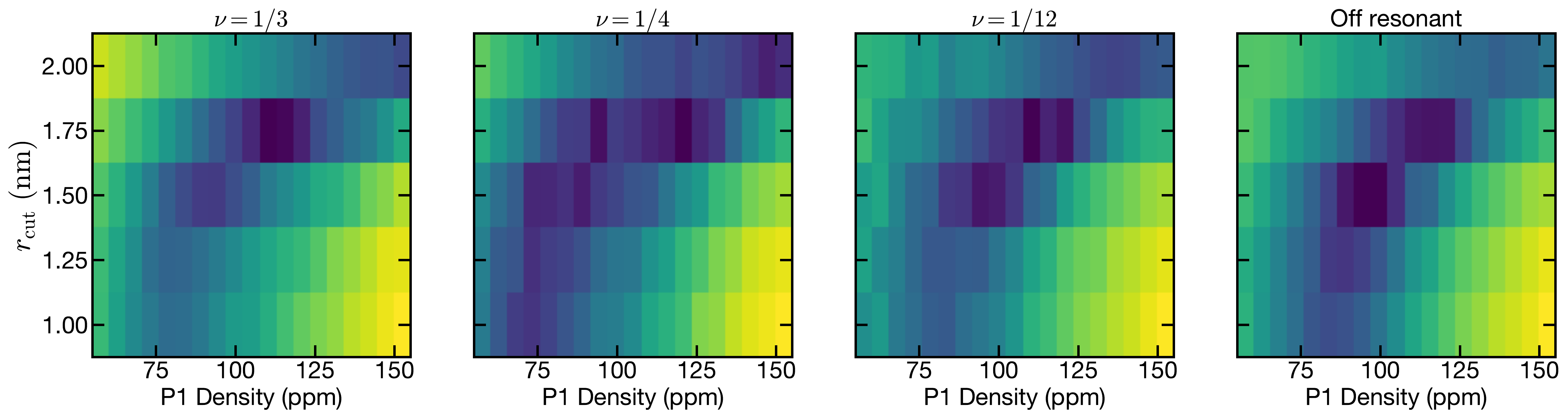}
    \caption{Total absolute residue as a function of both $r_{\mathrm{cut}}$ and P1 density $\rho_{\mathrm{P1}}$. Lighter colors corresponds to higher total residues.
    For all cases, the minima occurs around $(\rho_{\mathrm{P1}}, r_{\mathrm{cut}}) = (110~\mathrm{ppm}, 1.75~\mathrm{nm})$, providing a robust estimate of the P1 density in sample S1.
    }
    \label{fig:rhoP1_Basin}
\end{figure}

Having generated a many NV spectra for the region of interest $\rho_{P1} \in [50,150]$ and $r_{\mathrm{cut}} \in [1.0, 2.0]$, we find the best parameter pair $(\rho_{\mathrm{P1}}, r_{\mathrm{cut}})$ by minimizing the residuals between the computed and the experimental NV spectra.
For each situation (either off-resonant or on-resonance with each $\nu$ P1 subgroup), we obtain a minima for both parameters (Fig.~\ref{fig:rhoP1_Best}).
For sample S1, all minima occur around $(\rho_{\mathrm{P1}}, r_{\mathrm{cut}}) = (110~\mathrm{ppm}, 1.75~\mathrm{nm})$ (Fig.~\ref{fig:rhoP1_Basin}); from the size of the basis we extract the NV density of our system as $\rho_{\mathrm{NV}} = 110 \pm 10~\mathrm{ppm}$.
For sample S2, we consider the off-resonant case at $B= 383.2~\mathrm{G}$ and observe agreement with sample's S1 off-resonant spectra, leading to the same estimation of the density $\rho_{\mathrm{NV}} = 110 \pm 10~\mathrm{ppm}$.

\subsubsection{Double electron-electron resonance}

     \begin{figure}
         \centering
         \includegraphics[width = 6in]{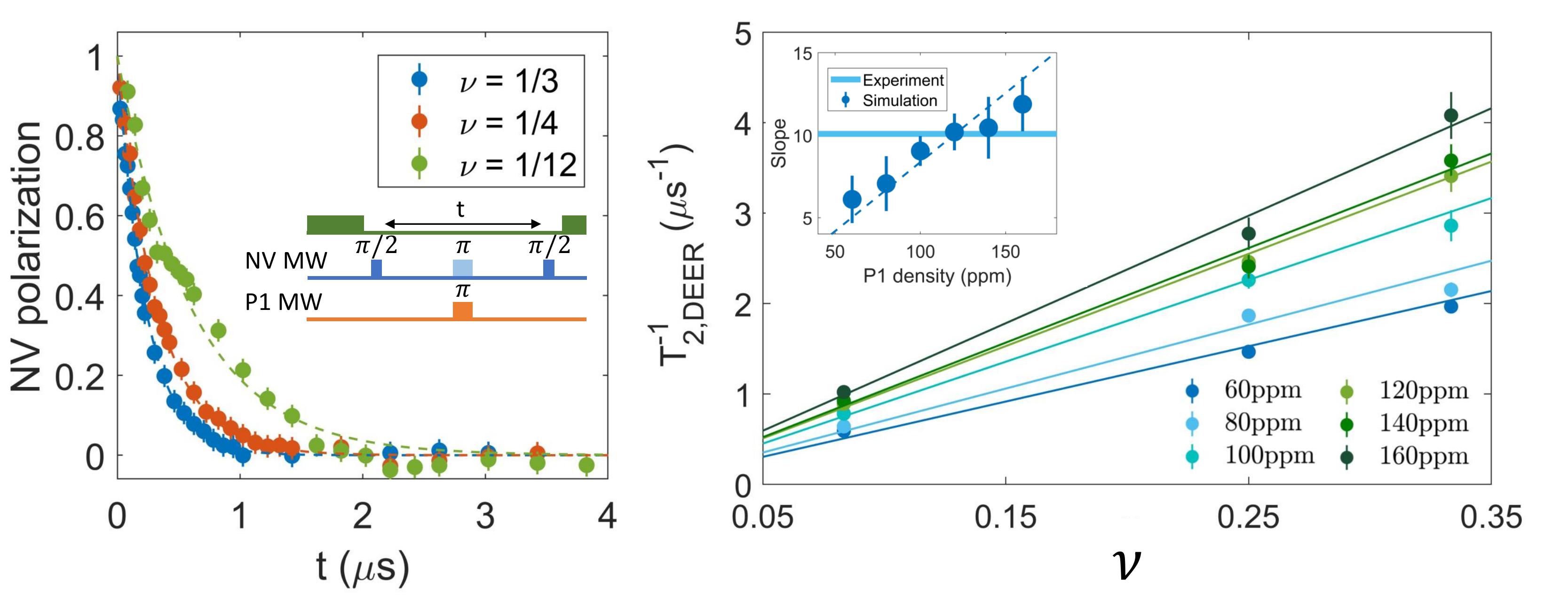}
         \caption{
           {\bf DEER spectroscopy to extract P1 density.}
           (a) DEER decay with 3 different subgroups $\nu = \frac{1}{3}, \frac{1}{4}, \frac{1}{12}$. The dashed lines corresponds to single exponential fit. Inset: DEER pulse sequence. (b) Simulation of DEER decay rate $T_{2,DEER}^{-1}$ as a function of P1 subgroups $\nu$ for different densities. Solid lines correspond to linear fit. Inset: Comparison of the slopes for experiment and numerical simulation to extract P1 density.
           }
         \label{fig:figS_DEER}
     \end{figure}
     
The second method to extract P1 density is double electron-electron resonance (DEER) spectroscopy, which consists of a normal spin-echo sequence on NV center, and an additional microwave $\pi$ flip on a specific subgroup of P1 centers with density ratio $\nu = \frac{1}{12}, \frac{1}{4}, \frac{1}{3}$ [Fig.~\ref{fig:figS_DEER}(a)].
DEER spectroscopy allows us to isolate the decoherence signal from the target P1 subgroup, thus provides a direct way to measure the Ising interaction strength between NV and P1 ensemble.
We perform the experiment on sample S1 with an applied external magnetic field $B_z = 171G$ along one of the NV axis.
By fitting the data to a single exponetial decay, we obtain a DEER coherence time $T_{2,\textrm{DEER}} = 0.69(5)$~$\mu$s, $0.32(1)$~$\mu$s, $0.23(1)$~$\mu$s for $\nu = \frac{1}{12}, \frac{1}{4}, \frac{1}{3}$ respectively (Fig.~\ref{fig:figS_DEER}(a)).

To quantitatively analyze the dependence of NV DEER coherence times on P1 density $\rho_{P1}$, we theoretically compute the corresponding spin dynamics using exact diagonalization of the effective many-body interacting Hamiltonian.
In particular, We simulate the DEER spin dynamics using 8 randomly positioned P1 centers surrounding a single NV center.
We average over $\sim 1000$ realizations of positional disorder, and obtain a smooth DEER coherence decay for different P1 densities and subgroup ratio $\nu$.
We fit each numerical curve to a single exponential decay and extract a simulated DEER time scale.
Fig.~\ref{fig:figS_DEER}(b) summarizes the DEER decay rate $\gamma_\textrm{D} = T_{2,\textrm{DEER}}^{-1}$ as a function of $\nu$ (effective density), where a linear dependence of $\gamma_\textrm{D}$ is identified for all P1 densities ranging from $60$~ppm to $160$~ppm.
We model the measured decay rate as $\gamma_\textrm{D}(\nu) = \gamma_0 + \nu \gamma_1(\rho_{P1})$.
The offset $\gamma_0$ represents the extrinsic decoherence; in simulations $\gamma_0 = 0$, but in experiments, additional sources of decoherence (e.g. magnetic impurities other than P1 centers) can lead to finite $\gamma_0$.
The slope $\gamma_1(\rho_{P1})$ characterizes the decay induced by NV-P1 Ising interaction, which exhibits a linear dependence on $\rho_{P1}$.
Comparing the slopes from experiment and simulation, we extract a P1 density $\rho_{P1}\sim 120$~ppm (Fig.~\ref{fig:figS_DEER}(b) inset).

\subsubsection{Electron paramagnetic resonance}

     \begin{figure}
         \centering
         \includegraphics[width = 6in]{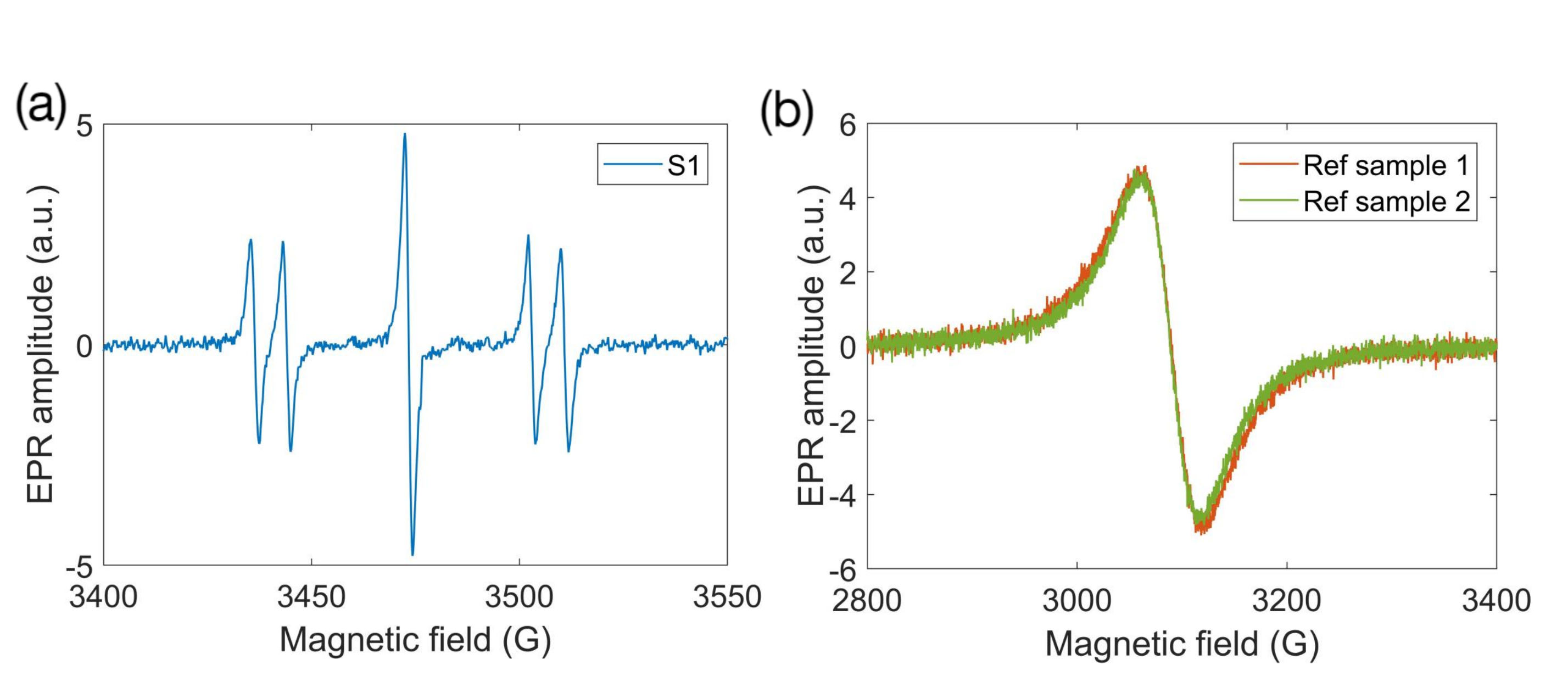}
         \caption{
           {\bf EPR spectroscopy to extract P1 density}
           (a) First derivative of EPR absorption spectrum for P1 centers in sample S1. (b) First derivative of EPR absorption spectra for electronic spins in two reference samples (CuSO$_4\cdot$ 5H$_2$O).
           }
         \label{fig:figS_EPR}
     \end{figure}
     
We also perform electron paramagnetic resonance (EPR) spectroscopy to calibrate the bulk P1 density in sample S1.
The sample is placed into a EPR microwave resonantor with a fixed applied microwave frequency.
By sweeping the strength of an external magnetic field, one can effectively tune the P1 spin energy splitting $\Delta$.
When $\Delta$ matches the applied microwave frequency, one observes an absorption of the microwave signal, whose strength is proportional to the total spin numbers inside the sample.
Fig.~\ref{fig:figS_EPR}(a) summarizes the first derivative of the microwave absorption signal from P1 centers in sample S1.

In order to extract an absolute P1 density, one needs to calibrate the strength of the EPR signal.
We follow the method described in Ref.~\cite{scott2016phenomenology}, where cupric sulfate pentahydrate crystals (CuSO$_4\cdot$ 5H$_2$O) are used as reference samples. 
In cupric sulfate pentahydrate crystal, there is a single $s=\frac{1}{2}$ electron that associated with every Cu$^{2+}$ ion, thus provides a clean and pronounce EPR absorption spectrum with well calibrated spin density.
We carefully choose two pieces of cupric sulfate pentahydrate crystals with similar shape and weight of sample S1, and obtain their EPR spectra Fig.~\ref{fig:figS_EPR}(b).
By comparing the double integral of the measured EPR spectra, we extract a P1 density $\rho_{P1}\sim 140$~ppm, which agrees well with the other two methods.

\subsection{Extraction of NV density}

     \begin{figure}
         \centering
         \includegraphics[width = 6in]{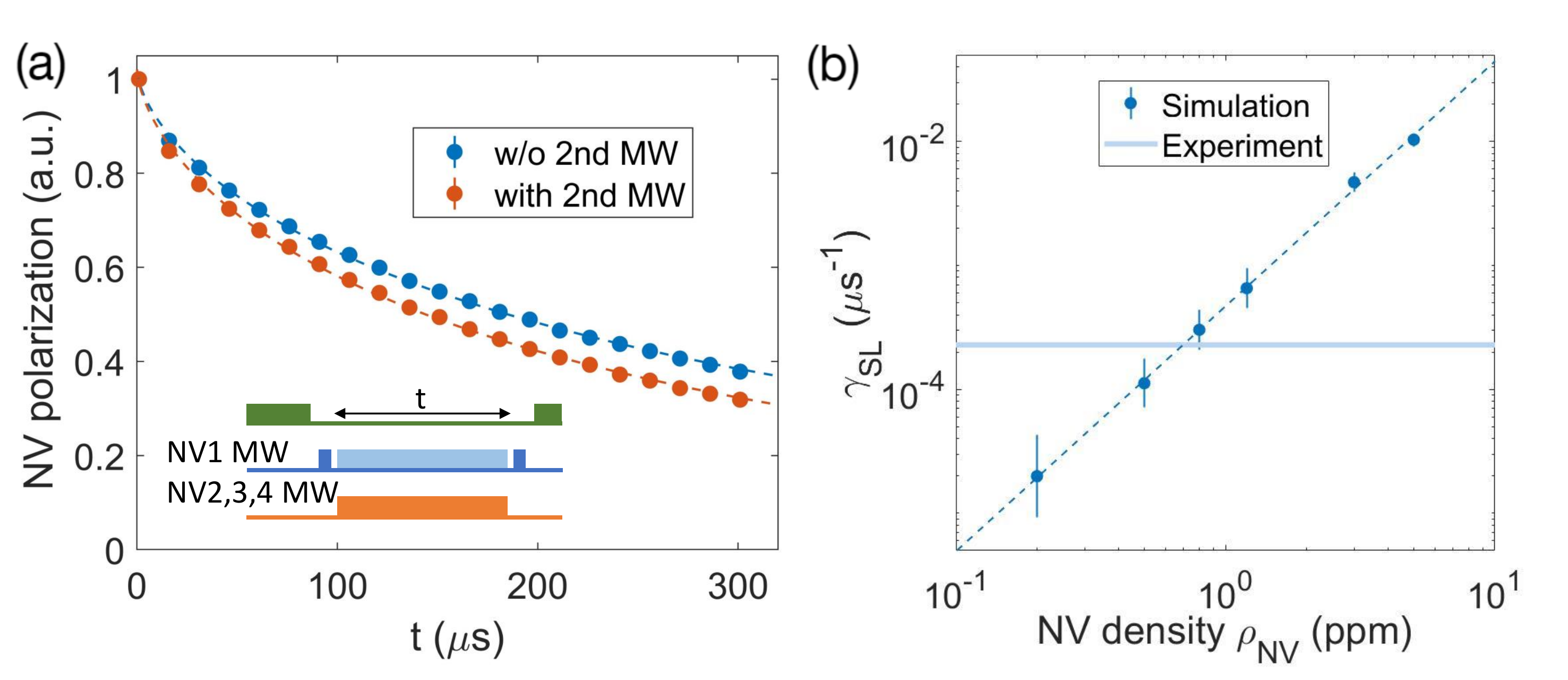}
         \caption{
           {\bf Extraction of NV density}
           (a) Spin locking measurement on NV1 with (orange) and without (blue) an additional microwave drive on NV2,3,4. The dashed lines are stretched exponential fit with power $\alpha = 0.65$. Inset: experimental pulse sequence. (b) Simulation of NV-NV dipolar induced decay rate $\gamma_{\mathrm{SL}}$ with different NV density $\rho_{NV}$. The dashed blue line is a power-law fit. Horizontal solid line is the value measured in experiment.
           }
         \label{fig:figS_NVdensity}
     \end{figure}

To extract NV density in the sample, we utilize a Hartmann-Hahn type spin-locking sequence \cite{hartmann1962nuclear} to directly measure the dipolar interaction between different NV groups.
In particular, by applying an external magnetic field $B\sim 358$~G along one of the NV axis, we isolate this specific NV subgroup (NV 1) from the other three degenerate NV subgroups (NV 2,3,4).
We then perform a standard spin locking sequence on $\ket{0}$ to $\ket{-1}$ transition of NV 1, which consists of an initial $\frac{\pi}{2}$ pulse along $\hat{x}$, followed by a continuous driving with strength $\Omega_1$ along $\hat{y}$ with time t, and a final $\frac{\pi}{2}$ pulse along $-\hat{x}$ to bring the NV back to $\ket{0}$ (Fig.~\ref{fig:figS_NVdensity}a inset).
Spin locking effectively decouples the NV center from magnetic noise in the environment, thus leads to a significant extension of NV coherence time.
We fit the data to a stretched exponential decay, and extract a spin locking time $T_{SL1} = 312(3)$~$\mu$s with stretched power $\alpha = 0.65(1)$ (Fig.~\ref{fig:figS_NVdensity}(1)).

To measure the dipolar interaction between NV groups, we apply an additional microwave to continuously rotate the other three degenerate NV groups (NV 2,3,4) with strength $\Omega_2$.
When $\Omega_2 = \Omega_1$, NV 2,3,4 become resonant with NV 1 in the rotating frame, leading to polarization exchange between the groups.
Indeed, we observe an enhanced decay from the experiment, $T_{SL2} = 242(3)$~$\mu$s (Fig.~\ref{fig:figS_NVdensity}(a)).
The additional decay rate from NV-NV dipolar interaction can be calculated as $\gamma_\mathrm{SL}=(T_\mathrm{SL2}^{-\alpha}-T_\mathrm{SL1}^{-\alpha})^{\frac{1}{\alpha}} = 2.28(3)\times 10^{-4} $~$\mu$s$^{-1}$.

To quantitatively analyze the dependence of $\gamma_\mathrm{SL}$ on total NV density $\rho_{NV}$, we theoretically simulate the corresponding spin dynamics using 8 randomly positioned group 2,3,4 NV spins surrounding a central NV in group 1.
We average over $\sim 1000$ realizations of positional disorder, and obtain a smooth spin locking signal for different NV densities.
Note that the simulation experiences a significant finite size effect at late time, therefore we fit only the early time dynamics to the stretched exponential functional form obtained from the experiment, and calculate $\gamma_\mathrm{SL}$ as a function of $\rho_{NV}$ (Fig.~\ref{fig:figS_NVdensity}b).
By comparing $\gamma_\mathrm{SL}$ from experiment to simulation, we extract $\rho_{NV} = 0.7\pm0.1$~ppm for sample S1.
For sample S2, we simply compare its NV fluorescence counts to sample S1 from the same experimental setup, and extract a NV density $\rho_{NV} = 0.30\pm0.04$~ppm.

\section{Continuous diffusive model}
\subsection{Detailed solution of the diffusion equation}
In the main text, we present the solution of the diffusion equation, which we use as the fitting functional form to analyze the experimental data and extract the diffusion coefficient. 
Here, we present the detailed derivation of these solutions. 

\subsubsection{Simple diffusion model}
In the simplest case, the diffusion equation is written as
\begin{equation}\label{eq:diff}
  \partial_t P(t,\bm{r}) = D \nabla^2 P(t,\bm{r}) - \frac{P(t,\bm{r})}{T_1} + Q(t,\bm{r}),
\end{equation}
where $P(t,\bm{r})$ is the polarization depending on both position $r$ and time $t$, $D$ is the diffusion coefficient, $T_1$ is the intrinsic depolarization time scale of our system, and $Q(t,\bm{r})$ corresponds to the polarization source (the NV). 
Considering our experimental geometry and polarization protocol, we assume:
\begin{equation}\label{eq:pol}
  Q(t,\bm{r}) = 
\begin{cases}
\frac{\Gamma}{(2\pi b^2)^{3/2}} e^{-r^2/(2b^2)} & -\tau_p<t<0 \\
0 & t > 0
\end{cases},
\end{equation}
where $\Gamma$ is the polarization rate, and $b$ reflects the range of the polarization transfer process from the NV and also guarantees that the polarization does not diverge at short times. 

To solve this partial differential equation (PDE), we follow the Green's function approach. 
In particular the corresponding impulse response problem is
\begin{equation}\label{eq:homoreal}
\begin{split}
  &\partial_t P(t,\bm{r}) = D \nabla^2 P(t,\bm{r}) - \frac{P(t,\bm{r})}{T_1},\\
  &P(t=t_0,\bm{r}) = \delta(\bm{r}-\bm{r}_0), 
\end{split}
\end{equation}
in real space, or
\begin{equation} \label{eq:homofourier}
\begin{split}
      & \partial_t P_{\bm{k}}(t) = -Dk^2P_{\bm{k}}(t)-\frac{P_{\bm{k}}(t)}{T_1},\\
      & P_{\bm{k}}(t=t_0) = 1,
\end{split}
\end{equation}
in Fourier space. 
With the solution of the above equation (Green's function) denoted as $G(t,t_0,\bm{r},\bm{r}_0)$, we can obtain the survival probability given our polarization scheme:
\begin{equation}\label{eq:convolution}
P(t,\bm{r}=0) = \int_{-\tau_{\mathrm{p}}}^0 dt_0 \int d\bm{r}_0 G(t,t_0,\bm{r},\bm{r}_0) Q(t_0,\bm{r}_0).
\end{equation}

In the undriven case, we can simply get
\begin{equation}\label{eq:sol}
  G(t,t_0;\bm{r},\bm{r}_0) = \frac{1}{[4\pi D(t-t_0)]^{3/2}}\mathrm{exp}\left[-\frac{(\bm{r}-\bm{r}_0)^2}{2D(t-t_0)}\right]\mathrm{exp}\left[-\frac{t-t_0}{T_1}\right]. 
\end{equation}
Correspondingly, the survival probability is
\begin{equation}
\label{eq:fit1}
  P (t,\bm{r}=0) = \frac{\Gamma e^{\frac{b^2}{D T_1}}}{ 4\pi D^{3/2}\sqrt{T_1}}\left\{F\left[\left(t + \frac{b^2}{D}\right)\frac{1}{T_1}\right] - F\left[\left(t+\tau_p+ \frac{b^2}{D}\right)\frac{1}{T_1}\right]\right\}, 
\end{equation}
where
\begin{equation}
  F(x) = \frac{1}{\sqrt{\pi}}\frac{ e^{-x}}{\sqrt{x}} - \mathrm{erfc} (\sqrt{x}).
\end{equation}
In the driven case, we have a different diffusion coefficient (denoted as $D^{\mathrm{dr}}$) and a different decay time (denoted as $T_1^{\mathrm{dr}}$) in Eqs.~\ref{eq:homoreal},~\ref{eq:homofourier} for $t>0$. 
With these modifications, the Green's function then reads
\begin{equation}\label{eq:sol_dr}
  G(t,t_0;\bm{r},\bm{r}_0) = \frac{1}{[4\pi (D^{\mathrm{dr}}t-Dt_0)]^{3/2}} \mathrm{exp}\left[-\frac{(\bm{r}-\bm{r}_0)^2}{2(D^{\mathrm{dr}}t-Dt_0)}\right]\mathrm{exp} \left[-\left(\frac{t}{T^{\mathrm{dr}}_1}-\frac{t}{T_1}\right) \right]. 
\end{equation}
Similarly, using Eqn.~\ref{eq:convolution}, we obtain the survival probability as
\begin{equation}
\label{eq:fit_dr}
  P (t,\bm{r}=0) = \frac{\Gamma e^{\frac{b^2}{D T_1}+(\frac{D^{\mathrm{dr}}}{D}-\frac{T_1}{T^{\mathrm{dr}}_1})\frac{t}{T_1}}}{ 4\pi D^{3/2}\sqrt{T_1}} \left\{ F\left[\left(\frac{D^{\mathrm{dr}}}{D}t + \frac{b^2}{D}\right)\frac{1}{T_1}\right]-e^{(\frac{T^{\mathrm{dr}}_1}{T_1}-\frac{D}{D^{\mathrm{dr}}})\frac{\tau_\mathrm{p}}{T^{\mathrm{dr}}_1}}F\left[\left(\frac{D^{\mathrm{dr}}}{D}t+\tau_\mathrm{p}+ \frac{b^2}{D}\right)\frac{1}{T_1}\right]\right\}.
\end{equation}

\subsubsection{Subleading correction}
As discussed in the main text, while Eq.~\ref{eq:homofourier} captures the simplest diffusion, a generic diffusive system is described by a more general form:
\begin{equation} \label{eq:homofourier}
\begin{split}
      & \partial_t P_{\bm{k}}(t) = -f(\bm{k})P_{\bm{k}}(t),\\
      & P_{\bm{k}}(t=t_0) = 1,
\end{split}
\end{equation}
where we neglect the homogeneous intrinsic decay, and as discussed in the main text, $f(\bm{k})$ is written as follows: 
\begin{itemize}
\item For short-range interacting system and long-range interacting system ($1/r^\alpha$ interaction in $d$ dimensions) with $\alpha\ge d+4$
\begin{equation}\label{eq:fourierk4}
  f(\bm{k}) = Dk^2+Ck^4+\cdots; 
\end{equation}
\item For long-range interacting system ($1/r^\alpha$ interaction in $d$ dimensions) with $d+2<\alpha< d+4$
\begin{equation}\label{eq:fourierk3}
  f(\bm{k}) = Dk^2+C_{\mathrm{lr}}k^3+\cdots. 
\end{equation}
\end{itemize}

With this diffusion equation in Fourier space, we can obtain the survival probability as
\begin{equation}\label{eq:GreensLR}
G(t_0,t)=\frac{1}{(2\pi)^d}\int d^d \bm{k} e^{-f(k)(t-t_0)}.
\end{equation}
At late times, $k^2$ term in Eqns.~\ref{eq:fourierk4}-\ref{eq:fourierk3} dominates the dynamics, while the subleading correction can be evaluated by expanding the exponential in Eqn.~\ref{eq:GreensLR} as follows.
\begin{itemize}
    \item When $d=3$ and $\alpha\ge 7$,
    \begin{equation}\label{eq:Sup_k4}
    G\approx\frac{1}{(2\pi)^3}\int d^3 \bm{k} e^{-D k^2(t-t_0)}\{1+D' k^4(t-t_0)\}=\frac{1}{[4\pi D (t-t_0)]^{3/2}}+\frac{15C}{32\pi^{3/2}D^{7/2}(t-t_0)^{5/2}}.
    \end{equation}
    \item When $d=3$ and $\alpha=6$ (our experimental setup),
    \begin{equation}\label{eq:Sup_k3}
    G\approx\frac{1}{(2\pi)^3}\int d^3 \bm{k} e^{-D k^2(t-t_0)}\{1+C_{\mathrm{lr}} k^3(t-t_0)\}=\frac{1}{[4\pi D (t-t_0)]^{3/2}}+\frac{C_{\mathrm{lr}}}{2\pi^2D^3(t-t_0)^2}.
    \end{equation}
\end{itemize}
Here we hasten to emphasize that the subleading correction $\sim (t-t_0)^{-2}$ distinguishes our long-range interacting system from the normal diffusion whose subleading correction $\sim (t-t_0)^{-5/2}$.
We also remark that, similar to the previous subsection, we can also convolve the above survival probability with the polarization process to obtain the correct form for our experimental signal.

\section{Rate equation model}

In this section we derive our semi-classical model using two different formalisms: the master equation and Fermi's golden rule.

\subsection{Master equation approach}

In developing a master equation approach to the polarization transfer rate in our spin system, we begin by isolating a single pair of spins whose polarization dynamics we wish to study.
Let us denote them by $S$ and $P$ (here we restrict our analysis of the NV center to the two lowest levels of interest in the polarization transfer dynamics dynamics)
For each of the spins, let us associate Pauli operators $\sz,\smi, \spl$.
The two-spin Hamiltonian can be written as:
\begin{equation}
    H = (\Delta + \delta_S)\frac{\sz_S}{2} + (\Delta + \delta_P)\frac{\sz_P}{2} + J_{zz} \sz_S \sz_P + J_{\perp} (\spl_{S} \smi_P + \smi_S \spl_P)
\end{equation}
where we already focus on the approximate energy conserving terms $J_{zz},J_{\perp}, \delta^S, \delta^P \ll \Delta$.
The only two states that can exhibit dynamics due to the interactions live in the zero magnetization subspace $\{ \ket{\uparrow_S\downarrow_P} = \ket{A}, \ket{\downarrow_S\uparrow_P} = \ket{B}\}$ with Hamiltonian: 
\begin{equation}
    H_{sub} = \delta \proj{A} + J_\perp \left[\projj{A}{B} + \projj{B}{A}\right]~,
\end{equation}
where $\delta$ accounts for the energy mismatch between the two levels.
In isolation, the population of the two states would coherently oscillate with a well-defined frequency; the presence of a bath of P1s and optical pumping leads to additional decoherence dynamics that modify the dynamics strikingly. 
These can be self-consistently included within the density matrix formalism, by adding a off-diagonal decoherence decay rate and optical pumping to an additional level. 

\begin{figure}
    \centering
    \includegraphics[width = 4.8in]{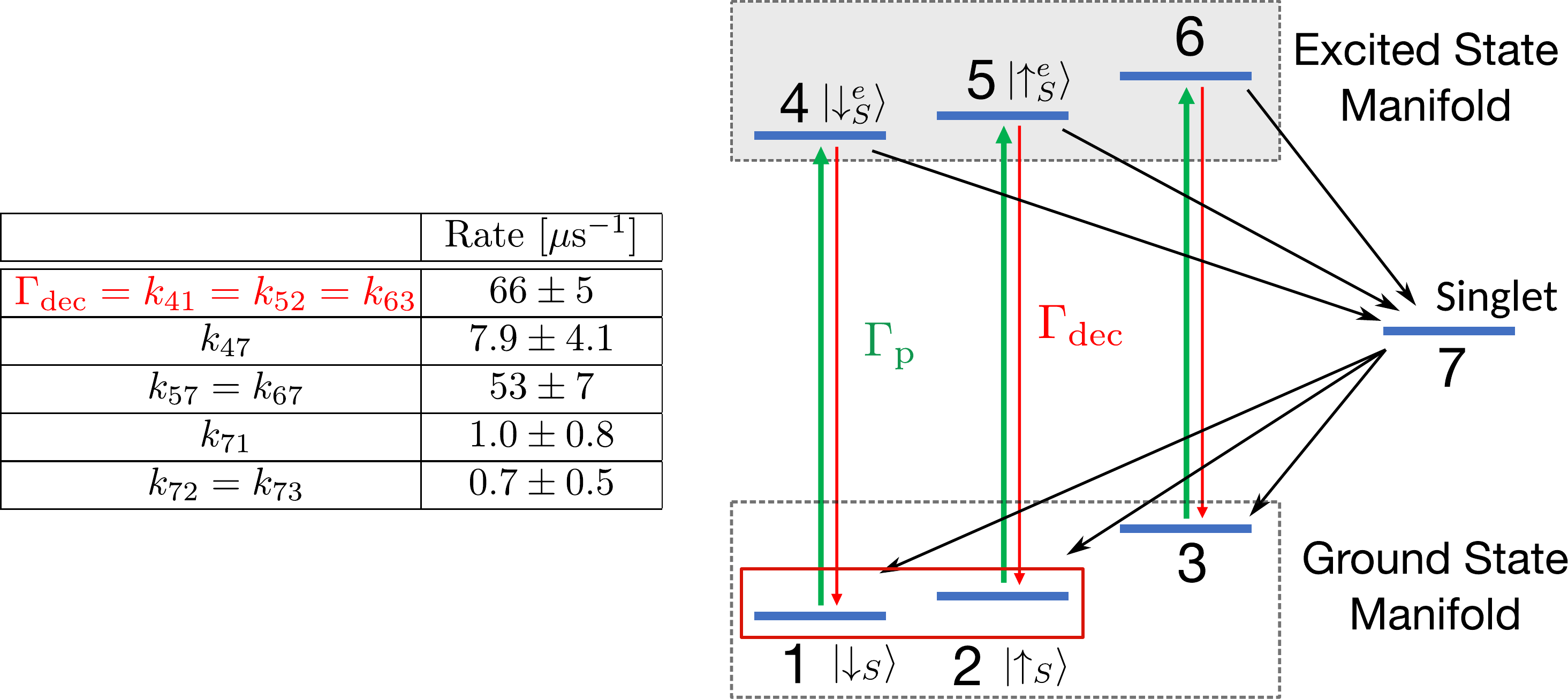}
    \caption{Diagram of the NV internal structure with $\ket{m_s=0} = \ket{\downarrow_S}$ and $\ket{m_s = -\frac{1}{2}} = \ket{\uparrow_S}$ levels highlighted.
    The NV level structure is composed of a spin-1 ground state and excited manifold as well as a single level.
    Rates between the different NV levels (table) as measured in Refs.~\cite{robledo2011spin, tetienne2012magnetic}. 
    }
    \label{fig:NVRates}
\end{figure}

\emph{Optical pumping to another state}---In simulating the dynamics of the experiment, one important feature is the polarization of the NV via its internal structure.
Briefly, the full structure of the NV includes two excited spin-$1$ manifolds, as well as a singlet level, Fig.~\ref{fig:NVRates}.
The various decay rates between the different states (independently studied in Refs.~\cite{robledo2011spin, tetienne2012magnetic}) leads to a preferential polarization of the $\ket{m_s=0}$ state in the ground state manifold, under spin-conserving optical polarization from the ground state to the excited states---let the rate of this process be $\Gamma_{\mathrm{p}}$ and the natural decay rate to be $\Gamma_{\mathrm{dec}}$.


To study this effect, let us consider optical pumping of the ground state levels $\ket{\uparrow_S}$ and $\ket{\downarrow_S}$ to the corresponding excited states $\ket{\uparrow^e_S}$ and $\ket{\downarrow^e_S}$. 
In the full Hilbert space, these induce transition $\ket{A} = \ket{\uparrow_S \downarrow_P} \leftrightarrow \ket{\uparrow_S^e \downarrow_P} = \ket{C}$ and $\ket{B} = \ket{\downarrow_S \uparrow_P} \leftrightarrow \ket{\downarrow_S^e \uparrow_P} = \ket{D}$.
The corresponding Linbladian quantum jump term for both pumping and decay are given by:
\begin{align}
\dot{\rho}_{\mathrm{opt}+\mathrm{dec}} &= - \frac{\Gamma_{\mathrm{p}}}{2} \left[ \proj{A}\rho +\rho\proj{A} -\proj{C}\rho_{AA} + \proj{B}\rho + \rho \proj{B} - 2 \proj{D}\rho_{BB} \right] \\
& - \frac{\Gamma_{\mathrm{dec}}}{2} \left[ \proj{C} \rho + \rho \proj{C} - 2\proj{A} \rho_{CC} + \proj{D} \rho + \rho \proj{D} - 2\proj{B} \rho_{DD} \right]
\end{align}
which becomes a bit more insightful in matrix form:
\begin{equation}
  \dot{\rho}_{\mathrm{opt}+\mathrm{dec}} =
  \left[
  \begin{array}{cc|cc}
    -\Gamma_{\mathrm{p}}\rho_{AA} + \Gamma_{\mathrm{dec}} \rho_{CC} & - \Gamma_{\mathrm{p}} \rho_{AB} & -\bar{\gamma} \rho_{AC} & -\bar{\gamma} \rho_{AD}\\
    -\Gamma_{\mathrm{p}} \rho_{BA} & -\Gamma_{\mathrm{p}}\rho_{BB} + \Gamma_{\mathrm{dec}} \rho_{DD} & -\bar{\gamma} \rho_{BC} & -\bar{\gamma} \rho_{BD}\\\hline
    -\bar{\gamma} \rho_{CA} & -\bar{\gamma}\rho_{CB} & \Gamma_{\mathrm{p}} \rho_{AA} - \Gamma_{\mathrm{dec}} \rho_{CC} & -\Gamma_{\mathrm{dec}} \rho_{CD}\\
    -\bar{\gamma} \rho_{DA} & -\bar{\gamma}\rho_{DB} & -\Gamma_{\mathrm{dec}} \rho_{DC} & \Gamma_{\mathrm{p}} \rho_{BB} - \Gamma_{\mathrm{dec}} \rho_{DD}
  \end{array}
  \right] \label{eq:DecAndExc}
\end{equation}
where $\bar{\gamma} = (\Gamma_{\mathrm{p}} + \Gamma_{\mathrm{dec}}) / 2$.

Immediately, we observe that the off-diagonal corrections with the $\ket{C}$ and $\ket{D}$ states are simply decaying. Since they start at zero, they remain zero and do not affect the dynamics of the system.
The pumping only affects the dynamics between $\ket{A}$ and $\ket{B}$ by inducing an additional decoherence of off-diagonal $\rho_{AB}$ term. The remaining dynamics affect only the diagonal component, which correspond to the populations in each of the levels.

This highlights that the presence of the complex structure of the NV center can be accounted by the diagonal components of the density matrix, up to an additional decoherence rate causes by the pumping to the excited manifold.

\emph{Extrinsic decoherence}---By contrast, adding the extrinsic decoherence rate arising from other spins in the system is much simpler and corresponds to an additional decay of the off-diagonal terms with rate $\gamma$.

Putting everything together, the equations of motion are given by:
\begin{align}
    \dot{\rho}_{AA} &= -iJ_\perp (\rho_{BA}-\rho_{AB}) - \Gamma_{\mathrm{p}}\rho_{AA} + \Gamma_{\mathrm{dec}} \rho_{CC}\\
    \dot{\rho}_{BB} &= -iJ_\perp (\rho_{AB}-\rho_{BA}) - \Gamma_{\mathrm{p}}\rho_{BB} + \Gamma_{\mathrm{dec}} \rho_{DD}\\
    \dot{\rho}_{CC} &= \Gamma_{\mathrm{p}}\rho_{AA} - \Gamma_{\mathrm{dec}} \rho_{CC}\\
    \dot{\rho}_{DD} &= \Gamma_{\mathrm{p}}\rho_{BB} - \Gamma_{\mathrm{dec}} \rho_{DD}\\
    \dot{\rho}_{AB} &= (i\delta -\gamma - \Gamma_{\mathrm{exc}}) \rho_{AB} - iJ_\perp(\rho_{BB} - \rho_{AA}) = [\dot{\rho}_{BA}]^*
\end{align}
while the remaining terms are zero. Adiabatically eliminating the coherence between $\ket{A}$ and $\ket{B}$, we get a modified set of equations for $\rho_{AA}$ and $\rho_{BB}$:
\begin{align}
    \dot{\rho}_{AA} &= -2|J_\perp|^2 (\rho_{AA}-\rho_{BB}) \frac{\gamma + \Gamma_{\mathrm{p}} }{(\gamma + \Gamma_{\mathrm{p}})^2 + \delta^2} - \Gamma_{\mathrm{p}}\rho_{AA} + \Gamma_{\mathrm{dec}} \rho_{CC}\\
    \dot{\rho}_{BB} &= -2|J_\perp|^2 (\rho_{BB}-\rho_{AA}) \frac{\gamma + \Gamma_{\mathrm{p}} }{(\gamma + \Gamma_{\mathrm{p}})^2 + \delta^2}  - \Gamma_{\mathrm{p}}\rho_{BB} + \Gamma_{\mathrm{dec}} \rho_{DD}
\end{align}

Finally, we assume that the density matrix remains diagonal, $\rho_{AA} \approx \rho^S_{\uparrow\uparrow} \rho^P_{\downarrow \downarrow}$ and $\rho_{BB} \approx \rho^S_{\downarrow\downarrow} \rho^P_{\uparrow \uparrow}$.
In a similar way we assume $\dot{\rho}_{AA}$ captures the polarization transfer rate so $\dot{\rho}_{AA} \approx \dot{\rho}^S_{\uparrow \uparrow} = - \dot{\rho}^P_{\downarrow\downarrow}$.

\subsection{Fermi's golden rule}

A different way to derive our semi-classical model is through Fermi's golden rule; polarization exchange corresponds to decay of a single spin to a bath composed of all other spin in the system. 
Owing to the presence of strong disorder (both on on-site fields and position), the spectrum of the bath modes should exbhibit important structure---peaked around the energy difference of each spin and with some broadening $\gamma$ induced by interactions.  

A more precise analysis of the decay closely follows the analysis of decay of an atom in electromagnetic field.
Focusing on a two level spin $\ket{s} = \in \{\ket{\uparrow}, \ket{\downarrow}\}$ and a set of bath modes $\ket{k}$, the Hilbert space of the system undergoing decay can be written as  $\{ \ket{\uparrow,0} = \ket{e}, \ket{\downarrow, k} = \ket{g_k} \}$, interacting via the Hamiltonian:

\begin{align}
  H = (\Delta + \delta) \proj{e} + \sum_k \epsilon_k \proj{g_k} + \sum_{k} J  \big[ \projj{e}{g_k} + \projj{g_k}{a}\big]~
\end{align}
where $\Delta +\delta$ corresponds to the splitting of the spin of interest and $\epsilon_k$ the energy of the mode $k$ of the bath.
Moving into the the interaction picture of $\ket{e}$ and $\ket{g_k}$:
\begin{align}
  H_{int} =\sum_{k} J_k \left[e^{-i((\Delta + \delta)-\epsilon_k)t}\projj{e}{g_k} + e^{i((\Delta + \delta)-\epsilon_k)t}\projj{g_k}{e}\right]~.
\end{align}



In the case when either of the spins is being pumped (like the NV must be during polarization), there will be an additional decoherence channel proportional to the strength of the pumping $\Gamma_{\mathrm{p}}$.
Including this contribution is most straightforwardly done via the density matrix $\rho$, where it emerges as a decay of the off-diagonal component.
\begin{align} 
  \dot{\rho}_{ee} &= -i \sum_k J \left[ e^{-i((\Delta + \delta) -\epsilon_k)t} \rho_{g_ke} - e^{i((\Delta + \delta) - \epsilon_k)t} \rho_{eg_k}\right] \label{eq:temp1}\\ 
  \dot{\rho}_{eg_k} &= -iJ e^{-i((\Delta + \delta) -\epsilon_k) t} (\rho_{g_kg_k} -\rho_{ee}) - \Gamma_{\mathrm{p}}\rho_{eg_k}
\end{align}

Formally integrating the second equation assuming zero coherence at $t=0$, $\rho_{eg_k}(t=0)=0$ yields:
\begin{align}
  \rho_{eg_k} = -iJ \int_0^tdt'~(\rho_{g_kg_k}(t') - \rho_{ee}(t')) e^{-i((\Delta + \delta)-\epsilon_k)t'} e^{-\Gamma_{\mathrm{p}} (t-t')}
\end{align}
Inserting into the Eqn.~\ref{eq:temp1} and focusing on the first term, we have:
\begin{align}
   e^{-i((\Delta + \delta) -\epsilon_k)t} & \int_0^tdt'~[\rho_{g_kg_k}(t') - \rho_{ee}(t')] e^{i((\Delta + \delta)-\epsilon_k)t'} e^{-\Gamma_{\mathrm{p}} (t-t')}\\
   &\approx \sum_k [\rho_{g_kg_k}(t) - \rho_{ee}(t)] \int_0^t dt'~ e^{[-i((\Delta + \delta) -\epsilon_k) - \Gamma_{\mathrm{p}}](t-t')}\\
   &\approx\lim_{t\to\infty} \sum_k (\rho_{g_kg_k}(t) - \rho_{ee}(t)) \frac{1-e^{[-i((\Delta + \delta) -\epsilon_k) - \Gamma_{\mathrm{p}}]t}}{i((\Delta + \delta) - \epsilon_k) + \Gamma_{\mathrm{p}}} \\
  &\approx\sum_k (\rho_{g_kg_k}(t) - \rho_{ee}(t)) \frac{1}{i((\Delta + \delta) - \epsilon_k) + \Gamma_{\mathrm{p}}}\\
  &\approx(\rho_{BB} - \rho_{AA}) \int_{-\infty}^\infty d\epsilon  \frac{\rho(\epsilon)}{i((\Delta + \delta) - \epsilon_k) + \Gamma_{\mathrm{p}}} \label{eq:finalInt} 
\end{align}
where, we have have taken $\rho_{g_kg_k}$ to be slowly varying across different modes $g_k$ around the center frequency of the bath modes and thus the average $\overline{\rho_{g_kg_k}}$.
Physically, this corresponds to coupling to a single other spin, where the bath modes correspond to a broadening of the spin energy levels and their occupation is determined by the state of the spin (either in $\ket{\uparrow}$ or $\ket{\downarrow}$).
Considering the interaction of multiple such modes corresponds to summing over many independent channels as described above.

$\rho(\epsilon)$ is the density of states of the bath modes, allowing us to transform the sum into an integral, which is a necessary input in our theory.
Motivated by the usual broadening in atomic physics, we take $\rho(\epsilon)$ to a Lorentzian, that in the rotating frame, is centered around $\Delta$, with FWHM $2\gamma$:
\begin{align}
  \rho(\epsilon) = \frac{1}{\pi} \frac{\gamma}{\gamma^2 + (\epsilon - \Delta)^2}~.
\end{align}
As such $\delta$ alone captures the energy mismatch between the spin and the center of the bath.
Solving for the integral in Eqn.~\ref{eq:finalInt} and including the second term (which is the complex conjugate of the first), we arrive at the formula for polarization transfer:
\begin{align}
  \dot{\rho}_{ee} &= |J|^2 (\overline{\rho_{g_kg_k}} - \rho_{ee}) 2 \Re\left[ \frac{1}{(\gamma + \Gamma_{\mathrm{p}}) + i\delta}\right] = -2|J|^2  \frac{\gamma + \Gamma_{\mathrm{p}}}{(\gamma + \Gamma_{\mathrm{p}})^2 + \delta^2} (\rho_{ee}-\overline{\rho_{g_kg_k}})~. \label{eq:Rate_Fermi}
\end{align}
Analogous to the master equation case, we assume that the density matrix is separable between the two spins involved analyzed and thus $\rho_{ee} = \rho^{(1)}_{\uparrow \uparrow} \rho^{(2)}_{\downarrow \downarrow}$ while $\overline{\rho_{g_kg_k}} = \rho^{(1)}_{\downarrow \downarrow} \rho^{(2)}_{\uparrow \uparrow}$ and $\dot{\rho}_{ee} \approx \dot{\rho}^{(1)}_{\uparrow \uparrow} = - \dot{\rho}^{(2)}_{\downarrow \downarrow}$.

Generalizing to many different spins corresponds to summing over the different bath spins that the first spin can decay to---each spin gives rise to a decay channel with slightly broadened levels and interacting with different couplings $J$.
Labelling the bath spins with $j$, we arrive at the total depolarization of the initial spin as:
\begin{align}
    \dot{\rho}^{(1)}_{\uparrow \uparrow} = \sum_j -2|J_j|^2  \frac{\gamma + \Gamma_{\mathrm{p}}}{(\gamma + \Gamma_{\mathrm{p}})^2 + \delta^2} (\rho^{(1)}_{\uparrow \uparrow} \rho^{(j)}_{\downarrow \downarrow} - \rho^{(1)}_{\downarrow \downarrow} \rho^{(j)}_{\uparrow \uparrow})
\end{align}
under pumping of spin $(1)$ of strength $\Gamma_{\mathrm{p}}$.



\section{Long-range diffusion on the lattice}

In Fig.~E14 of the Extended Data, we studied the survival probability in a  three dimensional disorderless lattice with short and long-range hoppings.

In particular we consider the following hopping problem:
\begin{equation}
    \partial_t P_i = \sum_j f(|\bm{r}_j-\bm{r}_i|) (P_j - P_i),
\end{equation}
where $P_i$ is the polarization of site $i$, $\bm{r}_i$ is its position and $f(r)$ describes the strength of the hopping between sites a distance $r$ apart.

We focus on two different hopping functions: a short-range $f_{\mathrm{sr}}(r)$ which is only non-zero for nearest neighbor sites (where it takes the value $\Gamma$), and a long-range $f_{\mathrm{lr}}(r) = \Gamma / r^\alpha$. 
Since we are interested in the intermediate regimes of power-laws where the approach to the diffusion behavior is modified due to the long-range interactions, we consider $\alpha = 6$ in $d=3$.

The problem is more straightforwardly solved in Fourier space (assuming periodic boundary conditions) where it becomes:
\begin{equation}
    \partial_t P_{\bm{k}} = P_{\bm{k}} \underbrace{ \sum_{\Delta \bm{r}} f(|\Delta \bm{r}|) (e^{i\bm{k}\cdot \Delta \bm{r}} -1)}_{f_{\bm{k}} }
\end{equation}
The problem is diagonal in Fourier space and requires only solving for the decay $f_{\bm{k}}$ of each Fourier mode $\bm{k}$ which can be easily computed in the discrete lattice via brute force or Fourier Transform.

Starting with unit polarization at the center site $\bm{r}_0 = (0,0,0)$ (which corresponds to uniform distribution in the discrete Fourier space), we compute the decay of each Fourier mode as:
\begin{equation}
    P_{\bm{k}} = \frac{1}{V} e^{f_{\bm{k}} t}~,
\end{equation}
where $V$ is the volume of the system (in this case the number of lattice sites)

The survival probability is given by the population in the original lattice point $\bm{r}_0$ which can be computed as the sum of the populations on all Fourier modes:
\begin{equation}
    S_p(t) = \sum_{\bm{k}} P_{\bm{k}}~.
\end{equation}

For a better comparison, we normalize the data by the natural timescale in the system $a^2/D$, where $a$ is the lattice constant and $D$ is the diffusion coefficient.
The diffusion coefficient can be trivially obtained for $f_{\mathrm{sr}}$ by computing the $k^2$ expansion of $f_{\bm{k}}$, while for $f_{\mathrm{lr}}$ one must perform a $1/L = V^{-1/3}$ expansion of finite-size values of the diffusion coefficient to extract the correct approach in the thermodynamic limit.

\bibliography{bibliography}


\title{Methods and Extended Data:\\
Emergent hydrodynamics in a strongly interacting dipolar spin ensemble}


\author{Chong Zu}
\thanks{These authors contributed equally to this work.}
\affiliation{Department of Physics, University of California, Berkeley, CA 94720, USA}

\author{Francisco Machado}
\thanks{These authors contributed equally to this work.}
\affiliation{Department of Physics, University of California, Berkeley, CA 94720, USA}

\author{Bingtian Ye}
\thanks{These authors contributed equally to this work.}
\affiliation{Department of Physics, University of California, Berkeley, CA 94720, USA}

\author{Soonwon Choi}
\affiliation{Department of Physics, University of California, Berkeley, CA 94720, USA}

\author{Bryce Kobrin}
\affiliation{Department of Physics, University of California, Berkeley, CA 94720, USA}
\affiliation{Materials Science Division, Lawrence Berkeley National Laboratory, Berkeley, CA 94720, USA}

\author{Thomas Mittiga}
\affiliation{Department of Physics, University of California, Berkeley, CA 94720, USA}
\affiliation{Materials Science Division, Lawrence Berkeley National Laboratory, Berkeley, CA 94720, USA}

\author{Satcher Hsieh}
\affiliation{Department of Physics, University of California, Berkeley, CA 94720, USA}
\affiliation{Materials Science Division, Lawrence Berkeley National Laboratory, Berkeley, CA 94720, USA}

\author{Prabudhya Bhattacharyya}
\affiliation{Department of Physics, University of California, Berkeley, CA 94720, USA}
\affiliation{Materials Science Division, Lawrence Berkeley National Laboratory, Berkeley, CA 94720, USA}

\author{Matthew Markham}
\affiliation{Element Six, Harwell, OX11 0QR, United Kingdom}

\author{Dan Twitchen}
\affiliation{Element Six, Harwell, OX11 0QR, United Kingdom}

\author{Andrey Jarmola}
\affiliation{Department of Physics, University of California, Berkeley, CA 94720, USA}
\affiliation{U.S. Army Research Laboratory, Adelphi, Maryland 20783, USA}

\author{Dmitry Budker}
\affiliation{Department of Physics, University of California, Berkeley, CA 94720, USA}
\affiliation{Helmholtz Institut Mainz, Johannes Gutenberg Universitat Mainz, 55128 Mainz, Germany}

\author{Chris R. Laumann}
\affiliation{Department of Physics, Boston University, Boston, MA 02215, USA}

\author{Joel E. Moore}
\affiliation{Department of Physics, University of California, Berkeley, CA 94720, USA}
\affiliation{Materials Science Division, Lawrence Berkeley National Laboratory, Berkeley, CA 94720, USA}

\author{Norman Y. Yao}
\affiliation{Department of Physics, University of California, Berkeley, CA 94720, USA}
\affiliation{Materials Science Division, Lawrence Berkeley National Laboratory, Berkeley, CA 94720, USA}

\date{\today}

\maketitle

\section{Methods}

\subsection{Experimental description}
\subsubsection{Sample preparation}
Sample S1 and S2 used in this work are synthetic type-Ib single crystal diamonds (Element Six) with intrinsic substitutional nitrogen (P1) concentration $\sim 110$~ppm.
To create NV centers, the samples were first irradiated with electrons (S1: $14$~MeV energy and $1\times10^{18}$~cm$^{-2}$ dosage, S2: $2$~MeV energy and $1\times10^{18}$~cm$^{-2}$ dosage) to generate vacancies.
After irradiation, the diamonds were annealed in vacuum ($10^{-6}$~Torr) with temperature $>800^\circ$C.
During annealing, the vacancies become mobile and probabilistically form NV centers with intrinsic P1 centers.
We calibrate P1 and NV densities independently (see SI \cite{SI}), and extract $\rho_{P1} \sim 110$~ppm and $\rho_{NV}\sim 0.7$~ppm for sample S1, and $\rho_{P1} \sim 110$~ppm and $\rho_{NV}\sim 0.3$~ppm for sample S2.

\subsubsection{Experimental pulse sequence}

In experiments, the measurement sequence is performed with a differential readout to mitigate the effect of NV and P1 charge dynamics under laser pumping and reliably extract the spin population difference between NV $\ket{0}$ and $\ket{-1}$ states (Fig.~\ref{fig:figS_seq}) \cite{mrozek2015longitudinal,choi2017depolarization, mittiga2018imaging,aslam2013photo,manson2018nv,jayakumar2016optical,bluvstein2019identifying}.
In particular, we first let the NV charge dynamics reach a steady state by waiting for $1$~ms in dark (I).
To reset the P1 polarization before each sequence, we also apply 8 consecutive $\frac{\pi}{2}$-pulses on the resonant P1 subgroup, with a pulse spacing $10$~$\mathrm{\mu}$s, more than two orders of magnitude longer than the Ramsey coherence time [$0.032\pm 0.005~\mathrm{\mu}$s].
After charge equilibration and P1 reset, we apply a laser polarization pulse with time $\tau_\mathrm{p}$ (II), followed by the spin manipulation pulses and time evolution (III).
In this work, we use up to three different microwave drives which target different transitions:
the NV $\ket{0} \leftrightarrow \ket{-1}$ (and thus the resonant P1 subgroup $\ket{\frac{1}{2}} \leftrightarrow \ket{-\frac{1}{2}}$),
the NV $\ket{0} \leftrightarrow \ket{1}$, 
and the non-resonant P1 subgroup $\ket{\frac{1}{2}} \leftrightarrow \ket{-\frac{1}{2}}$.
At the end, we detect the NV fluorescence via a second laser illumination (IV).
By repeating the same procedure with an additional NV $\ket{0} \leftrightarrow \ket{-1}$ $\pi$-pulse before readout, we measure the fluorescence of an orthogonal spin state, and can use the difference between the two measurement to extract the NV spin polarization \cite{choi2017depolarization,mrozek2015longitudinal}.
Since this general procedure is applied to all experiments in this work, in the main text we highlight the pulse sequences corresponding to the first II, III and IV regions.

\subsection{P1 hyperfine interaction}

The host substitutional nitrogen atom of the P1 center has non-zero nuclear spin, which leads to a strong hyperfine interaction between its electronic and nuclear spin. 
Since $^{14}$N isotope accounts for $\sim99.6\%$ of all the nitrogen atoms (natural abundance), in the following analysis, we consider each P1 as having a host $^{14}$N with nuclear spin $I=1$. 
The P1-$^{14}$N Hamiltonian under an applied external magnetic field $B$ can be written as
\begin{equation}
\begin{split}
H = & B(\gamma_e P_z + \gamma_n I_z) + Q(\sin\theta I_x+\cos\theta I_z)^2 + A_\| P_z I_z + A_\perp (P_xI_x + P_yI_y)
\end{split}
\end{equation}
where $Q = (2\pi)\times-4.95~\mathrm{MHz}$ is the nuclear quadruple splitting, 
$\gamma_e = (2\pi)\times2.8~\mathrm{MHz/G}$ and $\gamma_n = (2\pi)\times-0.307~\mathrm{kHz/G}$ are P1 electronic and $^{14}$N nuclear gyromagnetic ratio respectively, 
$\{I_x,I_y,I_z\}$ are the nuclear spin operators.
Due to Jahn-Teller distortion \cite{hall2016detection}, the P1 electron can occupy one of the four crystal axes with equal probability. 
This distortion also defines the axis of the $^{14}$N quadrupole interaction, whose effect is captured by the angle $\theta$ (defined as the angle between the external magnetic field $B$ and the P1 axis of interest).
$A_{\parallel}$ and $A_{\perp}$ characterize the hyperfine coupling strength between P1 electron and nuclear spins, whose values also depend on P1 axis.
When the P1 axis is parallel to the external magnetic field $B$ (probability $p_1 = \frac{1}{4}$, $\theta = 0$), $A_{\parallel,1} = (2\pi)\times114$~MHz and $A_{\perp,1} = (2\pi)\times81$~MHz. 
When the P1 axis is parallel to any of the other three crystal axes [probability $p_2 = \frac{3}{4}$, $\theta = \mathrm{arccos}(-\frac{1}{3})$], $A_{\parallel,2} = (2\pi)\times85$~MHz and $A_{\perp,2} = (2\pi)\times99$~MHz \cite{hall2016detection}.

\subsubsection{Five distinct P1 subgroups}
In our experiment, we work at $B\sim 500$~G, where the P1 electron splitting $\lvert \Delta_{P1} \rvert = \lvert \gamma_e B \rvert \gg \lvert A_{\perp} \rvert$ , and $\lvert A_{\parallel} \rvert \gg \lvert Q \rvert$, and thus we can neglect the transverse terms in the Hamiltonian:
\begin{equation}
H = B(\gamma_e P_z + \gamma_n I_z) +Q\cos^2\theta I_z^2 + A_\| P_z I_z.
\end{equation}
When the nuclear spin is in the $\ket{m_I = 0}$ state ($p_N = \frac{1}{3}$), the four different Jahn-Teller orientations for the P1 have nearly degenerate transitions at frequency $\sim \gamma_e B$, which corresponds to the P1 subgroup  $\nu = \frac{1}{3}$. 
When the nuclear spin is in the $\ket{m_I = \pm 1}$ state and the P1 axis is along with the external magnetic field direction, the P1 energy splitting is given by $\sim \Delta_{P1} \pm A_{\parallel,1}$, with a probability $\nu = p_1 p_N = \frac{1}{12}$ of the total P1 concentration. 
When the P1 occupies any of the other 3 axes, the different hyperfine couplings lead to a different energy splitting: $\sim \Delta_{P1} \pm A_{\parallel,2}$, with $\nu = (1-p_1) p_N = \frac{1}{4}$ of the total P1 concentration. 
Taken together, there are five distinct P1 groups, with effective ratios $\nu \in \{ \frac{1}{12}, \frac{1}{4}, \frac{1}{3}, \frac{1}{4}, \frac{1}{12} \}$ of the total P1 density.

\subsubsection{Nuclear mediated leakage channel}
To control the strength of the onsite random field, we continuously drive an off-resonant P1 subgroup with a microwave field to ``average out'' a portion of the Ising interactions.
However, this microwave may induce transitions within the resonant P1 subgroup which lead to additional unwanted depolarization channels.
Such channels can be divided in two different categories:
1) the microwave field off-resonantly drives the resonant P1 subgroup, effectively enhancing the intrinsic depolarization decay;
2) the microwave resonantly drive the P1-$^{14}$N system, depolarizing the P1 electron through different \emph{nuclear spin} states.
While the former channel can be captured by including a modified $T_1$ in our analysis, the latter requires a more careful treatment. 
Here, we show that, although the polarization leakage happens at certain driving frequencies (`$\nu=\frac{1}{3}$' peak in Fig.~3c), the other peaks (`$\nu=\frac{1}{12}$', `$\nu=\frac{1}{4}$' and `$F$') do not suffer from such leakage. 
This corroborates our understanding that the change of polarization dynamics in Fig.~3d purely originates from reducing strength of onsite random fields. 

To study the leakage mechanism, we first consider the Hamiltonian of the P1-$^{14}$N in the presence of a microwave field [$H_{\mathrm{MW}} = \Omega \cos(\omega t)$] with strength $\Omega$ and frequency $\omega$.
In the rotating frame of microwave frequency $\omega$ and after the secular approximation,
the system is described by the following effective Hamiltonian:
\begin{align}\label{eq:HLeakRWA}
    H = (B \gamma_e-\omega) P_z + (\gamma_n B + A_\| P_z) I_z + Q\left[ \sin\theta I_x+\cos\theta I_z\right]^2 + \Omega P_x ~.
\end{align}
The first two terms, in conjunction with the different Janh-Teller orientations, yield the aforementioned five peak structure.
The addition of the nuclear quadrupole interaction mixes the nuclear spin states of the system, which, in conjunction with the microwave drive of the P1 electronic spin, leads to new depolarization channels for our states of interest, i.e. $|P_z=+\frac{1}{2},I_z=+1\rangle$ and $|P_z=-\frac{1}{2},I_z=+1\rangle$ in the $\nu=\frac{1}{4}$ subgroup. 
In this case, when the second driving frequency is resonant to the $\nu=\frac{1}{3}$ P1 subgroup ($B\gamma_e \approx \omega$), $|P_z=+\frac{1}{2},I_z=\pm1\rangle$ and $|P_z=-\frac{1}{2},I_z=\mp1\rangle$ are degenerate.
The presence of the nuclear quadrupole interaction $Q\sin^2\theta I_x^2$ mixes $|P_z=+\frac{1}{2},I_z=\pm1\rangle$ with $|P_z=+\frac{1}{2},I_z=\mp1\rangle$ which is then connected to $|P_z=-\frac{1}{2},I_z=\mp1\rangle$ via the microwave drive ($\Omega_{\mathrm{drive}} P_x$).
The coupling strength between the two states can be estimated using second-order perturbation theory as $\Omega \frac{Q\sin^2 \theta}{A_{\|}}\sim (2\pi)\times0.6 \mathrm{MHz}$ for the drive $\Omega_{\mathrm{drive}} = (2\pi)\times11.7~\mathrm{MHz}$ considered in the main text.
The resulting depolarization dominates the dynamics and leads to the fast decay seen in Fig.~3b.
A similar process occurs when $\omega \approx \gamma_e B+\frac{A_{\|}}{2}$, $|P_z=+\frac{1}{2},I_z=+1\rangle$ and $|P_z=-\frac{1}{2},I_z=0\rangle$ are degenerate, and thus microwave drive induces polarization leakage due to the quadrupole mixing of the levels.

Fortunately, these are the only two frequencies where such leakage effects are dominant.
We verify this expectation by numerically simulating the dynamics of the P1-$^{14}$N system starting from the $|P_z=+\frac{1}{2},I_z=+1\rangle$ state.
Indeed we only observe fast decay when the driving frequency is near $\omega = \gamma_e B$ (dashed line in Fig.~3b) and $\omega = \gamma_e B+\frac{A_{\|}}{2}$. 
The above analysis further corroborates the flexibility of our platform and how the observed additional decay in Fig.~3d arises from a reduction of the strength of the onsite random fields which directly lead to faster transport.





\subsection{Coherence measurement of the P1 ensemble}

To directly measure the coherent properties of P1 system, we first apply a laser pulse with duration $\tau_\textrm{p}=1000$~$\mu$s to polarize the $\nu = \frac{1}{3}$ P1 subgroup, and then shelve the NV spin into the highly off-resonant $\ket{+1}$ sublevel for a fixed time $\tau_{\mathrm{w}} = 30$~$\mu$s (Fig.~\ref{fig:figS_Coherence}).
Within this shelving period, we apply various dynamical decoupling sequences (Ramsey, XY-8 and interaction decoupling) to measure the P1 spin-coherence time $T_2$.
While the XY-8 is designed to cancel the Ising portion of the Hamiltonian, 
we implement a recently introduced robust interaction decoupling sequence (termed DROID-60) which decouples the full dipolar interaction \cite{choi2020robust,zhou2020quantum}.
After unshelving the NV spin back to $\ket{0}$, we wait for $\tau_0 = 10~\mu$s $\sim \tau_{\textrm{th}}$ to let it locally thermalize with the nearby P1 spins, and detect the resulting NV polarization as a measure of local P1 polarization.

\subsection{Continuous diffusive model}
\label{sec:Met_Cont}
To verify and study the late-time hydrodynamics of the system, we build up a phenomenological diffusion equation for the polarization profile. 
By solving this equation, we derive the functional form of the survival probability measured in our experiment. 
This allows us to extract the diffusion coefficient, as well as to study the subleading correction to diffusion in our disordered long-range interacting system. 

\subsubsection{Gaussian diffusion and associated survival probability}
In a diffusive system, the polarization dynamics should be, at leading order, captured by the following diffusion equation:
\begin{equation}\label{Methods--eq:diff}
  \partial_t P(t,\bm{r}) = D \nabla^2 P(t,\bm{r}) - \frac{P(t,\bm{r})}{T_1} + Q(t,\bm{r})~,
\end{equation}
where $P(t,\bm{r})$ is the polarization as a function of both time $t$ and position $\bm{r}$, 
$D$ is the diffusion coefficient, $T_1$ is the intrinsic depolarization timescale of our system, 
and $Q(t,\bm{r})$ corresponds to the polarization source (the NV). 
Considering our experimental geometry and polarization protocol, we assume the polarization process occurs over a short distance $b$ while the laser is on and can be modeled by:
\begin{equation}\label{eq:pol}
  Q(t,\vec{r}) = 
\begin{cases}
\frac{\Gamma}{(2\pi b^2)^{3/2}} e^{-r^2/(2b^2)} & -\tau_p<t<0 \\
0 & t > 0
\end{cases},
\end{equation}
where $\Gamma$ is the polarization rate. 
The presense of a finite $b$ reflects the range of the polarization transfer process from the NV to the P1 ensemble and guarantees that the polarization does not diverge at short times. 

This problem can be solved via a Green's function formalism (see SI \cite{SI}), which yields the solution:
\begin{equation}
\label{eq:fit1}
\begin{split}
  P &(t,\vec{r}=0) = \frac{\Gamma e^{\frac{b^2}{D T_1}}}{ 4\pi D^{\frac{3}{2}}\sqrt{T_1}}\times  \left\{F\left[\left(t + \frac{b^2}{D}\right)/T_1\right] - F\left[\left(t+\tau_p+ \frac{b^2}{D}\right)/T_1\right]\right\}, 
\end{split}
\end{equation}
where
\begin{equation}
  F(x) = \frac{1}{\sqrt{\pi}}\frac{ e^{-x}}{\sqrt{x}} - \mathrm{erfc} (\sqrt{x}).
\end{equation}

In Fig.~3b,d, we also consider the situation where an off-resonant P1 subgroup is driven with a microwave field to ``cancel" the Ising interaction and thus reduce the on-site field strength.
This leads to a faster diffusion with coefficient $D^{\mathrm{dr}}$.

To capture the hydrodynamics in this setup, we must incorporate this diffusion coefficient in Eqn.~(\ref{Methods--eq:diff}) after the polarization process ($t\ge 0$)---throughout polarization the system remains undriven and the diffusion coefficient is $D$. 
We also note that the decay timescale (denoted as $T^{\mathrm{dr}}_1$) may also get modified. 
Following the similar approach to analyze the undriven case \cite{SI}, we obtain the survival probability for the experiment as
\begin{equation}
\label{eq:fit_dr}
\begin{split}
  P (t,\vec{r}=0) &= \frac{\Gamma e^{\frac{b^2}{D T_1}}}{ 4\pi D^{3/2}\sqrt{T_1}} e^{\left(\frac{D^{\mathrm{dr}}}{D}-\frac{T_1}{T^{\mathrm{dr}}_1}\right)\frac{t}{T_1}}\times \\
  & \left\{ F\left[\left(\frac{D^{\mathrm{dr}}}{D}t + \frac{b^2}{D}\right)/T_1\right]\right. -e^{\left(\frac{T^{\mathrm{dr}}_1}{T_1}-\frac{D}{D^{\mathrm{dr}}} \right)\frac{\tau_{\mathrm{p}}}{T^{\mathrm{dr}}_1}}\left. F\left[\left(\frac{D^{\mathrm{dr}}}{D}t+\tau_p+ \frac{b^2}{D}\right)/T_1\right]\right\}.
\end{split}
\end{equation}

\subsubsection{Details of fitting with diffusion equation}
Up until now, we have been considering the dynamics of a single NV center surrounded by an ensemble of P1 spins.
However, the finite NV defect density in our sample leads to the polarization overlap between different NV-P1 systems.
Owing to the randomness in the NV center positions, this effect can be captured by a simple constant background.
In particular, since each NV is randomly placed in the system and we measure all NVs, the above overlap effect is spatially averaged and can be treated as a homogeneous background whose dynamics is governed only by depolarization.
Given the volume of each NV-P1 system, $V=1/\rho_{NV}$, the background polarization is simply written as:
\begin{equation}
\label{eq:background}
\begin{split}
P_{bg}(t) &= \frac{1}{V}\int_{-\tau_p}^0 \Gamma e^{-(t-t_0)/T_1} dt_0 = \rho_{NV}\Gamma T_1 \left(e^{-t/T_1}-e^{-(t+\tau_p)/T_1}\right).
\end{split}
\end{equation}

Combining the background $P_{\mathrm{bg}}$ with the survival probability $P(t,\vec{r}=0)$ (i.e. Eqn.~(\ref{eq:fit1},\ref{eq:fit_dr})) yields the experimental signal, from where we extract the diffusion coefficient $D$ and the range of polarization $b$ (using an independently calibrated $T_1$, Fig.~\ref{fig:ext_T1}).
To be specific, for the sample S1 at room temperature and for each $\nu$, we fit to a global value of $D$ and $b$, while letting $\Gamma$ change across different $\tau_\mathrm{p}$ (Fig.~\ref{fig:Supp_S1_DiffusionExtraction}).
In the driven experiment, we use the values of $b$, $\Gamma$, $D$ and $T_1$ from the undriven case, as well an independently calibrated $T^{\mathrm{dr}}_1$ (see below) to fit the data and extract $D^{\mathrm{dr}}$ as the sole fitting parameter (Fig.~3d).
For the sample S2 at low temperature (which has similar P1 density as sample S1) fit to $D, \Gamma$ and $b$ with $T_1$ being independently characterized from the late time decay (gray curve in Fig.~1b). 
Here, we emphasize that in the fitting functional form, $\rho_{NV}$ (which is independently measured--- see SI \cite{SI}) provides the necessary length scale that connects the measured decay timescales to the diffusion coefficient. 

\subsubsection{Uncertainty of the diffusion coefficient}

Since the extraction of the diffusion coefficient from the experimental data relies on independently assessed values like $T_1$ and $\rho_{NV}$, the uncertainty in the diffusion coefficient must reflect our uncertainty in these values as well.
To this end, we perform our fitting procedure while sampling these quantities according to their uncertainties.
This gives us a distribution of the diffusion coefficients from which we can extract a mean and a standard deviation to characterize the final value and its uncertainty.

This becomes important when understanding the effect of driving disorder on the diffusive dynamics: 
since the undriven diffusion coefficient $D$ is an input parameter to our extraction of the driven diffusion coefficient $D_{\mathrm{dr}}$, the uncertainty in $D_{\mathrm{dr}}$ includes our uncertainty in $D$ as well.
To understand whether the two dynamics have consistent diffusion coefficients, we should consider a fixed value of $D$ and analyze whether the extracted $D_{\mathrm{dr}}$ is consistent (given our uncertainty in the remaining parameters).
In Fig.~\ref{fig:diffusionComparison}, we explicitly consider the distribution of $D$ and $D_{\mathrm{dr}}$ from our fitting analysis in order to highlight the correlation between the two values.
Crucially, we observe that, for any fixed $D$, the corresponding extracted $D_{\mathrm{dr}}$ is significantly larger than $D$.
This analysis reflects the conclusion already highlighted in Fig.~3d---driving the other $\nu=1/4$ group modifies and hastens the diffusive dynamics in our system.

\subsubsection{Extraction of late time $T_1$ time}

One important parameter in extracting the dynamics of the survival probability $S_p(t)$ from the observed NV dynamics is the extrinsic polarization time $T_1$. 
Such timescale can depend on details of the sample, temperature and application of additional driven fields.
To this end, we extract it via the late time behavior of the polarization dynamics under different conditions; at late time, the local polarization profile has expanded and the observed dynamics is dominated by the slowly decay of the background polarization, Fig.~\ref{fig:ext_T1}.

\subsection{Derivation of geometric factor correction to diffusion}

While both Gaussian and exponential (Fig.~1 main text) polarization profiles exhibit the same diffusive scaling for the survival probabilities, the difference in their shapes modifies the relationship between the height and the width of the distribution and thus how one can extract the diffusion coefficient.
In particular, in the Gaussian case this relationship is given by $S_p^g(t) = P_{\mathrm{tot}} / (2\pi \langle r^2\rangle)^{3/2}$ while in the exponential case it is given by $S_p^e(t) = P_{\mathrm{tot}} /  [8\pi (\langle r^2\rangle)^{3/2}]$, where $P_{\mathrm{tot}}$ is the total polarization in the system.
Replacing with $\langle r^2\rangle = 6Dt$ and equating the two survival probabilities leads to a constant factor correction $g=2\pi^{1/3}$ between the two diffusion coefficients.
As such, the extracted diffusion coefficients (extracted assuming a Gaussian profile), should be corrected by multiplying its value by $g$.
Crucially, the corrected diffusion coefficients are in excellent agreement with the diffusion coefficients directly extracted via the growth of the mean-square displacement $\langle r^2\rangle$.

\subsection{Semi-classical model}

While the precise many-body quantum dynamics of our model are dictated by the underlying microscopic Hamiltonian, its calculation remains beyond the realm of possibility of current numerical techniques.
To this end we build a semi-classical model of the polarization dynamics based upon the the microscopic details of the Hamiltonian that allows us to quantitatively characterize the observed dynamics (e.g. extracting the diffusion coefficient), as well as investigate properties beyond the current experimental reach (e.g. studying the spatial profile of the polarization).
Crucially, this model provides a direct and quantitative connection between the underlying microscopic Hamiltonian and the emergent macroscopic dynamics.

At its core, our approach relies on the calculation of the polarization-transfer rate between any pair of spins.
Such rate can be calculated via two independent but complementary approaches: 
the first leverages the Fermi's Golden rule to compute the decay of one of the spin polarization into another spin (whose levels have been broadened due to interactions with a ``bath" of the remaining P1 spins, Fig.~1c);
the second isolates the dynamics of the pair and computes the polarization-transfer rate under decoherence generated by the remaining  ``bath" P1 centers.
The resulting polarization-transfer rate is given by Eqn.~(5).
For details of the derivation, see the SI \cite{SI}.

Armed with the rate of polarization transfer, we numerically study the polarization dynamics of the system by reducing the exponentially large quantum state of the system ($2^N$ coefficients for $N$ spins), into $2N$ coefficients that capture the individual populations $\rho_{i,\sigma}$ of each of the levels $\sigma$ of each spin $i$ of the system.
For the P1 $\sigma \in \{-\frac{1}{2}, +\frac{1}{2}\}$, while for NV $\sigma \in \{0, -1\}$.
One then obtains a differential equation for the populations:
\begin{equation}\label{eq:Sup_DifEq}
    \partial_t \rho_{i,\sigma} = \sum_{j} \Gamma_{i,j} (\rho_{i,\bar{\sigma}} \rho_{j,\sigma'} - \rho_{i,\sigma} \rho_{j,\bar{\sigma'}} ),
\end{equation}
where $\bar{\sigma}$ corresponds to the other level of the spin $i$ and $\sigma'$ corresponds to the level of spin $j$ such that interactions lead to a transition $\sigma \bar{\sigma'} \leftrightarrow \bar{\sigma} \sigma'$.

\subsubsection{Estimating the extrinsic decoherence rate $\gamma$}
We estimate the extrinsic decoherence rate $\gamma$ in the rate equation using the measured spin-echo coherence time of NV.
In particular, the spin-echo pulse sequence is depicted in Fig.~\ref{fig:figS_SpinEcho}: after polarizing the NV center via a green laser, a $\frac{\pi}{2}$-pulse prepares the NV spin into a coherent superposition of $\ket{m_s=0}$ and $\ket{m_s = -1}$, which is allowed to dephase during a time $t$. 
A $\pi$-pulse at the center of the sequence ``echos" out the on-site random field generated by the nearby P1 centers, thus provides a direct estimation of the extrinsic decoherence time of a single spin in the system.
We fit the spin-echo decay using a form $e^{-(t/T_2^{\mathrm{echo}})^{1.5}}$ \cite{bauch2020decoherence} and extract $T_2^{\mathrm{echo}} = 1.9\pm 0.1$~$\mu s$ ($\gamma \sim 1/T_2^{\mathrm{echo}} \approx 0.5~\mu s^{-1}$).

\subsubsection{Estimating the on-site random field distribution $\delta_i$}
The distribution of on-site random fields $\delta_i$ is directly determined using the intrinsic linewidth of the NV spin state.
In particular, we perform pulsed optically detected magnetic resonance (ODMR) sequence to measure the NV linewidth: after polarizing the NV center via a green laser, we apply a microwave $\pi$ pulse and sweep its frequency $\omega$ across the NV $\ket{0}$ to $\ket{-1}$ transition (Fig.~\ref{fig:figS_Linewidth} inset). 
To avoid microwave power broadening of NV transition, we choose a sufficient weak microwave $\pi$ pulse with duration $2$~$\mathrm{\mu}$s.
We then sample the on-site random field $\delta_i$ from the experimentally measured NV linewidth depicted in Fig.~\ref{fig:figS_Linewidth}.

\subsubsection{Simulating the polarization process}
While the above differential equation enables the study of polarization transfer within the spin system, to correctly capture the spin dynamics we must also take into account the polarization process via the NV center.
To this end, we extend the previous differential equation, Eqn.~(\ref{eq:Sup_DifEq}), to include the dynamics of the remaining levels of the NV center (in particular the $\ket{+1}$ state of the ground state manifold, as well as the $\ket{m_s = 0,\pm 1}$ states of the excited manifold and the singlet level).
Leveraging previous results on the decay between these states \cite{tetienne2012magnetic, robledo2011spin,SI}, we can simulate the internal polarization process of the NV center provided we add a pumping rate $\Gamma_{\mathrm{p}}$ which captures the laser pumping rate from the ground to the excited state manifold.
The precise value of $\Gamma_{\mathrm{p}}$ depends on the laser power applied as well as the geometry of the system.
Based on the analysis of the saturation of NV fluorescence with laser power, $\Gamma_{\mathrm{p}}$ is estimated to be $\sim 0.1~\mathrm{\mu s^{-1}}$. 
We emphasize that this pumping rate leads to an enhancement of the decoherence between the NV and the P1 centers, which can be accounted for by replacing $\gamma$ with $\gamma + \Gamma_{\mathrm{p}}$ in Eqn.~(5) when computing this polarization transfer rate.

\subsubsection{Agreement between semi-classical model and experimentally observed dynamics}

We first consider the depolarization dynamics in sample S1, as in Fig.~2b.
Using a single NV center surrounded by a finite number of P1 spins ($N=\{300,225,75\}$ for the groups $\nu=\{1/3,1/4,1/12\}$ in agreement with the measured spin defect density), we simulate the polarization protocol with $\Gamma_{\mathrm{p}}=0.1~\mathrm{\mu s^{-1}}$ for $\tau_{\mathrm{p}}$ and compute the subsequent polarization dynamics of the NV center given by the difference in populations in the $\ket{0}$ and $\ket{-1}$ states.
For the $\nu \in \{ 1/3, 1/4\}$ groups, we observe excellent agreement with the experimental data for over 4 orders of magnitude in $\tau_{\mathrm{p}}$ and throughout then entire experimental timescale using a linewidth $\gamma=0.5~\mathrm{\mu s^{-1}}$, Fig.~\ref{fig:Sup_Agreement}.
For the $\nu=1/12$ group, we observe good agreement, albeit with a smaller range of $\tau_{\mathrm{p}}$ and using $\gamma = 1.5~\mathrm{\mu s^{-1}}$. 
We believe this discrepancy arises from a much larger separation between the strength of the on-site fields and the flip-flop rate of the ensemble.

We now turn to sample S2 when resonant with $\nu = \frac{1}{3}$. 
Given the approximately equal P1 density of the sample, we consider the dynamics of a single NV defect surrounded by $N_{\mathrm{P1s}}= 300$ P1 centers.
Using $\Gamma_{\mathrm{p}} = 0.25~\mathrm{\mu s^{-1}}$ and $\gamma = 0.3~\mathrm{\mu s^{-1}}$, we observe excellent agreement throughout the entire dynamics, Fig.~\ref{fig:Sup_LTAgreement}. 

The excellent agreement observed in the NV polarization decay in both samples gives us confidence that our semi-classical model can capture the polarization dynamics in the sample and provide an accurate calculation of the diffusive properties of the spin ensemble.

\subsection{Computing the diffusion coefficient from the semi-classical description}
\label{sec:meth_diff}

One of the features of our semi-classical model is the ability to access the \emph{spatial} profile of the polarization, enabling an \emph{independent} characterization of the diffusive dynamics via the direct study of the spread of the entire polarization profile.
In particular, we leverage our semi-classical model to compute the mean-squared displacement of the polarization and observe the characteristic linear growth with time, Figs.~\ref{fig:r2_finiteSizeScaling} and \ref{fig:ext_Dr2}.
From this linear growth, we can directly extract the diffusion coefficient of the dynamics via
\begin{align} \label{eq:Sup_r2}
\langle r^2\rangle = 6 D_{\langle r^2 \rangle} t~,
\end{align}
where the coefficient $6 = 2d$ includes the information about the dimensionality $d$ of our system.

Focusing our analysis on the dynamics of the P1 centers, we can disregard the details of the NV and transform our semi-classical model into a set of linear differential equations on the polarization of the P1 centers $P_i = \rho_{i,\uparrow} - \rho_{j,\downarrow}$:
\begin{align}
    \dot{P}_i = \sum_{j} \Gamma_{ij} (P_j - P_i)
\end{align}
Crucially, this linearity condition enables us to map this set of  differential equations into a continuous time random walk of the polarization through the positions of the defects.
Such approach enables us to consider the polarization dynamics of much larger system sizes (up to $N \sim 4\times 10^4$ P1s) because we no longer need to build the dense transition rate matrix $\Gamma_{ij}$ and solve for the associated eigenvalue problem.

Our simulation protocol is then as follows. 
We first generate an ensemble of P1 spins with random positions $\bm{r}_i$ and onsite energies $\delta_i$ surrounding a P1 spin at position $\bm{r}_0 = (0,0,0)$. 
Starting with the polarization at $\bm{r}_0$ at $t=0$, we compute its dynamics through the system as follows:
When the polarization is on spin $i$ (at position $\bm{r}_i$),
the hopping rate away from spin $i$ is given by $\Gamma_{\mathrm{tot}} = \sum_{j} \Gamma_{ij}$; thus the polarization remains in spin $i$ for time $\delta t$ which is a Poisson random variable with mean $\Gamma_{\mathrm{tot}}^{-1}$.
The probability of hopping to spin $j$ is given by the branching ratio $\beta_{j} = \Gamma_{ij} / \Gamma_{\mathrm{tot}}$.
We repeat this process until a time $t_{\mathrm{max}} \sim 1000~\mathrm{\mu s}$ has elapsed.


\emph{Computing the mean-square displacement}---
For each random walk, we record the displacement squared of the polarization from its original position $\bm{r}_0$ as a function of time $t$.
Averaging over many random walks immediately yields the mean squared displacement of the polarization profile.

Unfortunately, this quantity is particularly sensitive to the system size considered [which sets an upper bound on the maximal $\langle r^2 \rangle (t)$], specially given the long-range nature of the transition rates, Fig.~\ref{fig:r2_finiteSizeScaling}.
To this end, we perform a finite size scaling, where we extrapolate the infinite system size behavior by studying $\langle r^2 \rangle(t)$ for different system sizes $N\in\{10^3, 15^3, 20^3, 25^3, 30^3\}$ and assuming that the system size effects are linear with the inverse lengthscale of the system $L^{-1} \sim N^{-1/3}$.
We find that this ansatz agrees with the observed numerical data, and leads to a long-lived linear increase in the $\langle r^2\rangle (t)$, from which we can extract the diffusion coefficient in our semi-classical description.
We estimate the error in our analysis by fitting Eqn.~(\ref{eq:Sup_r2}) to different early time regimes (from $0$ to $T_{\mathrm{max}} \in [30,300]~\mathrm{\mu s}$) and taking its value as half the range of extracted diffusion coefficients.

\subsection{Dynamical correction to diffusion}
\label{sec:theoryDerivation}

The experiments average over both NV centers within the sample and shot-to-shot fluctuations in the environment. 
Thus, the quantity of interest is the disorder-averaged diffusion kernel, or Green's function. 
In frequency and momentum space, this is given by:
\begin{equation}
    G(\bm{k}, \omega) = \frac{1}{-i\omega + Dk^2 + \Sigma(\bm{k}, \omega)}
\end{equation}
where $-i\omega + Dk^2$ describes the eventual diffusive  dynamics at asymptotically late times and long wavelengths while the self-energy $\Sigma(\bm{k},\omega)$ captures the corrections due to disorder averaging.
Focusing on the leading corrections in this limit, the small $\omega$ and $\bm{k}$ expansion of $\Sigma(\bm{k}, \omega)$ can be constrained by noting that it is (1) analytic, (2) isotropic, and (3) probability conserving:
\begin{align}
\label{eq:selfenergy}
    \Sigma(\bm{k}, \omega) =  D' k^2 + C k^4 + \ldots + (-i\omega) (\ell^2 k^2 + \ell'^4 k^4 + \ldots) + \ldots ~,
\end{align}
where $\ell^2 = C_{\mathrm{dyn}}$ in the main text.

A few remarks are in order.
First, the $D' k^2$ term simply corrects the bare diffusion coefficient, $D \to D + D'$, and can be absorbed into a new definition of $D$.
Second, the $Ck^4$ and $\ell^2 (-i \omega) k^2$ terms have the same scaling dimension near the diffusive fixed point (where $z=2$); they are accordingly the leading irrelevant corrections in the renormalization group sense. 
Third, the dynamical corrections (involving powers of $\omega$) do not appear in translationally invariant classical hopping systems, where the only corrections in $\Sigma$ arise due to the spatial Fourier transform of the hopping kernel.
This picture is modified in disordered systems, where a diagrammatic analysis of the disorder average generically yields the dynamical corrections. 

While Eq.~\eqref{eq:selfenergy} follows on very general symmetry grounds, for completeness we include a short derivation of the one-loop self-energy in a continuum model with a spatially random local diffusion coefficient in order to indicate how it is generated. 
The actual coefficient $\ell^2$ produced in this calculation depends on the UV-cutoff (as expected) and thus its detailed numerical form is of limited utility.
We assume that the diffusivity is $D(x) = D + \delta D(x)$ is Gaussian with 
\begin{align}
    \overline{\delta D(x) \delta D(x')} &= \Delta_D \delta(x-x')
\end{align}
The Martin-Siggia-Rose (MSR) \cite{martinsiggiarose} action generating the Green's function of the polarization $P(x,t)$ in a \emph{fixed} diffusion environment $D(x) $ is given by
\begin{align}
    S_D = i \int dt d^dx\, \hat{P}(x,t) (\partial_t - D \vec{\nabla}^2) P(x,t) + i \int dt d^dx\, (\vec{\nabla}\hat{P}(x,t)) \cdot (\vec{\nabla} P(x,t)) \delta D(x) 
\end{align}
where $\hat{P}(x,t)$ is the MSR conjugate response field to $P(x,t)$. 
Here, $\langle \hat{P}(x,t) P(x',t') \rangle_D = -i G_D(x', t'; x, t)$ provides the fixed environment Green's function of the polarization.
The action generating the disorder averaged Green's function can be obtained by integrating over the fluctuations in $\delta D(x)$. 
Technically, this is straightforward because in the MSR approach, the normalization $Z_D = \int DP D\hat{P} e^{S_D} = 1$ is independent of the realization of $D(x)$; see Ref.~\cite{Fisher1984} for more discussion.
We obtain the effective action
\begin{align}
    S &= S_0 + S_1 \\
    S_0 &= i \int dt d^dx\, \hat{P}(x,t) (\partial_t - D \vec{\nabla}^2) P(x,t) \\
    S_1 &= - \frac{\Delta_D}{2} \int dt dt' d^dx\, \left[(\vec{\nabla}\hat{P}(x,t)) \cdot (\vec{\nabla} P(x,t))\right]\left[ (\vec{\nabla}\hat{P}(x,t')) \cdot (\vec{\nabla} P(x,t')) \right]
\end{align}

We now compute $-i G(x,t) = \langle \hat{P}(0,0) P(x,t) \rangle$ using  perturbation theory to leading loop order.
The bare propagator, $-i G_0(\bm{k},\omega)$, can be taken as  an arrow pointing from $\hat{P}$ to $P$:
\begin{align}
\begin{tikzpicture}[]
\draw (0,0) edge[electron] node[above] {$\bm{k}, \omega$} (1,0);
\end{tikzpicture}
= \frac{-i}{-i \omega + D k^2} 
\end{align}
The interaction vertex ($S_1$) is
\begin{align*}
    \begin{tikzpicture}[baseline={([yshift=-.5ex]current bounding box.center)}]
    \draw (0,0) edge[electron] node[below left] {$\bm{q'}, \omega'$} (0.5,0);
    \draw (0.5,0) edge[electron] node[below right] {$\bm{q'}-\bm{k}, \omega'$} (1,0);
    \draw (0,1) edge[electron] node[above left] {$\bm{q}, \omega$} (0.5,1);
    \draw (0.5,1) edge[electron] node[above right] {$\bm{q}+\bm{k}, \omega$} (1,1);
    \draw (0.5,0) edge[int] node[right] {$\bm{k}$} (0.5,1);
    \end{tikzpicture}
    &= -\frac{\Delta_D}{2}(\bm{q}\cdot(\bm{q}+\bm{k}))(\bm{q}'\cdot(\bm{q}'-\bm{k}))
\end{align*}
We note that causality implies that any diagrams with closed $P$-loops are identically zero, since $G_0$ is retarded.
As usual, we can organize the perturbative expansion for the full propagator $-i G$ in terms of a self-energy $G^{-1} = G_0^{-1} + \Sigma$,
\begin{align}
    \begin{tikzpicture}[]
\draw (0,0) edge[electron,double] node[above] {$\bm{k}, \omega$} (1,0);
\end{tikzpicture}
&=     \begin{tikzpicture}[]
\draw (0,0) edge[electron] node[above] {$\bm{k}, \omega$} (1,0); 
\end{tikzpicture}
+  \begin{tikzpicture}[]
\draw (0,0) edge[electron] node[above] {$\bm{k}, \omega$} (1,0); 
\draw (1,0) edge[electron] node[above] {$\Sigma$} (2,0);
\draw[dashed] (1,0) arc (180:0:0.5);
\draw (2,0) edge[electron,double] (3,0);
\end{tikzpicture}
\end{align}
To one loop, the only diagram contributing to $\Sigma$ is 
\begin{align*}
    \Sigma^{(1)}(\bm{k}, \omega) &= \begin{tikzpicture}[baseline={([yshift=-.5ex]current bounding box.center)}]
    \draw (0,0) edge[electron] node[below] {$\bm{q}, \omega$} (1,0);
    \draw[dashed] (0,0) arc (180:0:0.5) (0.5,0.5) node[above] {$\bm{k}-\bm{q}$};
    \end{tikzpicture} 
    = -\frac{\Delta_D}{2} \int \frac{d^dq}{(2\pi)^d} \frac{ (\vec{q}\cdot \vec{k})^2}{-i\omega + D q^2} 
\end{align*}
This is clearly isotropic and quadratic in $\vec{k}$. It is straightforward to evaluate $\Sigma^{(1)}$ as an expansion in $-i \omega$ (in 3d):
\begin{align}
    \Sigma^{(1)} 
    &= - \frac{\Delta_D \Lambda^3}{36\pi^2 D} k^2 + \frac{\Delta_D \Lambda}{12 \pi^2 D^2} (-i \omega) k^2 + \cdots 
\end{align}
where $\Lambda$ is a UV momentum cutoff.
As expected we find a correction to the long-wavelength diffusion constant $D' k^2$ and a dynamical correction $\ell^2 (-i \omega) k^2$.

We now tease out the phenomenological role of the dynamical correction term for the observed diffusion.
Focusing on the short-range case (without $k^{\alpha-d}$ correction):
\begin{equation} \label{eq:Meth_dynCorr}
    G(\bm{k},\omega) = \frac{1}{-i\omega( 1 + \ell^2 k^2) + D k^2} \Rightarrow G(\bm{k},t) = \frac{e^{-D k^2 t / (1 + \ell^2 k^2)} }{1+ \ell^2 k^2}~.
\end{equation}
In the $t\to 0$ limit, the polarization profile approaches the Yukawa potential form $G(\bm{k},t=0) = \frac{1} {1+ \ell^2 k^2}$. Returning to real space in three dimensions,
\begin{align} \label{eq:t0Yukawa}
    G(\bm{r}, t=0) = \frac{e^{-r/\ell}}{4\pi \ell^2 r}~.
\end{align}
We note that this form should be interpreted as the shape to be expected of $G$ at early times in the crossover to Gaussian behavior. 
It does not reduce to a delta function because we have neglected the higher order $k$ and $\omega$ corrections which govern the short-distance, early-time dynamics. 

However, the presence of this dynamical correction has important late time effects as well.
From Eqn.~(\ref{eq:Meth_dynCorr}), one observes that the large $k$ modes now decay with a \emph{constant} rate.
This implies that the short-distance singularity of the Yukawa-potential decays only after the timescale  $\ell^2/D$.

We end this section with a few remarks.
First, at even later times, the weight of the singularity decays exponentially while the diffusive behavior of the survival probability decays as a power-law in time, $S_p(t) \sim t^{-d/2}$, dominating the late-time physics.
This is in agreement with the presence of the late-time diffusive fixed point.
Second, such a singularity cannot be experimentally observable owing to the short-range cutoff of our system: we measure the polarization of the NV centers which must remain finite.
One can understand this singularity as being regularized by the short-range details arising from the discrete nature of our randomly positioned spin system.
Finally, while the $D'k^4$ term has the same scaling dimension as $(-i \omega) \ell^2k^2$, its inclusion in the analysis above does not change the qualitative statements, nor does it alter the nature of the early time behavior (it only appears in the exponent of Eqn.~\eqref{eq:Meth_dynCorr} and does not change the $t\to0$ behavior).

\subsection{Evidence for non-Gaussian diffusion under the two sources of disorder: positional disorder and random onsite fields}

In this section we present evidence that highlights the role of disorder in the observed long-lived exponential polarization profile, whether it arises from positional disorder or random onsite fields.
To this end, we consider two different models where we include the two effects separately:
first, we consider positional disorder of P1 centers with same density as sample S2 in the manuscript, but without onsite random fields;
second, we consider no positional disorder by placing the P1 defects in a cubic lattice (with matching density as that considered in sample S2) and include the same strength of onsite random fields.

Our results are summarized in Fig.~\ref{fig:DifWithDis}, where for both models (albeit with different timescales owing to the different average hopping rates) we observe that the polarization profile remains non-Gaussian throughout the time considered (fact made clearer when considering the rescaled polarization profile in the right column).
These results agree with the field theoretical calculation of the source of the dynamical correction, as this term arises from the presence of a spatially varying diffusion coefficient which arises from either positional disorder or a onsite random fields in the sample.

\subsection{Extracting dynamical correction $C_\mathrm{dyn}$}

To extract the strength of dynamical correction, we consider the same methodology when extracting the diffusion coefficient from the rate equation model, but instead of computing the mean squared displacement, we focus on the details of the polarization profile.
From our analytical derivation, the dynamics of such polarization profile are dominated by a diffusive and a disorder-generated term at late enough times. 
The value of $C_\mathrm{dyn} = \ell^2$ can be then obtained by fitting the resulting the resulting evolution to the dynamics under the diffusion correction (including the dynamical correction), starting from a Yukawa form for the polarization profile (Fig.~1 in main text).
To minimize finite size and time biases, we consider the dynamics between $t_\mathrm{min}$ and $t_\mathrm{max}$, the latter set to the time when the polarization per spin at the edge is larger than $3\times 10^{-6}$.
The reported value of $\ell$ corresponds to an average over the values extracted for different $t_{\mathrm{min}}$ over the range where the values are consistent (last 3 data points in Fig.~\ref{fig:ellExtraction}).
We summarize the extracted $\ell$ in Table~\ref{tab:ell}.

\subsection{Long-range modifications to diffusion}
In building a description that goes beyond the leading order term, we first consider the problem in Fourier space, where it can be generically cast as:
\begin{equation} \label{eq:supp_diffk}
       \partial_t P_{\bm{k}}(t) = -f(\bm{k})P_{\bm{k}}(t) , 
\end{equation}
where we have ignored the source and depolarization terms.
Note that the diffusive nature of the dynamics arises as a leading order term $Dk^2$ in $f(\bm{k})$, as discussed in Eqn.~(\ref{Methods--eq:diff}).
In isotropic short-range interacting system, $f(\bm{k})$ can be expanded in a series of even powers in $k$:
\begin{equation}
  f(\bm{k}) = Dk^2+Ck^4+\cdots .
\end{equation}

When the polarization transfer rate is long-range (decays as $\sim 1/r^\alpha$ at large distances $r$) in a $d$-dimensional system, this simple expansion breaks down owing a divergent $(\alpha-d)$-order moment---this to a non-analyticity in $f(\bm{k})$.
We now show how the long-range transfer $h(\bm{r})$ rate induces such terms in the decay rate of the Fourier modes
\begin{equation}
\begin{split}
  f(\bm{k}) =\int [1-\cos(\bm{k}\cdot\bm{r})] h(\bm{r}) ~ d^{d}\bm{r}.
\end{split}
\end{equation}

In particular, we consider a transfer rate where beyond some short-range cutoff $r_0$ the transfer decays simply decays as a power-law $\sim 1/r^\alpha$ for $\alpha > d$.
Focusing on the long-range part (since any non-singular short-range contribution just leads to even terms) we have:
\begin{equation}
\begin{split}
  f_{\mathrm{lr}}(\bm{k}) &=\int \frac{1-\cos(\bm{k}\cdot\bm{r})}{r^\alpha} d^d\bm{r}
  =\frac{\pi^{\frac{d}{2}}}{\Gamma(\frac{d}{2})}\int \frac{1-\cos(kr\cos\theta)}{r^\alpha} r^{d-1}~d\cos\theta ~dr = \frac{2\pi^{\frac{d}{2}}}{\Gamma(\frac{d}{2})}\int_{r_0}^\infty \left[ 1-\frac{\mathrm{sin}(kr)}{kr}\right] r^{d-\alpha-1} dr\\ & 
  = \frac{2\pi^{\frac{d}{2}}}{\Gamma(\frac{d}{2})}\left\{\Gamma(d-\alpha-1)\sin\left(\frac{\alpha + 1- d}{2}\pi\right) k^{\alpha-d}+\frac{r_0^{d-\alpha}}{d-\alpha}\left[ -1+{}_{1}F_{2}\left( \frac{d-\alpha}{2};\frac{3}{2},1+\frac{d-\alpha}{2};-\frac{k^2 r_0^2}{4}\right) \right] \right\}\\&
  = \frac{2\pi^{\frac{d}{2}}}{\Gamma(\frac{d}{2})}\left\{\Gamma(d-\alpha-1)\mathrm{sin}\left(\frac{\alpha + 1- d}{2}\pi\right) k^{\alpha-d}+\left[ \frac{r_0^{d-\alpha+2}}{6(\alpha-d-2)}k^2+\frac{r_0^{d-\alpha+4}}{120(d-\alpha+4)}k^4+\mathcal{O}(k^6)\right]\right\}, 
\end{split}  
\end{equation}
where $\frac{2\pi^{\frac{d}{2}}}{\Gamma(\frac{d}{2})}$ is the surface area of a $d$-dimensional ball with unit radius, and ${}_{1}F_{2}(a_1;b_1,b_2;x)$ is the generalized hypergeometric function, which can be expanded as a power series of the variable $x$. 
The interpretation is simple: while the last term only contains terms with even power of $k$ (similar to diffusion in short-range systems), the first term ($\sim k^{\alpha-d}$) captures the long-range nature, making the dynamics qualitatively different from short-range interacting systems. 

Therefore, with the presence of long-range interaction, $f(\bm{k})$ in general can be written as: 
\begin{equation}
  f(\bm{k}) = 
Dk^2+ C_{\mathrm{lr}} k^{\alpha - d} + Ck^4+\cdots
\end{equation}
where $D$, $C_{\mathrm{lr}}$ and $C$ are model-dependent coefficients. 
This result immediately highlights three import regimes:
When $d<\alpha<d+2$, the leading power is no longer the $k^2$ term and instead a $k^{\alpha-d}$ term becomes the leading contribution---the system is no longer diffusive and enters the Lévy-flight regime \cite{levy1954theorie,shlesinger1995levy}.
When $d+2<\alpha<d+4$, the leading order term remains the diffusive term but the sub-leading correction that control the approach to diffusion is set by a $k^{\alpha-d}$ term.
When $d+4< \alpha$, neither the leading nor the sub-leading term arises from the long-range transfer rate and the dynamics do not deviate significantly from the short-range case.

We now emphasize that, in the window of intermediate power-laws, $d+2<\alpha <d+4$, the survival probability exhibits a qualitatively different approach ($\sim t^{-\alpha/2+1}$) to the diffusive fixed point. 
For the 3D systems we focus on, at late times the survival probability $P(\bm{r}=0,t)$ should follow:
\begin{itemize}
    \item When $\alpha=6$ (our experimental platform),
    \begin{equation}\label{eq:Sup_k3}
    P(t,\vec{r}=0)=\frac{1}{(4\pi D t)^{3/2}}+\frac{C_{\mathrm{lr}}}{2\pi^2D^3t^2}.
    \end{equation}
    \item When $\alpha\ge 7$,
    \begin{equation}\label{eq:Sup_k4}
    P(t,\vec{r}=0)=\frac{1}{(4\pi D t)^{3/2}}+\frac{15C}{32\pi^{3/2}D^{7/2}t^{5/2}};
    \end{equation}
\end{itemize}
Clearly, the $\sim t^{-2}$ approach is distinct from conventional diffusion where the approach is proportional to $\sim t^{-5/2}$ (Fig.~\ref{fig:FigS_LongRange}).

We also remark that, similar to a previous discussion, we should convolve the above survival probability with the polarization process (both polarization duration and short-range details) to obtain the correct form for our experimental signal. 

\subsection{Additional analysis of experimental data}


In this subsection, we complement our analysis of the depolarization dynamics in Fig.~1b by considering the sensitivity of our fit to the range of data considered, both when the extrinsic depolarization time $T_1$ is independently determined (Fig.~\ref{fig:ext_T1}) or allowed to be a free parameter.

Following the prescription in the ``Continuous diffusive model'' section, we complement the diffusive description in Eqn.~(\ref{Methods--eq:diff}) with the different higher corrections to diffusion: $C_{\mathrm{lr}}k^3$, $C_{\mathrm{lr}}k^3 + Ck^4$, $Ck^4$, $Ck^4 + C' k^6$ and $C_{\mathrm{dyn}} k^2 \partial_t$.

Summarized in Fig.~\ref{fig:ext_FitsfixedT1}, we consider a fixed $T_1 = 2.6~\mathrm{ms}$ and fit the sample S2 data across different time ranges (red shaded region).
As highlighted in Fig.~1b, it is difficult to determine whether the inclusion of the $C_{\mathrm{lr}}k^3$ modification explains the deviations from simple diffusion: 
the late time dynamics are equally well captured when using different modifications, while the extension to early time is not systematically improved upon its addition.

\bibliography{ref_v14.bib}

\newpage
\newpage 

\renewcommand{\theequation}{E\arabic{equation}}
\renewcommand{\thefigure}{E\arabic{figure}}
\renewcommand{\thetable}{E\arabic{table}}

\section{Extended Data}

\renewcommand{\arraystretch}{2} 
\begin{table*}[h!]
\setlength{\tabcolsep}{0.7em} 
  \centering
 \begin{tabular}{c||c|c|c} 
    \hline \hline
    \makecell{$\ell~(\mathrm{nm})$}& \makecell{$\nu = 1/3$} & \makecell{$\nu = 1/4$} & \makecell{$\nu = 1/12$} \\ \hline \hline
    \makecell{Sample S1} & $12.9 \pm 1.4$  & $13.6 \pm 1.4$ & $20.3 \pm 1.9$ \\ \hline
    \makecell{Sample S2} & $14.3 \pm 1.6$  & $15.1 \pm 1.5$ & $19.3 \pm 1.9$ \\ 
    \hline \hline 
  \end{tabular}
 \caption{
   Extracted $\ell$ from the spin polarization dynamics for the different sample considered (S1 and S2) and the different P1 subgroups.
   With decreasing density, we observe a corresponding increase in the lengthscale $\ell$.
   Crucially, $\ell$ remains always larger than the P1-P1 distance highlighting that its value is not a simple consequence of the discrete nature of the spins in our system.
 }
     \label{tab:ell}
\end{table*} 




\begin{figure}[h!]
         \centering
         \includegraphics[width = 4.5in]{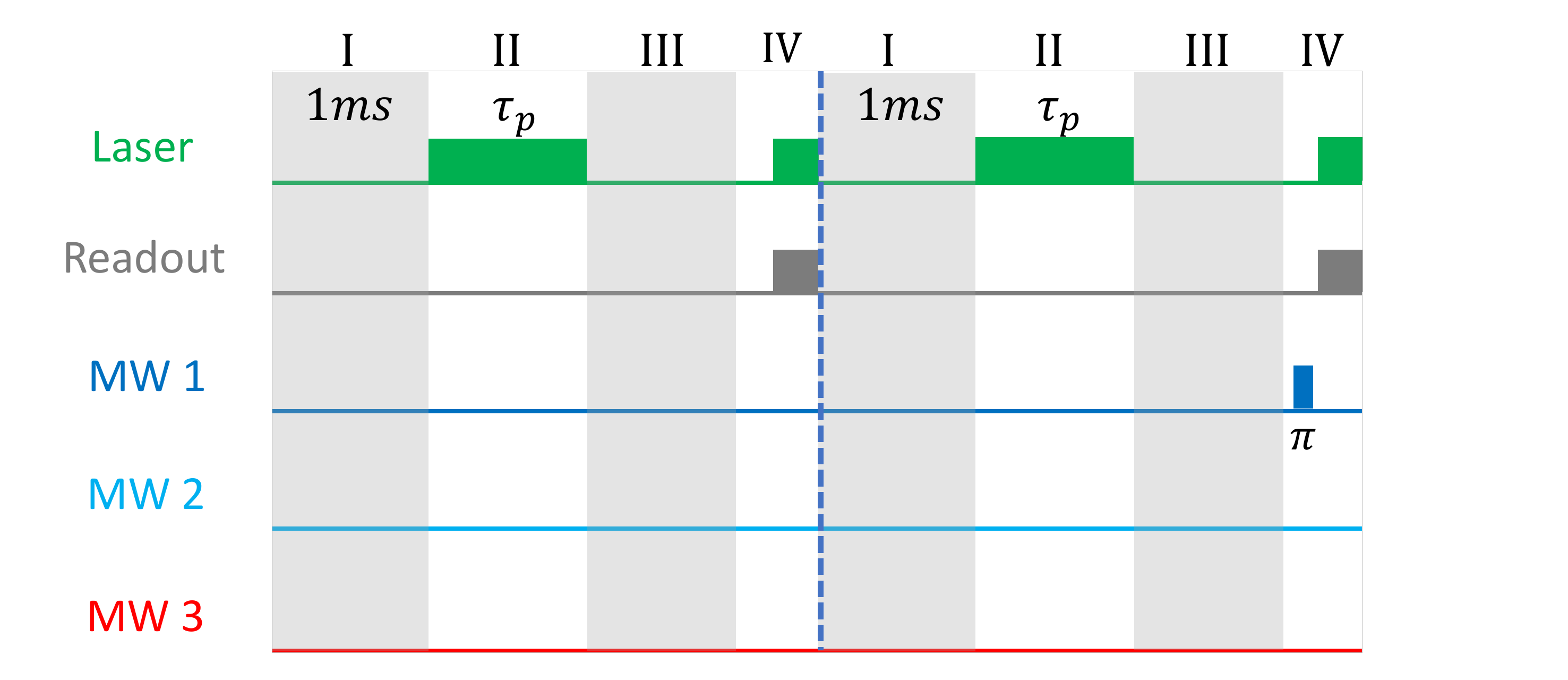}
         \caption{
           {\bf Experimental sequence to measure spin dynamics}
           I: charge equilibrium and spin polarization reset ($1$~ms); II: Spin polarization ($\tau_\mathrm{p}$); III: Spin manipulating pulses and time evolution; IV: spin state readout.
           }
         \label{fig:figS_seq}
     \end{figure}

\begin{figure*}[h!]
    \centering
    \includegraphics[width = 6.2in]{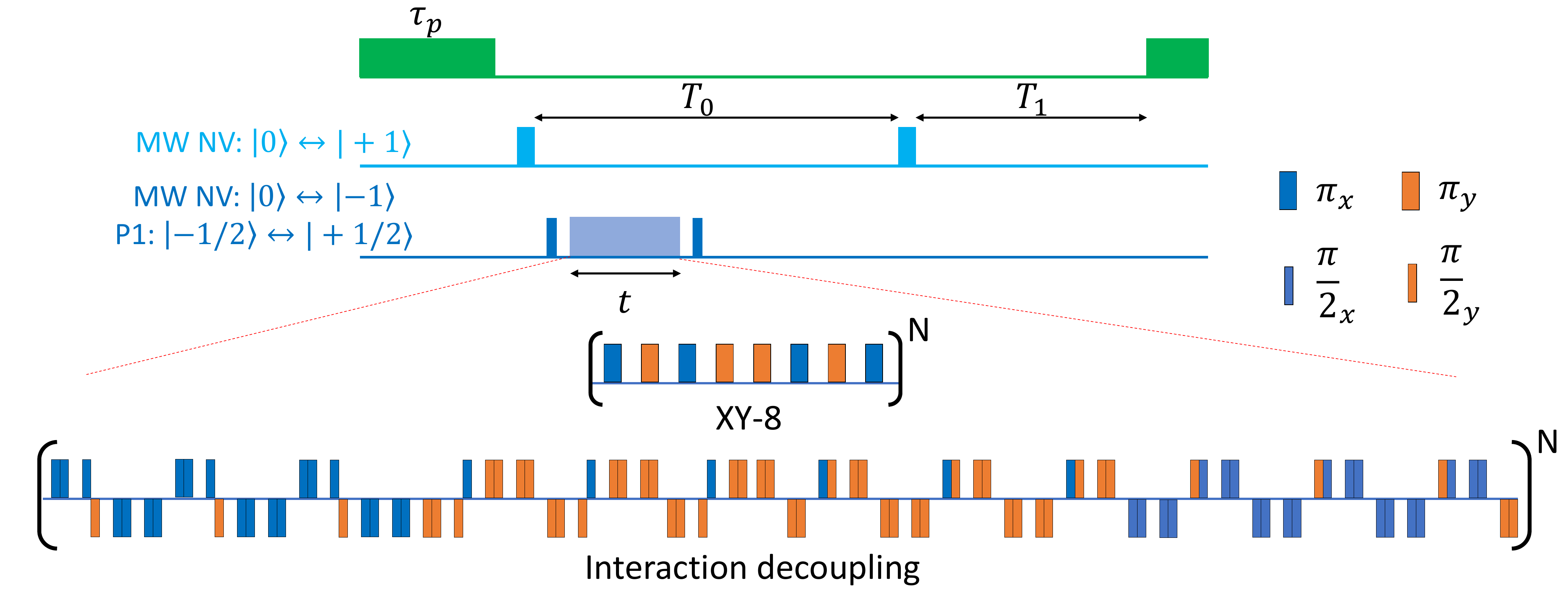}
    \caption{
    {\bf Experimental sequence to measure the coherence of P1 ensemble.}
    For XY-8 and interaction decoupling sequences, we fix the interval between pulses to be $\tau = 10$~ns, and increase the number of repetition N. The P1 $\pi$ ($\frac{\pi}{2}$) pulse duration is set to $36$~ns ($18$~ns). In the interaction decoupling sequences, the pulses at bottom side correspond to rotations along $-\hat{x}$ (blue) and $-\hat{y}$ (orange) axes.
    }
    \label{fig:figS_Coherence}
\end{figure*}

\begin{figure*}
    \centering
    \includegraphics[width = 6.8in]{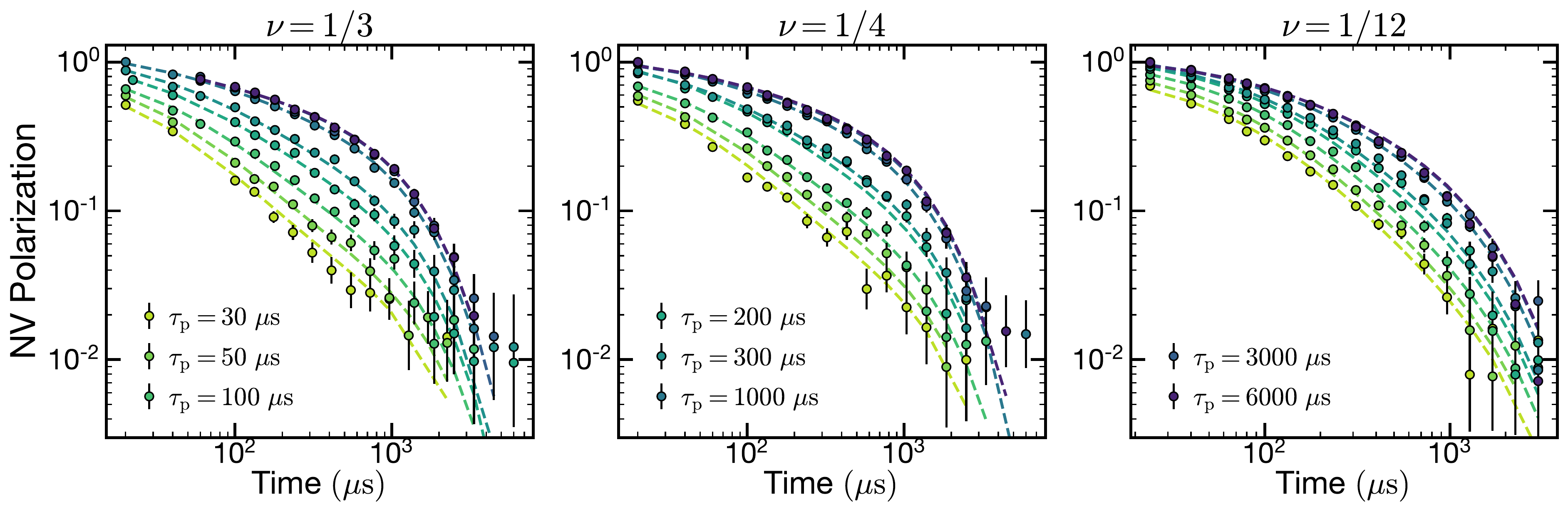}
    \caption{
      Fitting of the depolarization data in sample S1 for different groups $\nu$ (different panels) and different pumping times $\tau_{\mathrm{p}}$ (different colors).
      For each group,
      we fit the experimental data across all $\tau_{\mathrm{p}}$ data to a diffusive model in Eqn.~(\ref{eq:fit1}) with an additional background $P_{bg}$.
      From this procedure we extract both $D$ and $b$, as well as, a $\tau_{\mathrm{p}}$-dependent $\Gamma$, which captures the reduction in efficiency of the NV-P1 polarization transfer owing to the saturation of polarization near the NV.
    }
    \label{fig:Supp_S1_DiffusionExtraction}
\end{figure*}

\begin{figure}
    \centering
    \includegraphics[width = 5in]{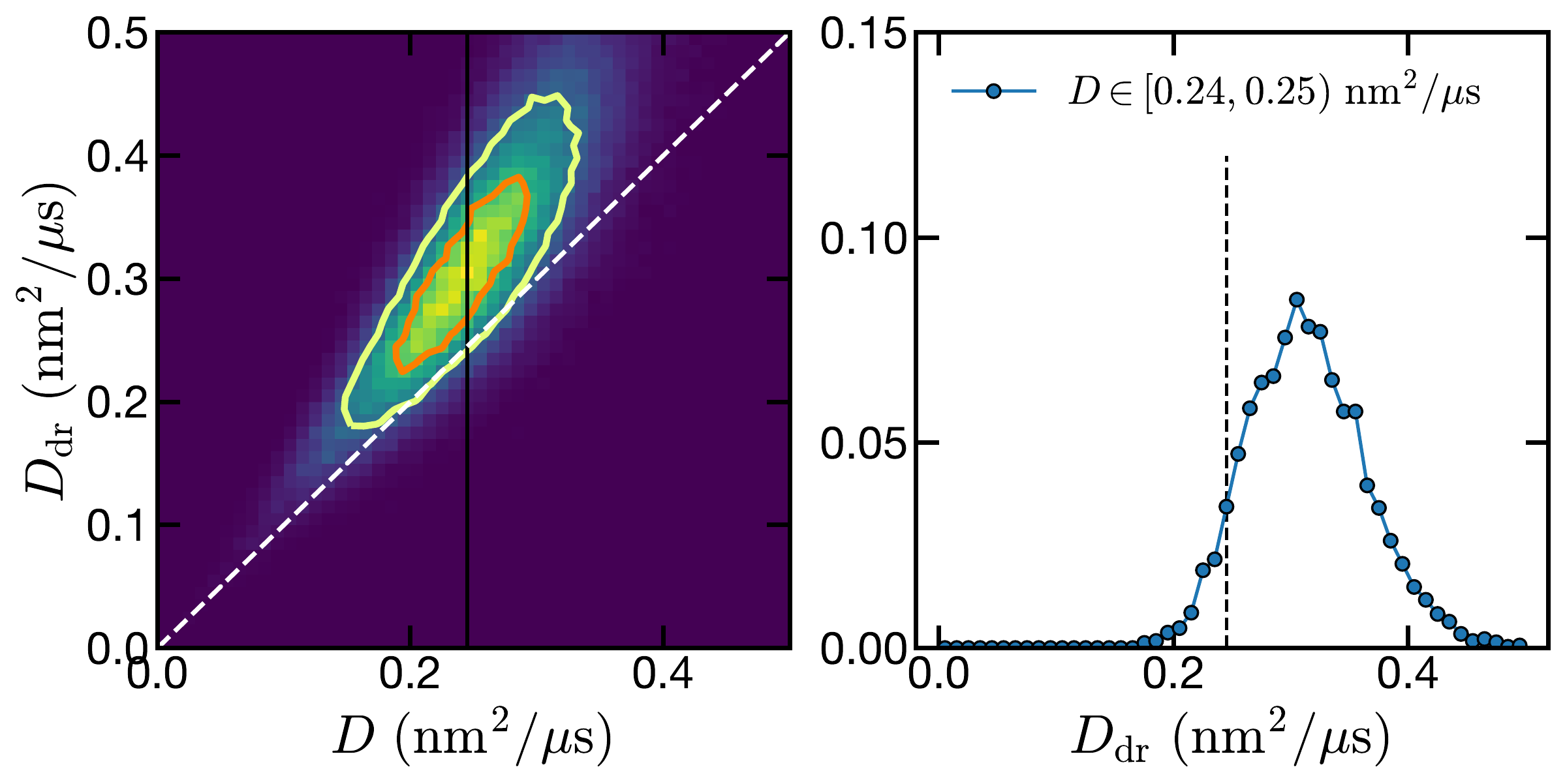}
    \caption{(Left Panel) Distribution of diffusion coefficients from our fitting procedure (lighter color corresponds to higher probability). Given a diffusion coefficient $D$, we observe that the extracted diffusion coefficient under a drive $D_{\mathrm{dr}}$ of the other $\nu=1/4$ group is consistently larger, highlighting that the drive modifies the diffusive dynamics by changing the onsite disorder of the system.
    (Right panel) Distribution of $D_{\mathrm{dr}}$ for $D \in [0.24, 0.25)$. Dashed vertical line corresponds to the value of $D$.
    }
    \label{fig:diffusionComparison}
\end{figure}

\begin{figure}
    \centering
    \includegraphics[width = 6.7in]{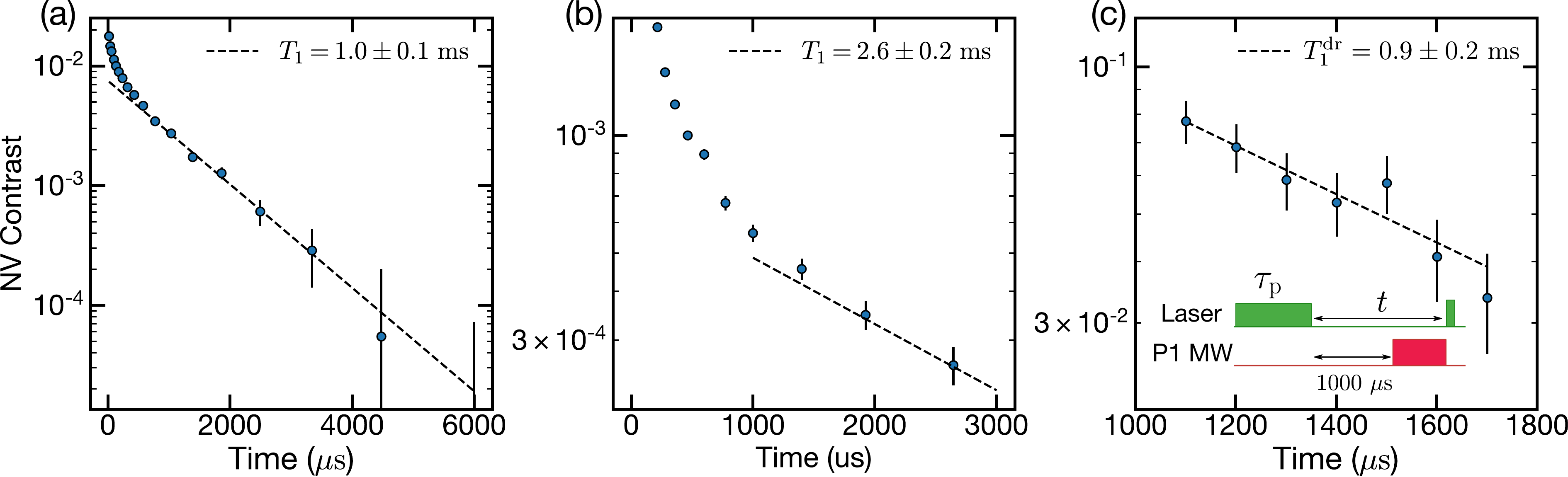}
    \caption{
        {\bf Measurement of late time extrinsic decay time $T_1$ for different samples and under driving. }
        (a) and (b) Extraction of the extrinsic depolarization time of samples S1 and S2 at room and low temperature ($25~\mathrm{K}$) and after polarizing for $\tau_{\mathrm{p}}=1000~\mathrm{\mu s}$ and $\tau_\mathrm{p} = 30~\mathrm{\mu s}$, respectively.
        The late time behavior follows an exponential decay with timescale given by $1.0\pm0.1$~ms and $2.6\pm0.2$~ms respectively.
        (c) To extract the modified intrinsic depolarization time $T_1^{\mathrm{dr}}$ of $\nu = \frac{1}{4}$ P1 subgroup with the presence of a strong microwave driving $\Omega = 11.7$~MHz on the other $\nu = \frac{1}{4}$ subgroup, we apply the following pulse sequence: after a laser pumping time $\tau_\mathrm{p}=1000$~$\mu$s, we wait for $1$~ms so that the initial P1 spatial polarization profile diffuses to a nearly homogeneous background which decays with intrinsic depolarization time of P1 centers.
        We then turn on a continuous microwave driving on the other $\nu = \frac{1}{4}$ P1 subgroup, and measure the resulting background decay; the resulting timescale is given $T_1^{\mathrm{dr}} = 0.9\pm0.2$~ms. 
       }
    \label{fig:ext_T1}
\end{figure}

     \begin{figure}
         \centering
         \includegraphics[width = 4.5in]{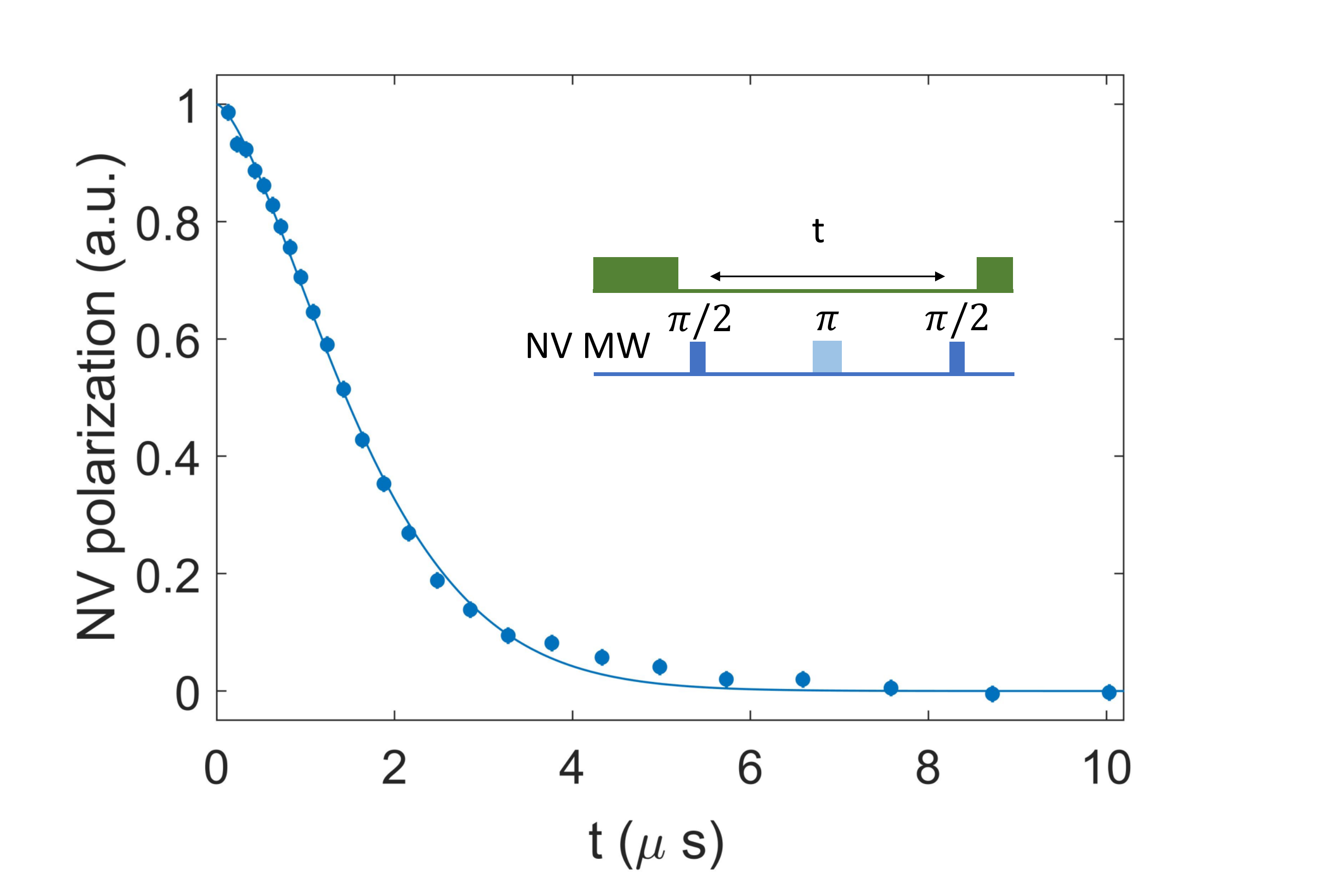}
         \caption{
           {\bf Spin-echo coherence measurement of NV center.}
           Spin-echo measurement of NV center when off-resonant with any P1 centers. The solid line corresponds to a fit $e^{-(t/T_2^{echo})^{1.5}}$, with $T_2^{echo} = 1.9\pm0.1$~$\mu s$. Inset: experimental pulse sequence. 
           }
         \label{fig:figS_SpinEcho}
     \end{figure}

\begin{figure}
    \centering
    \includegraphics[width = 3.0in]{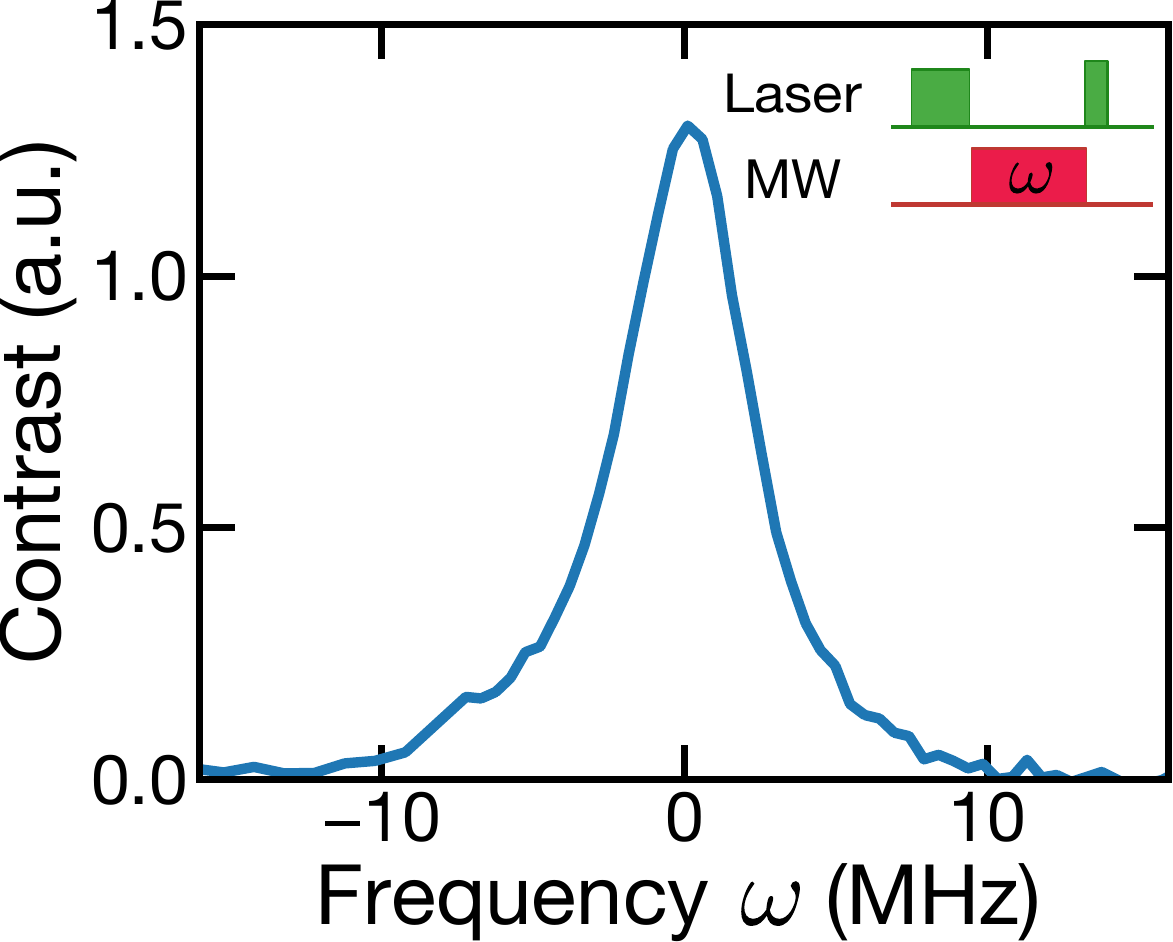}
    \caption{NV ODMR spectrum (shifted by the resonance frequency of $1.552~\mathrm{GHz}$) used to sample the onsite-random field $\delta_i$. (inset) ODMR pulse sequence.}
    \label{fig:figS_Linewidth}
\end{figure}

\begin{figure}
    \centering
    \includegraphics[width = 6.5in]{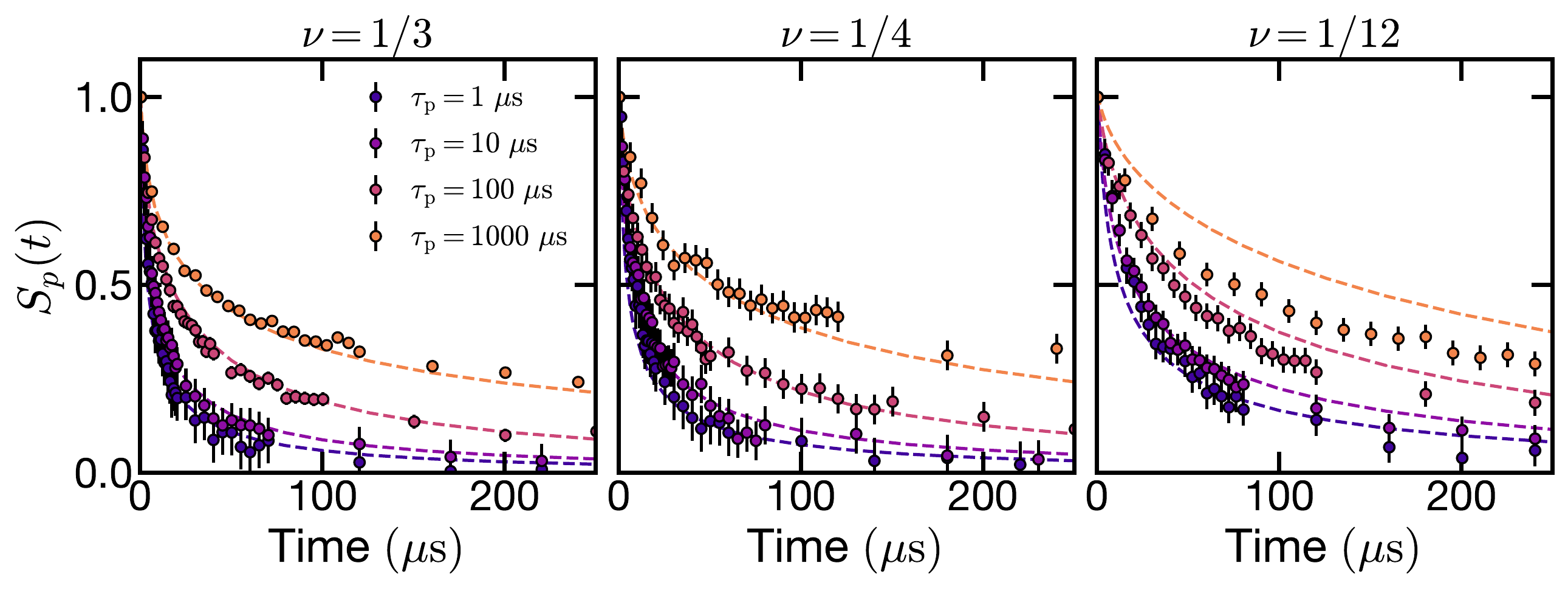}
    \caption{
    NV depolarization decay in sample S1 for different polarization times $\tau_{\mathrm{p}}$ and when resonant with different P1 subgroups.
    Dashed lines correspond to the ensemble averaged polarization dynamics of a single NV defect surrounded by a finite number of P1 centers ($N_{\mathrm{P1s}}=\{300,225, 75\}$ for $\nu = \{1/3,1/4,1/12\}$, in agreement with the defect density ratio in the sample) using a pumping rate $\Gamma_{\mathrm{p}} = 0.1~\mathrm{\mu s^{-1}}$.
    We observe excellent agreement for the P1 subgroups $\nu = \{1/3, 1/4\}$ with linewidth $\gamma = 0.5~\mathrm{\mu s^{-1}}$, for over 4 orders in magnitude in $\tau_p$ and for the entirety of the experimentally accessible timescales.
    For $\nu=1/12$, we observe agreement in a smaller range of $\tau_{\mathrm{p}}$ using $\gamma=1.5~\mathrm{\mu s^{-1}}$.
    }
    \label{fig:Sup_Agreement}
\end{figure}

\begin{figure}
    \includegraphics[width = 2.8in]{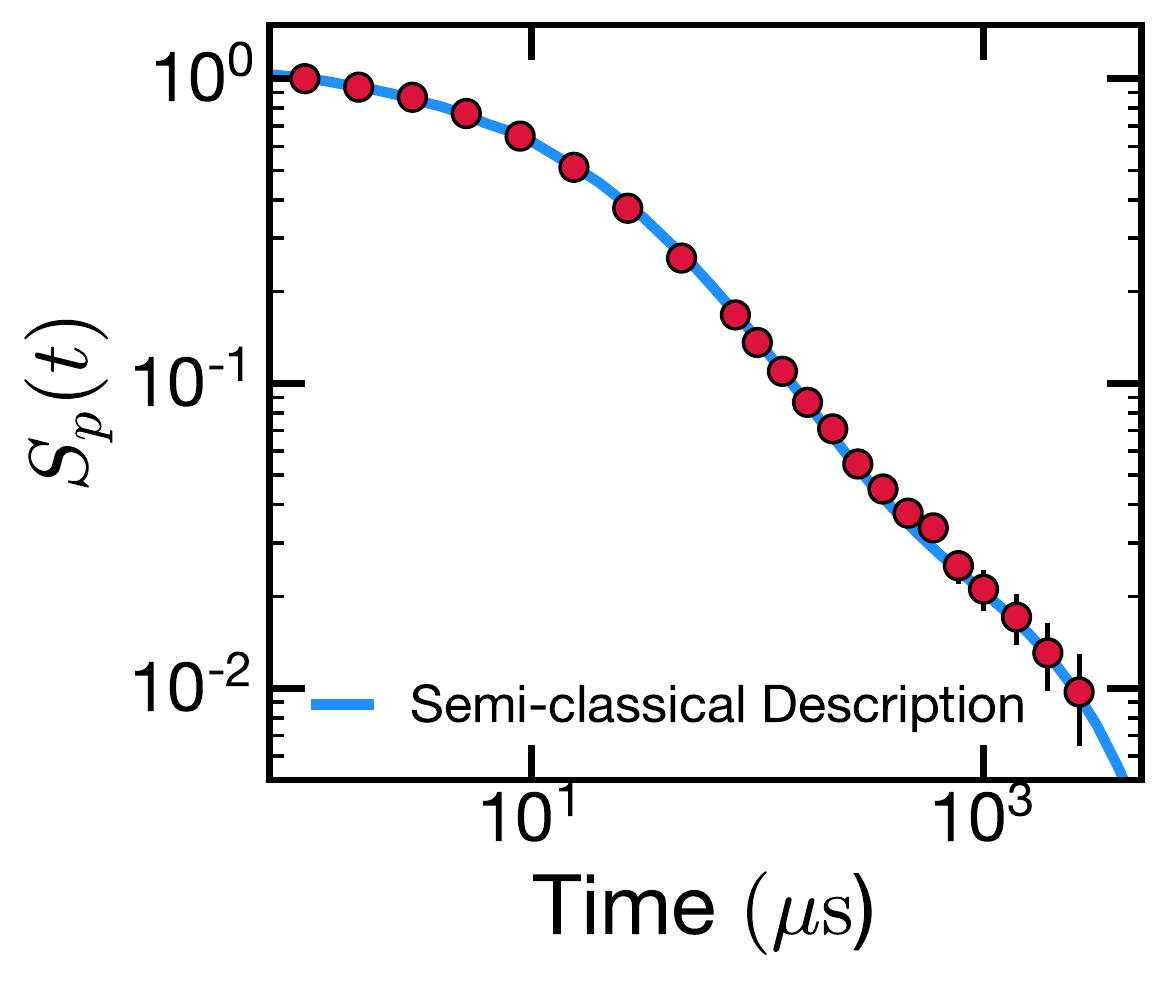}
    \caption{Depolarization dynamics for $\nu=1/3$ and $\tau_p = 30~\mathrm{\mu s}$ in sample S2 at $T=25~\mathrm{K}$.
    The semi-classical description accurately captures the entirety of the depolarization dynamics with $N_{\mathrm{P1s}}=300$, $\gamma = 0.3~\mathrm{\mu s^{-1}}$ and $\Gamma_{\mathrm{p}} = 0.25~\mathrm{\mu s^{-1}}$.
    }
    \label{fig:Sup_LTAgreement}
\end{figure}

\begin{figure}
    \centering
    \includegraphics[width=6.2in]{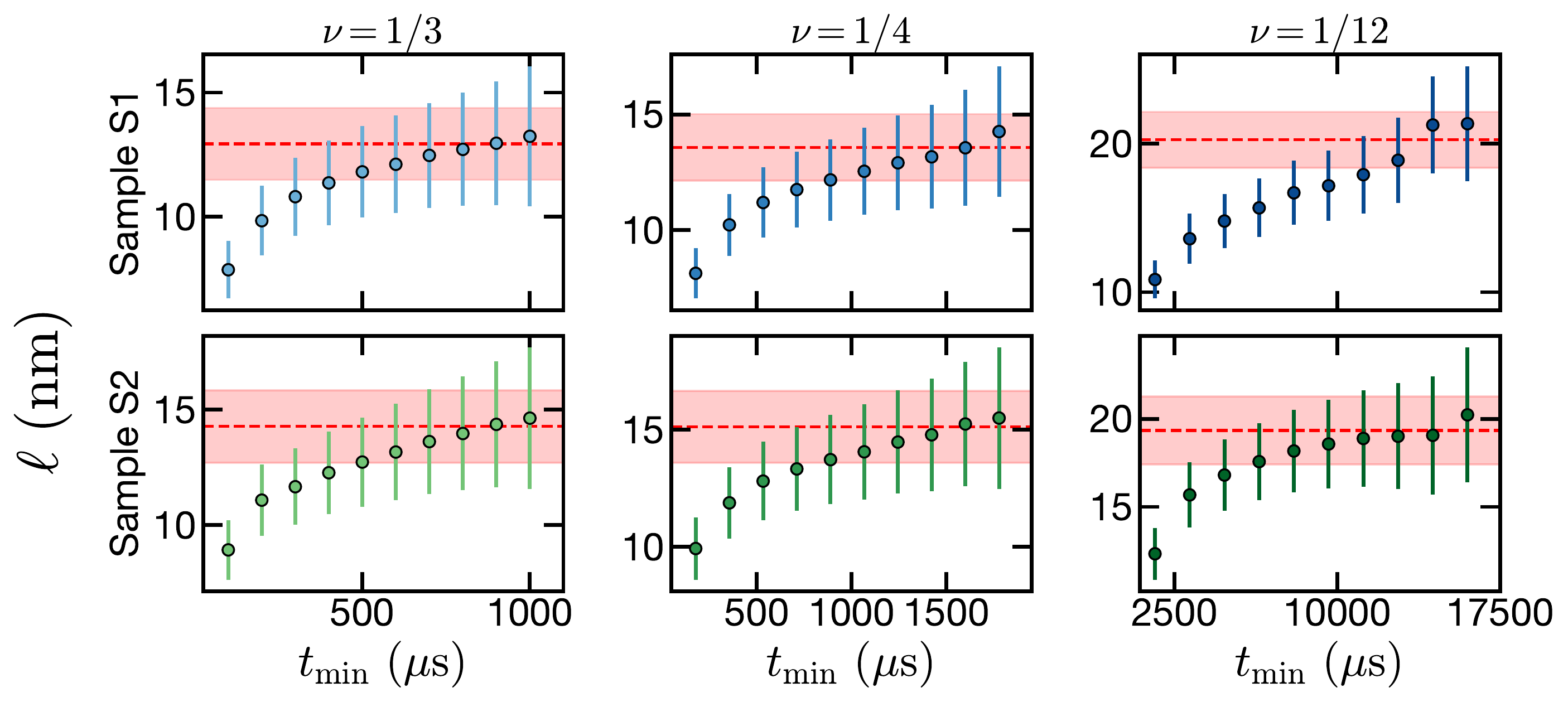}
    \caption{
    Extracted $\ell$ for different samples and different P1 groups as a function of the early time cut-off $t_\mathrm{min}$.
    Averaging over the last 3 data points, where the $\ell$ are consistent, yields the reported value of $\ell$---red dashed line corresponds to the final value while the shaded area the associated uncertainty.
    }
    \label{fig:ellExtraction}
\end{figure}

\begin{figure}
    \centering
    \includegraphics[width = 6.8in]{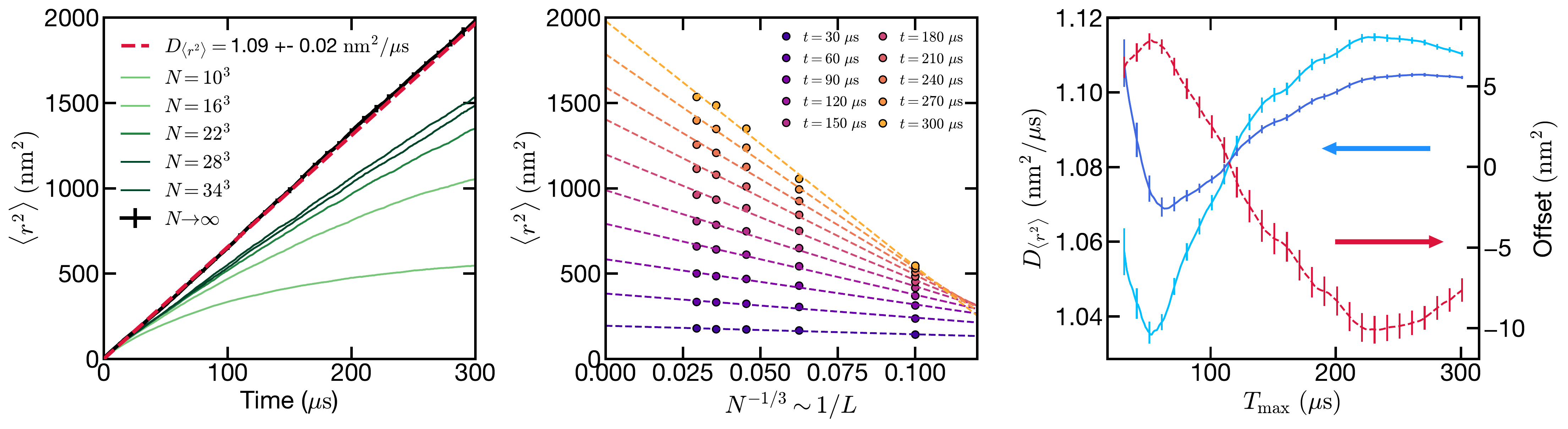}
    \caption{
    Summary of extraction of mean-square displacement and diffusion coefficient.
    (left) Growth of $\langle r^2\rangle$ for different system sizes $N$ and the infinite system scaling (black line). 
    (middle) Finite size scaling of $\langle r^2\rangle$ to $N\to \infty$ assuming a linear in $L^{-1} \sim N^{-1/3}$ correction for representative values of $t$.
    (right) Fitting the early-time growth of $\langle r^2\rangle$ up to different times $T_{\mathrm{max}} \in [30,300]$ leads to slightly different values of the diffusion coefficient, whether including a constant offset (light blue) or not (dark blue).
    Considering the fit without an offset, the final diffusion coefficient is taken to be the average with an uncertainty given by half the range of diffusion coefficients.
    }
    \label{fig:r2_finiteSizeScaling}
\end{figure}

\begin{figure}
    \centering
    \includegraphics[width = 7.2in]{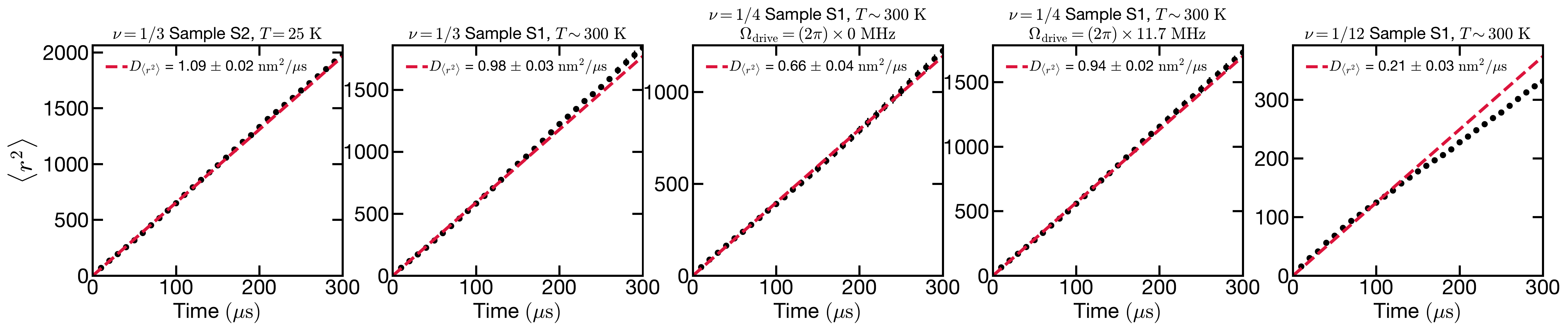}
    \caption{Using the semi-classical model to study the different experimental conditions using the parameters discussed in Methods.
    Computing the growth of $\langle r^2\rangle$ we extract the diffusion coefficient, which is in great agreement with the experimentally extracted values after correcting for the non-Gaussian polarization profile (Table I in main text). 
    }
    \label{fig:ext_Dr2}
\end{figure}

\begin{figure}
    \centering
    \includegraphics[width = 6in]{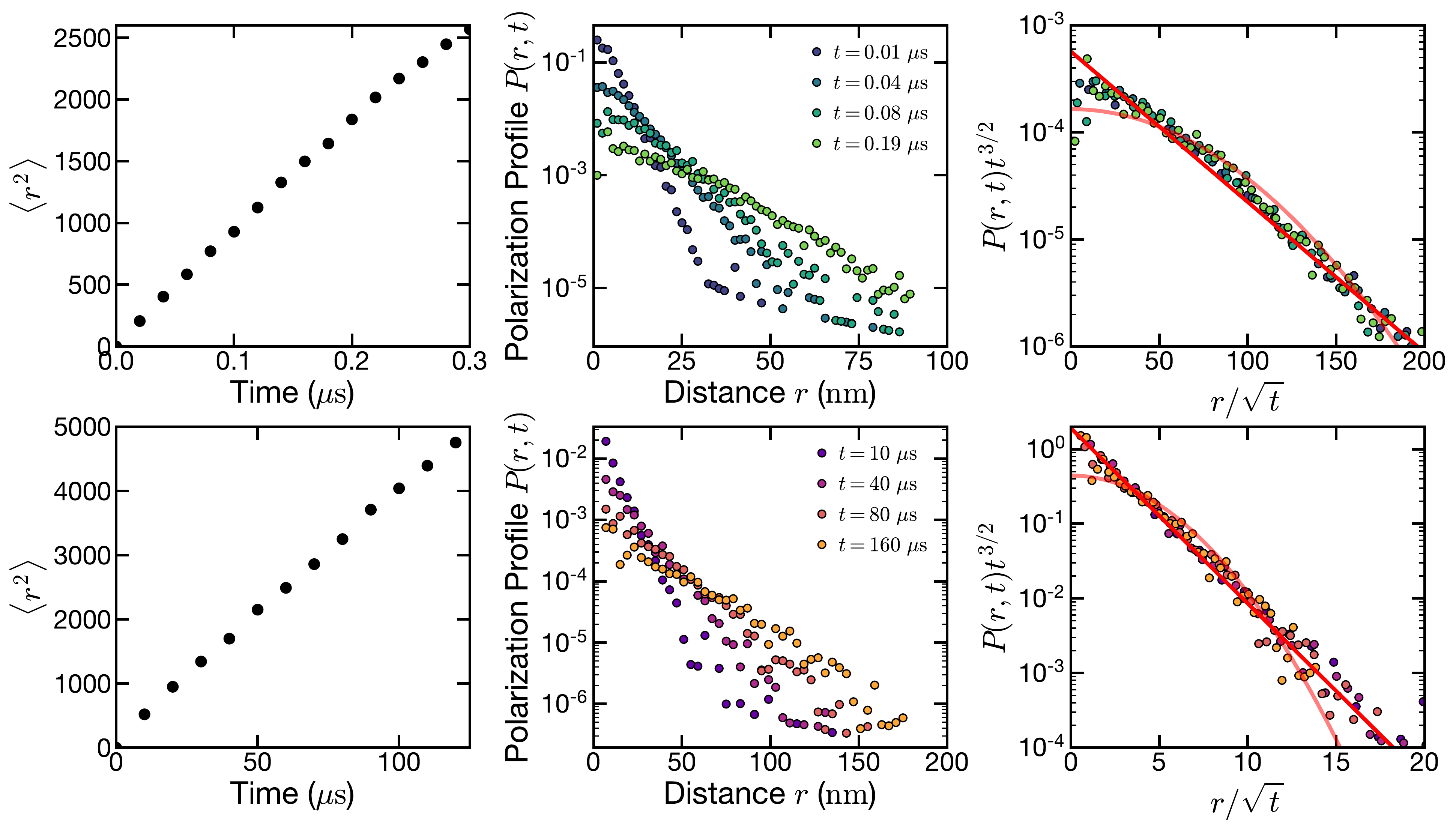}
    \caption{Diffusive dynamics of a long-range system $\alpha = 6$ for two types of disorder: in the top row we consider positional disorder without onsite fields, in the bottom row we consider random onsite fields with spins placed within a cubic lattice. 
    In either case of disorder we observe, an early time linear growth of the mean square displacement (left column) that highlights the diffusive nature of the dynamics, but the polarization profile remains non-Gaussian (middle column) throughout the evolution considered.
    This latter point is highlighted by rescaling the polarization profile $P(r,t)$ according to the diffusive dynamical exponent (right column), and observing that the collapse of different time traces agree with a simple exponential profile (Gaussian profile plotted for comparison).
    }
    \label{fig:DifWithDis}
\end{figure}

\begin{figure*}
    \centering
    \includegraphics[width = 2.8in]{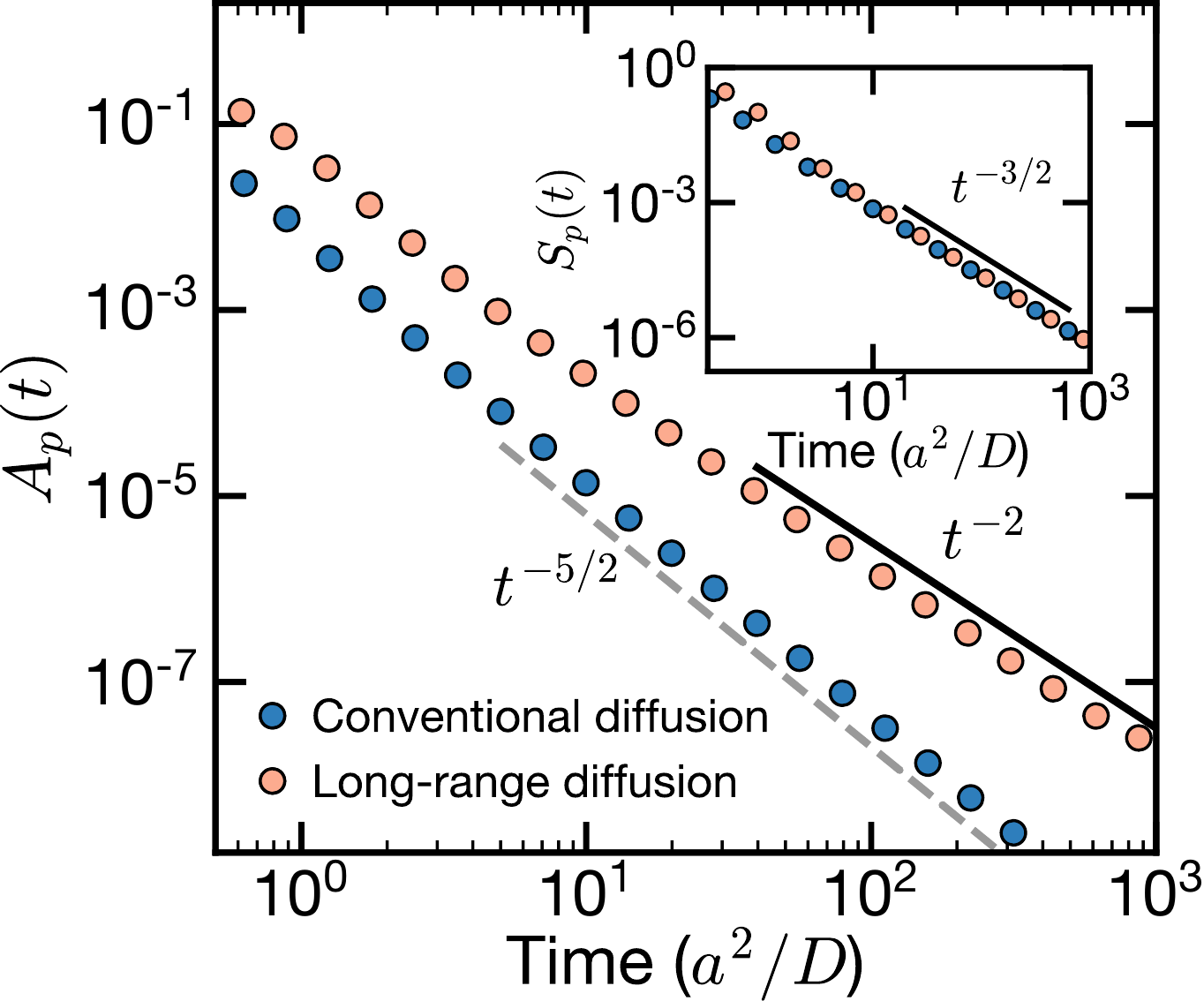}
    \caption{
    The presence of a long-range $k^3$-term parametrically modifies the approach, $A_p(t) = S_p(t)-(4\pi Dt)^{-3/2}$, to the late-time Gaussian fixed point, as high-lighted in a three-dimensional, disorder-less numerical simulation, with lattice constant $a$ and diffusion coefficient $D$).
    }
    \label{fig:FigS_LongRange}
\end{figure*}


\begin{figure*}
    \centering
    \includegraphics[width = 7in]{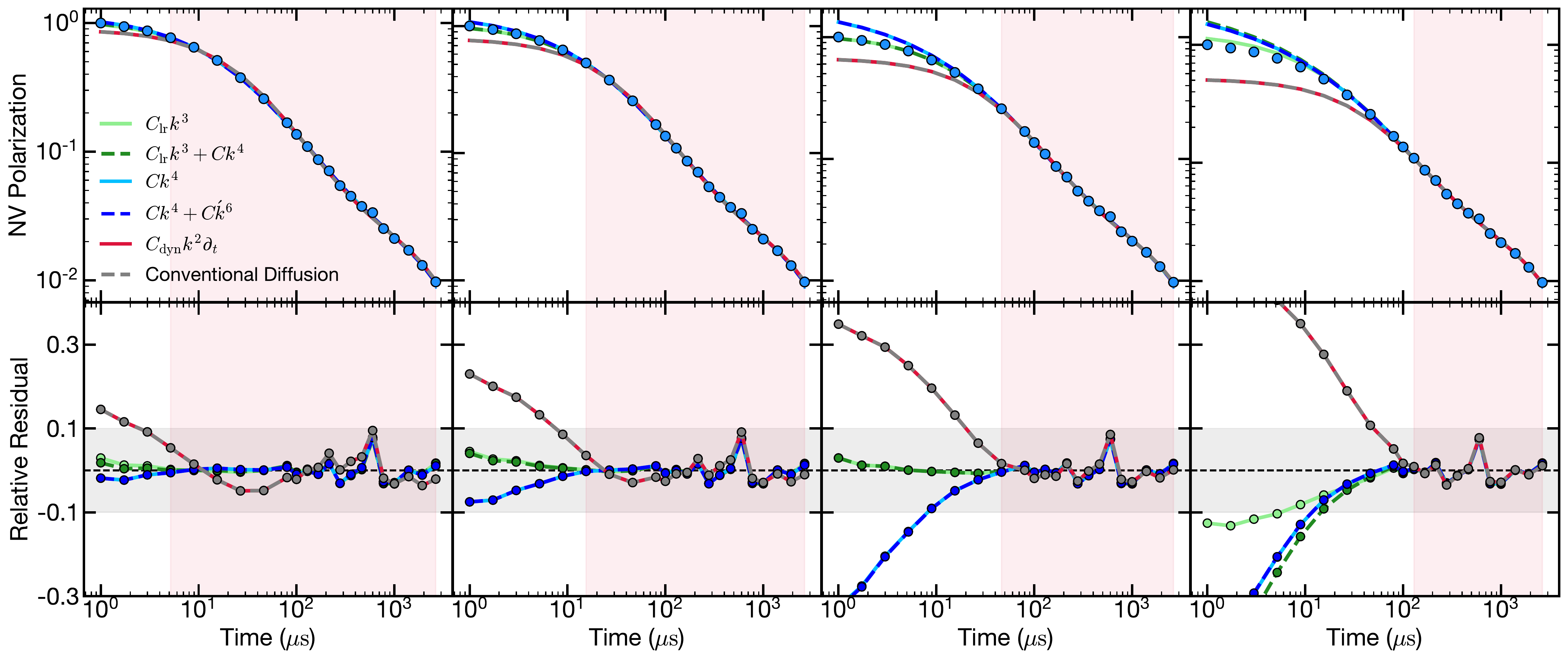}
    \caption{
        Fitting of the diffusive description with different terms and fixed $T_1 = 2.6~\mathrm{ms}$ in sample S2 with $\tau_\mathrm{p}=30~\mathrm{\mu s}$.
        Different columns represent fitting to a different range of the data (highlighted by the red shaded region).
        The inclusion of more terms in the diffusive description allows for a better fit of the data, however the improvement in the fitting range is only significant when the fitting regimes includes early time data ($\lesssim 30~\mathrm{\mu s}$), as highlighted in the second row of the relative residuals.
        All data are presented with logarithmically spaced y axis, except in the grey shaded region where a linear regime is used to highlight the fluctuations of the residuals around $0$.
        Fits in Fig.~1b correspond to the third column.
    }    
    \label{fig:ext_FitsfixedT1}
\end{figure*}